\documentclass[a4paper,fleqn,10pt]{article}
\pdfoutput=1
\usepackage{amsmath}
\usepackage{amssymb}
\usepackage{array}
\usepackage{calc}
\usepackage{longtable}
\usepackage{multirow}
\usepackage{pstricks}
\usepackage{graphicx}
\usepackage{xspace}
\usepackage{mciteplus}
\usepackage[a4paper,pdfborder={0 0 0}]{hyperref}
\usepackage[format=hang,labelfont=bf,hypcap=true,skip=0pt]{caption}
\usepackage{subfig}
\usepackage{sectsty}
\allsectionsfont{\sffamily}
\subsubsectionfont{\mdseries\itshape\large}
\let\spreprint\empty
\newcommand{\preprint}[1]{\def\spreprint{\protect#1}}
\let\sinstitute\empty
\newcommand{\institute}[1]{\def\sinstitute{\protect#1}}
\makeatletter
\renewcommand{\maketitle}{\begingroup
  \null\thispagestyle{empty}%
    \ifx\spreprint\empty
      \vskip 5ex
    \else
      \flushright\large\spreprint\vskip 2ex
    \fi
    \vskip 5ex
    \flushleft
      {\sffamily\bfseries\huge\@title}\vskip 2ex
      \@author\vskip 2ex
      \ifx\sinstitute\empty
      \else
        {\small\sinstitute}
      \fi
    \vskip 5ex
  \endgroup
}
\makeatother
\renewenvironment{abstract}{\begin{center}
  {\large\sffamily\bfseries Abstract: }
  \begin{minipage}[t]{0.75\textwidth}
}{\end{minipage}\end{center}\vskip 10ex}

\numberwithin{equation}{section}
\newcommand{\MCatNLO}{M\protect\scalebox{0.8}{C}@N\protect\scalebox{0.8}{LO}\xspace}
\newcommand{\MENLOPS}{ME\protect\scalebox{0.8}{NLO}PS\xspace}
\newcommand{\POWHEG}{P\protect\scalebox{0.8}{OWHEG}\xspace}

\newcommand{\Herwig}{H\protect\scalebox{0.8}{ERWIG}\xspace}

\newcommand{\Pythia}{P\protect\scalebox{0.8}{YTHIA}\xspace}

\newcommand{\BlackHat}{BlackHat\xspace}
\newcommand{\MCFM}{MCFM\xspace}
\newcommand{\Sherpa}{S\protect\scalebox{0.8}{HERPA}\xspace}

\newcommand{\Amegic}{A\protect\scalebox{0.8}{MEGIC++}\xspace}

\newcommand{\DO}{D\O\xspace}
\long\def\symbolfootnote[#1]#2{\begingroup%
\def\thefootnote{\fnsymbol{footnote}}\footnote[#1]{#2}\endgroup}
\newcommand{\abs}[1]{\left| #1\right|}
\newcommand{\rbr}[1]{\left( #1\right)}
\newcommand{\abr}[1]{\langle #1\rangle}
\newcommand{\cbr}[1]{\left\{ #1\right\}}
\newcommand{\sbr}[1]{\left[ #1\right]}
\newcommand{\im}{\imath}
\newcommand{\jm}{\jmath}
\newcommand{\args}[1]{\{\vec{#1}\}}

\newcommand{\done}{{\rm d}}
\newcommand{\order}{\mathcal{O}}
\newcommand{\mc}[1]{\mathcal{#1}}
\newcommand{\mr}[1]{\mathrm{#1}}
\newcommand{\dst}{\displaystyle}

\newcommand{\bmap}[3]{b_{#1,#2}(#3)}
\newcommand{\rmap}[3]{r_{\widetilde{#1},\tilde{#2}}(#3)}
\newcommand{\bea}{\begin{eqnarray}}
\newcommand{\eea}{\end{eqnarray}}
\newcommand{\bi}{\begin{itemize}}
\newcommand{\ei}{\end{itemize}}

\hypersetup{
  pdfauthor={Stefan Hoeche, Frank Krauss,
    Marek Schoenherr, Frank Siegert},
  pdftitle={NLO matrix elements and truncated showers},
  pdfkeywords={QCD, NLO, Matrix Element, Parton Shower,
    Truncated Shower, CKKW}
}
\preprint{ZU-TH 13/10\\IPPP/10/73\\DCPT/10/146\\CERN-PH-TH/2010-195\\
  MCNET/10/17}
\author{Stefan H{\"o}che$^1$, Frank Krauss$^{2,3}$,
  Marek Sch{\"o}nherr$^4$, Frank Siegert$^{2,5}$}
\title{NLO matrix elements and truncated showers}
\institute{$^1$ Institut f{\"u}r Theoretische Physik, 
  Universit{\"a}t Z{\"u}rich, CH-8057 Zurich, Switzerland\\
  $^2$ Institute for Particle Physics Phenomenology,
  Durham University, Durham DH1 3LE, UK\\
  $^3$ PH-TH, CERN, CH-1211 Geneva 23, Switzerland\\
  $^4$ Institut f{\"u}r Kern- und Teilchenphysik,
  Technische Universit{\"a}t Dresden, D-01062, Dresden, Germany\\
  $^5$ Department of Physics \& Astronomy,
  University College London, London WC13 6BT, UK\\}
\begin{document}
\maketitle
\begin{abstract}
  In this publication, an algorithm is presented that combines the ME+PS approach 
to merge sequences of tree-level matrix elements into inclusive event
samples~\cite{Hoeche:2009rj} with the \POWHEG method, which combines exact 
next-to-leading order matrix element results with the parton 
shower~\cite{Nason:2004rx,*Frixione:2007vw}. It was developed in parallel 
to the \MENLOPS technique discussed in~\cite{Hamilton:2010wh} and has been 
implemented in the event generator \Sherpa~\cite{Gleisberg:2003xi,*Gleisberg:2008ta}.
The benefits of this approach are exemplified by some first predictions for 
a number of processes, namely the production of jets in $e^+e^-$-annihilation,
in deep-inelastic $ep$ scattering, in association with single $W$, $Z$ 
or Higgs bosons, and with vector boson pairs at hadron colliders.

\end{abstract}
\section{Introduction}

In the past decade, the incorporation of higher-order corrections into 
parton-shower simulations has been in the centre of formal improvements 
of existing general-purpose Monte-Carlo event generators like
\Herwig~\cite{Corcella:2000bw,*Corcella:2002jc}, 
\Pythia~\cite{Sjostrand:1995iq,*Sjostrand:2006za}, and 
\Sherpa~\cite{Gleisberg:2003xi,*Gleisberg:2008ta}.  
As parton showers approximate higher-order matrix elements only in the soft 
and collinear limits of the real-emission phase space, they do inherently 
not yield a good prediction in the hard domain. This deficiency can lead 
to rather large discrepancies between Monte-Carlo predictions and experimental
data and should therefore be corrected. In this endeavour, two somewhat 
orthogonal approaches have been pursued.

To improve the description of hard QCD radiation, ``merging algorithms''
(ME+PS), have been proposed~\cite{Catani:2001cc,*Krauss:2002up}, which were 
shown to be correct up to next-to-leading logarithmic accuracy in 
$e^+e^-$-annihilation into hadrons.  
These methods were improved and extended in a series of 
publications~\cite{Lonnblad:2001iq,*Lavesson:2007uu,Hamilton:2009ne}.
A reformulation which generically maintains the logarithmic 
accuracy of the parton shower was achieved in~\cite{Hoeche:2009rj}\footnote{
  The algorithm has been further extended to include QED 
  effects~\cite{Hoeche:2009xc} and multi-scale problems~\cite{Carli:2010cg}.}, 
providing also the means to classify the various other methods 
and implementations according to their formal accuracy.  
It makes use of ``truncated'' parton showers, which are a concept initially 
introduced in~\cite{Nason:2004rx} in the context of the \POWHEG method.
An alternative merging technique, known as ``MLM merging'', was suggested
in~\cite{Mangano:2001xp,*Mangano:2006rw} and described in more detail 
and compared with the other algorithms in~\cite{Hoche:2006ph,*Alwall:2007fs}.

While the ME+PS methods succeed at improving the simulation of multijet 
events, they do not address the apparent problem that the total cross 
sections of the simulated inclusive samples are still of leading-order 
accuracy.  However, this accuracy often is insufficient for tests of physics 
within and searches beyond the Standard Model. Examples for such situations 
range from luminosity measurements at the LHC through the production of $W$ 
bosons to the determination of Yukawa couplings of the Higgs boson, once it 
is found.  There has been a number of proposals of how to include at least 
the full next-to-leading order results in the simulation, but only two 
of them have been fully worked out and implemented in publicly available 
programs.  

The first method, \MCatNLO, pioneered in~\cite{Frixione:2002ik}, and 
implemented for a large number of processes~\cite{Frixione:2003ei,*Frixione:2005vw,*Frixione:2008yi,*Weydert:2009vr,*Torrielli:2010aw} relies 
on using the parton-shower kernels and their universal behaviour in the soft and 
collinear limits to subtract the infrared divergences of real-emission 
contributions to the NLO cross section.  The parton shower then starts from either
a Born-like configuration or from a configuration determined by the residual 
real correction contribution of the NLO calculation.  By construction, there
is some dependence on the details of the actual shower algorithm, which, to
a certain extent, up to now seemed to limit the versatility of the method.  
This dependence was overcome by the second method, \POWHEG, which was 
initially presented in~\cite{Nason:2004rx,*Frixione:2007vw}.  This technique 
essentially is an improvement of an ingenious reweighting method, known for 
nearly two decades and applied individually to a plethora of 
processes~\cite{Seymour:1994df,*Seymour:1994we,*Corcella:1998rs,*Miu:1998ju,*Corcella:1999gs,*Norrbin:2000uu}.
To promote it to full NLO accuracy, it was supplemented with a 
local, phase-space dependent $K$-factor.  The \POWHEG method has been worked 
out for a number of processes~\cite{Nason:2006hfa,*Frixione:2007nw,*Frixione:2007nu,
  *Alioli:2008gx,*Alioli:2008tz,*Hamilton:2008pd,*Alioli:2009je,*Nason:2009ai,*Hamilton:2009za}, 
using different parton-shower algorithms.  A framework 
incorporating the core technology, independent of the specific 
parton-shower implementation and the matrix elements for the processes 
in question has been published in~\cite{Alioli:2010xd}.  It should be noted, 
though, that the \POWHEG method relies on truncated parton showers in order
to maintain the logarithmic accuracy in the radiation pattern, and it is this 
constraint, which introduces a parton-shower dependence.  This 
has been an open issue and it was discussed in some of the above works,
since older Fortran versions of \Pythia and \Herwig typically do not provide 
truncated parton showers.  Although it was typically
assumed that the omission of truncation does not lead to sizable effects,
the findings in~\cite{Hamilton:2009ne} suggest that this depends significantly
on the parton-shower model and the respective mismatch of its ordering parameter 
with the definition of the hardness scale.

Having at hand two, somewhat orthogonal, methods (ME+PS and \POWHEG) to 
improve both the hard QCD radiation activity {\em and} the total event rate 
in a given process, the question naturally arises whether it is possible to 
combine both into an even more powerful approach.  This is the topic 
addressed by this publication, resulting in a practical algorithm for merging 
both techniques. This algorithm has been implemented into the multi-purpose
event generator
\Sherpa~\cite{Gleisberg:2003xi,*Gleisberg:2008ta}, and will be made
publicly available in a future version of the program.  In a parallel 
development, Hamilton and Nason~\cite{Hamilton:2010wh} suggested an identical 
method; however, their actual implementation only approximates the formal
result. Due to the formal equivalence of both proposals, we will generally
refer to the new technique as the \MENLOPS approach.
For a detailed presentation of the method, the reader is referred to 
Sec.~\ref{Sec::MENLOPS}, while Sec.~\ref{sec:prerequisites} is used to introduce
the notation employed in this paper.  In Sec.~\ref{Sec::Results}, some first 
predictions of the \Sherpa implementation are exhibited, exemplifying the 
improvements that can be achieved.  These results and their interpretation 
solidify the findings of~\cite{Hamilton:2010wh}. Section~\ref{Sec::Conclusions}
contains our conclusions.

\section{Improving parton showers with higher-order matrix elements}
\label{sec:prerequisites}
This section is devoted to the introduction of a formalism that allows 
to describe, on the same footing, the two basic methods for correcting 
parton-shower algorithms with real-emission matrix elements.  A common notation 
will be established, which will later allow to merge the \POWHEG method 
as reformulated in~\cite{Hoeche:2010pf}, with the ME+PS method proposed 
in~\cite{Hoeche:2009rj}.  Like in~\cite{Hoeche:2010pf} the discussion is 
restricted to a dipole-like factorisation~\cite{Catani:1996vz,*Catani:2002hc}.
While the ME+PS approach allows to incorporate matrix-element information 
for arbitrary final states, in the \POWHEG method only the first emission 
is fully corrected.
To compare, and, ultimately, to combine both algorithms, only the expressions 
for the differential cross section describing the first emission off a given 
core interaction must therefore be worked out. This is where the combination 
takes place.

Denoting sets of $n$ particles in a $2\to (n-2)$ process by 
$\args{a}=\{a_1,\ldots,a_n\}$, while their respective flavours and momenta
are specified separately through $\args{f}=\{f_1,\ldots,f_n\}$ and
$\args{p}=\{p_1,\ldots,p_n\}$, the generic expression for a fully differential 
Born-level cross section can be written as a sum over all contributing 
flavour combinations:
\begin{equation}
  \done\sigma_B(\args{p})\,=\;
  \sum_{\args{f\,}}\done\sigma_B(\args{a})\;,
  \qquad\text{where}\qquad
  \done\sigma_B(\args{a})\,=\;\done\Phi_B(\args{p})\,\mr{B}(\args{a})\;,
\end{equation}
The individual terms in the sum are given by
\begin{equation}
  \begin{split}
  \mr{B}(\args{a})\,=&\;\mc{L}(\args{a})\,\mc{B}(\args{a})\;,
  &\mc{B}(\args{a})\,=&\;
  \frac{1}{F(\args{a})}\,\frac{1}{S(\args{f\,})}\,
  \abs{\mc{M}_B}^2(\args{a})\;,\\
  \done\Phi_B(\args{p})\,=&\;\frac{\done x_1}{x_1}\frac{\done x_2}{x_2}\,
  \done{\it\Phi}_B(\args{p})\;,
  &\mc{L}(\args{a};\mu^2)\,=&\;x_1f_{f_1}(x_1,\mu^2)\;x_2f_{f_2}(x_2,\mu^2)\;,
  \end{split}
\end{equation}
where $\abs{\mc{M}_B}^2(\args{a})$ denotes the partonic matrix element 
squared, $\done{\it\Phi}_B(\args{a})$ is the corresponding differential 
$n$-particle partonic phase-space element, $S(\args{f})$ is the symmetry 
factor, $F(\args{a})$ is the flux factor, and $\mc{L}$ is the parton 
luminosity.  In the case of leptonic initial states, and ignoring QED 
initial-state radiation, the parton distribution functions $f(x,\mu^2)$ are 
replaced by $\delta(1-x)$.

In a similar fashion, the real-emission part of the QCD next-to-leading order
cross section can be written as a sum, depending on parton configurations
$\{a_1,\ldots,a_{n+1}\}$, by replacing the Born-level matrix elements $\mc{B}$ 
with the real-emission matrix elements $\mc{R}$ and the Born-level phase space
$\done\Phi_B$ with the real-emission phase-space $\done\Phi_R$.

Furthermore, it is useful to introduce a notation for mappings from 
real-emission parton configurations to Born-level parton configurations and 
vice versa.  They are given by (cf.~\cite{Hoeche:2010pf})
\begin{align}\label{eq:parton_maps}
  \bmap{ij}{k}{\args{a}}\,=&\;\left\{\begin{array}{c}
    \args{f\,}\setminus\{f_i,f_j\}\cup\{f_{\widetilde{\im\jm}}\}\\
    \args{p\,}\to\args{\tilde{p\;}}
    \end{array}\right.
  &&\text{and}
  &\rmap{\im\jm}{k}{f_i,\Phi_{R|B}\,;\args{a}}\,=&\;
  \left\{\begin{array}{c}
    \args{f\,}\setminus\{f_{\widetilde{\im\jm}}\}\cup\{f_i,f_j\}\\
    \args{\tilde{p\;}}\to\args{p\,}
    \end{array}\right.\;.
\end{align}
The map $\bmap{ij}{k}{a}$ combines the partons $a_i$ and $a_j$ into a 
common ``mother'' parton $a_{\widetilde{\im\jm}}$, in the presence of the 
spectator $a_k$ by defining a new flavour $f_{\widetilde{\im\jm}}$ and by 
redefining the particle momenta. The inverse map, $\rmap{\im\jm}{k}{a}$
determines the parton configuration of a real-emission subprocess from
a Born parton configuration and a related branching process 
$\widetilde{\im\jm},\tilde{k}\to ij,k$. The radiative variables $\Phi_{R|B}$ 
are thereby employed to turn the $n$-parton momentum configuration into an 
$(n+1)$-parton momentum configuration.

Following~\cite{Hoeche:2010pf}, the real-emission matrix elements, 
$\mc{R}(\args{a})$, can be decomposed into a number of terms $\mc{R}_{ij,k}$ as
\begin{equation}\label{eq:def_rho}
  \mc{R}_{ij,k}(\args{a})\,:=\;\rho_{ij,k}(\args{a})\,\mc{R}(\args{a})\;,
  \quad\text{where}\quad
  \rho_{ij,k}(\args{a})\,=\;\frac{\mc{D}_{ij,k}(\args{a})}{
    \sum_{\{m,n\}}\sum_{l\neq m,n}\mc{D}_{mn,l}(\args{a})}\;,
\end{equation}
where $\mc{D}_{ij,k}$ are the dipole terms defined ibidem.
Therefore, the real-emission differential cross section can be rewritten 
as a sum of trivially factorised contributions
\begin{equation}\label{eq:split_real_xs}
  \done\sigma_R(\args{a})\,=\;\sum_{\{i,j\}}\sum_{k\neq i,j}
    \done\sigma_B(\bmap{ij}{k}{\args{a}})\,
    \done\sigma_{R|B}^{ij,k}(\args{a})\;,
\end{equation}
where
\begin{equation}\label{eq:def_real_dipole}
  \done\sigma_{R|B}^{ij,k}(\args{a})\,=\;
    \done\Phi_{R|B}^{ij,k}(\args{p})\,
    \frac{\mr{R}_{ij,k}(\args{a})}{\mr{B}(\bmap{ij}{k}{\args{a}})}\;.
\end{equation}
and where $\mr{R}_{ij,k}=\mc{L}\,\mc{R}_{ij,k}$, cf.~\cite{Hoeche:2010pf}.
The no-branching probability of a parton-shower algorithm
is derived, based on Eq.~\eqref{eq:def_real_dipole} and the decomposition of 
the dipole terms $\mc{D}_{ij,k}$ in the parton-shower approximation. It reads
\begin{equation}\label{eq:nbp_ps}
  \Delta^{\rm(PS)}(t',t'';\args{a})\,=\;\prod_{\{\widetilde{\im\jm},\tilde{k}\}}
  \Delta^{\rm(PS)}_{\widetilde{\im\jm},\tilde{k}}(t',t'';\args{a})\;,
\end{equation}
where the dipole-dependent no-branching probabilities are given by\footnote{
  Note that, if initial-state partons are involved in the splitting,
  $\Delta^{\rm(PS)}_{\widetilde{\im\jm},\tilde{k}}$ is a constrained no-branching 
  probability, as $x_{1/2}\leq 1$ before and after the branching process.
}
\begin{equation}\label{eq:nbp_ps_ijk}
  \begin{split}
  \Delta^{\rm(PS)}_{\widetilde{\im\jm},\tilde{k}}(t',t'';\args{a})\,=&\;
  \exp\left\{-\sum_{f_i=q,g}
    \int_{t'}^{t''}\frac{\done t}{t}\,\int_{z_{\rm min}}^{z_{\rm max}}\done z\,
    \int_0^{2\pi}\frac{\done\phi}{2\pi}\,J_{ij,k}(t,z,\phi)\,
  \right.\\&\qquad\qquad\qquad\times\left.
    \frac{1}{S_{ij}}\,\frac{\alpha_s}{2\pi}\,
    \mc{K}_{ij,k}(t,z,\phi)\,
    \frac{\mc{L}(\rmap{\im\jm}{k}{f_i,t,z,\phi;\args{a}};t)}
         {\mc{L}(\args{a\,};t)}\,\right\}\;.
  \end{split}
\end{equation}
The quantities $\mc{K}_{ij,k}$ are the parton-shower evolution kernels, 
which depend on the parton flavours $f_i$, $f_j$ and $f_k$ and on the radiative 
phase space $\Phi_{R|B}^{ij,k}=\{t,z,\phi\}$, cf.~\cite{Hoeche:2010pf}. 
The variable $t$, with $t\varpropto2\,p_ip_j$, is the evolution parameter
while $z$ is called the splitting variable of the parton-shower model and
$J_{ij,k}$ is the Jacobian factor associated with the transformation of 
integration variables.

\subsection{The \texorpdfstring{\POWHEG}{POWHEG} approach}
\label{Sec::PowHEG}
Using a simple corrective weight, it is possible to modify the parton-shower 
such that it produces the $\order(\alpha_s)$ radiation pattern of the 
matrix element~\cite{Hoeche:2010pf}. The corresponding dipole-dependent
no-branching probability reads
\begin{equation}\label{eq:nbp_mecorr_ijk}
  \begin{split}
  \Delta^{\rm(ME)}_{\widetilde{\im\jm},\tilde{k}}(t',t'';\args{a})\,=&\;
    \exp\left\{-\sum_{f_i=q,g}
    \frac{1}{16\pi^2}\,
    \int_{t'}^{t''}\done t\,\int_{z_{\rm min}}^{z_{\rm max}}\done z\,
    \int_0^{2\pi}\frac{\done\phi}{2\pi}\,J_{ij,k}(t,z,\phi)
    \right.\\&\qquad\qquad\qquad\times\left.
    \frac{1}{S_{ij}}\,
    \frac{S(\rmap{\im\jm}{k}{f_i;\args{f\,}})}{S(\args{f\,})}\,
    \frac{\mr{R}_{ij,k}(\rmap{\im\jm}{k}{f_i,t,z,\phi;\args{a}})}{
      \mr{B}(\args{a})}\,\right\}\;,
  \end{split}
\end{equation}
where the actual matrix elements $\mr{R}$ and $\mr{B}$, which also include
the respective parton luminosity factors, replace the parton shower kernel
$\mc{K}$ and the parton luminosities of the equation above, 
Eq.~\eqref{eq:nbp_ps_ijk}.

The key point of the \POWHEG method is, to supplement Monte-Carlo event samples
from such matrix-element corrected parton showers with an approximate 
next-to-leading order weight to arrive at NLO accuracy. This is achieved by 
multiplying Eq.~\eqref{eq:nbp_mecorr_ijk} with the local $K$-factor 
$\bar{\mr{B}}(\args{a})/\mr{B}(\args{a})$, where
\begin{equation}\label{eq:def_bbar}
  \begin{split}
  \bar{\mr{B}}(\args{a})\,=&\;
    \mr{B}(\args{a})+\tilde{\mr{V}}(\args{a})+\mr{I}(\args{a})\\
  &\quad+
    \sum_{\{\widetilde{\im\jm},\tilde{k}\}}
    \sum_{f_i=q,g}
    \int\done\Phi_{R|B}^{ij,k}
  \sbr{\vphantom{\int}\,\mr{R}_{ij,k}(\rmap{\im\jm}{k}{\args{a}})
    -\mr{S}_{ij,k}(\rmap{\im\jm}{k}{\args{a}})\,}\;.
  \end{split}
\end{equation}
In this expression, $\tilde{\mr{V}}(\args{a})$ is the NLO virtual contribution, 
including the collinear mass factorisation counterterms, while $\mr{S}_{ij,k}(\rmap{\im\jm}{k}{a})$
and $\mr{I}(\args{a})$ denote real and integrated subtraction terms, 
respectively.

This yields the following master formula for the value of an infrared and 
collinear safe observable, $O$
\begin{equation}\label{eq:master_powheg}
  \begin{split}
  \abr{O}^{\rm (POWHEG)}\,=&\;\sum_{\args{f\,}}\int\done\Phi_B(\args{p})\,
  \bar{\mr{B}}(\args{a})\left[\,\vphantom{\int_A^B}
    \underbrace{\Delta^{\rm(ME)}(t_0,\mu^2;\args{a})}_{\text{unresolved}}\,
    O(\args{a})\right.\\
  &\qquad+\,
    \sum_{\{\widetilde{\im\jm},\tilde{k}\}}
    \sum_{f_i=q,g}\frac{1}{16\pi^2}
    \int_{t_0}^{\mu^2}\done t\,\int_{z_{\rm min}}^{z_{\rm max}}\done z\,
    \int_0^{2\pi}\frac{\done\phi}{2\pi}\,J_{ij,k}(t,z,\phi)\;
    O(\rmap{\im\jm}{k}{\args{a}})\\
  &\qquad\qquad\times\,
    \underbrace{\frac{1}{S_{ij}}\,
    \frac{S(\rmap{\im\jm}{k}{\args{f\,}})}{S(\args{f\,})}\,
    \frac{\mr{R}_{ij,k}(\rmap{\im\jm}{k}{\args{a}})}{
      \mr{B}(\args{a})}\;
    \Delta^{\rm(ME)}(t,\mu^2;\args{a})}_{\text{resolved}}\,
    \left.\vphantom{\Bigg)_{\int}^{\int}}\right]\;.
  \end{split}
\end{equation}
Note that the second term in the square bracket describes resolved emissions, 
simulated by the matrix-element corrected parton shower, while the first term
incorporates unresolved emissions and virtual corrections.

To reveal the fixed order properties of \eqref{eq:master_powheg} it is useful 
to inspect its expansion keeping terms up to $\order(\alpha_s)$.
\begin{equation}\label{eq:powheg_expansion}
  \begin{split}
&\abr{O}^{\rm (POWHEG)}
\,=\;\sum_{\args{f\,}}\int\done\Phi_B(\args{p})\,
    \left(\mr{B}+\tilde{\mr{V}}+\mr{I}\right)(\args{a})\;O(\args{a})\\
  &\qquad+\sum_{\args{f\,}}
    \sum_{\{\widetilde{\im\jm},\tilde{k}\}}
    \sum_{f_i=q,g}
    \int\done\Phi_B(\args{p})\,\done\Phi_{R|B}^{ij,k}
  \sbr{\vphantom{\int}\,\mr{R}_{ij,k}(\rmap{\im\jm}{k}{\args{a}})\,\Theta(t_0\!-t)
    -\mr{S}_{ij,k}(\rmap{\im\jm}{k}{\args{a}})\,}O(\args{a})\\
  &\qquad+\sum_{\args{f\,}}
    \sum_{\{\widetilde{\im\jm},\tilde{k}\}}
    \sum_{f_i=q,g}
    \int\done\Phi_B(\args{p})\,\done\Phi_{R|B}^{ij,k}\;
    \mr{R}_{ij,k}(\rmap{\im\jm}{k}{\args{a}})\,\Theta(t-t_0)\; O(\rmap{\im\jm}{k}{\args{a}})
      \;+\;\order(\alpha_s^2)\;.
  \end{split}
\end{equation}
It is imperative to note that neither the presence nor the precise 
form of $\Delta^{\rm(ME)}$ in the resolved emission term influence the fixed-order
accuracy of the method at $\order(\alpha_s)$. Similarily, different choices of 
scales at which $\alpha_s$ is evaluated in $\mr{R}$, the $\mr{\bar{B}}$-function, 
and the matrix-element corrected parton shower emission terms contribute at 
$\order(\alpha_s^2)$ because 
$\alpha_s(\mu_1)=\alpha_s(\mu_2)\left(1+\order(\alpha_s)\right)$~\cite{Nason:2004rx,*Frixione:2007vw}.

In order to obtain the correct leading logarithmic behaviour of the real-emission term, 
$\alpha_s$ and the parton luminosities in $\Delta^{\rm(ME)}$ and the resolved emission
term of Eq.~\eqref{eq:master_powheg} must be evaluated at scale $k_T^2$.
NLL accuracy can be restored for processes with no more than three coloured partons
by means of the replacement~\cite{Catani:1989ne,*Catani:1990rr}
\begin{equation}
  \alpha_s\to\alpha_s\cbr{\,1+\frac{\alpha_s}{2\pi}\sbr{
    \rbr{\frac{67}{18}-\frac{\pi^2}{6}}C_A-\frac{5}{9}n_f}\,}\;,
\end{equation}
where the $\overline{\rm MS}$ expression of $\alpha_s$ should be used.
In our Monte-Carlo implementation of the \POWHEG method, we follow this approach.

\subsection{The ME+PS approach}
\label{Sec::meps}
The ME+PS approach for the inclusion of matrix-element corrections into the
parton-shower relies on a twofold generation of radiative corrections:
Through an ordinary parton shower on the one hand and through real-radiation 
tree-level matrix elements on the other hand. In contrast to the \POWHEG
method, which only corrects the first emission off the core interaction, 
the ME+PS technique can be employed for arbitrary higher-order tree-level configurations.

A first algorithm to achieve this was presented in~\cite{Catani:2001cc,*Krauss:2002up}.  
The solution there is based on separating the radiative phase space into a 
region of soft/collinear emissions, the parton-shower (PS) region, and a 
region of hard emissions, the matrix-element (ME) region.  By demanding each 
region to be filled by the respective way of generating radiation and some
reweighting double counting and other problems can be avoided.  The
original formulation has been tremendously improved in~\cite{Hoeche:2009rj},
by realising that, formally, separate splitting kernels in the ME and PS regions
can be defined, which add up to the full splitting kernel:
\begin{align}\label{eq:kernel_meps}
  \mc{K}^{\rm ME}_{ij,k}(t,z,\phi)=&\;\mc{K}_{ij,k}(t,z,\phi)\;
    \Theta\rbr{\vphantom{\sum}Q_{ij,k}(t,z,\phi)-Q_{\rm cut}}\\
  \mc{K}^{\rm PS}_{ij,k}(t,z,\phi)=&\;\mc{K}_{ij,k}(t,z,\phi)\;
    \Theta\rbr{\vphantom{\sum}Q_{\rm cut}-Q_{ij,k}(t,z,\phi)}\;.
\end{align}
The functional form of the separation criterion $Q_{ij,k}$ is in principle 
arbitrary as long as it identifies soft and collinear divergences in the 
real-radiation matrix elements.  The approach, fully outlined 
in~\cite{Hoeche:2009rj}, then replaces the splitting kernels in 
the ME region by the ratio of the real-emission and Born matrix elements, 
just like this is done in a matrix-element corrected parton shower.
However, in contrast to a reweighting technique, only emission terms 
are modified and no correction is applied to the no-emission probabilities. 

The ME+PS technique can be implemented in a master formula for the first emission,
describing the expectation value of an arbitrary infrared safe observable $O$, 
similar to the \POWHEG case:
\begin{equation}\label{eq:master_meps}
  \begin{split}
  &\abr{O}^{\rm (MEPS)}\,=\;\sum_{\args{f\,}}\int\done\Phi_B(\args{p})\,
  \mr{B}(\args{a})\left[\,\vphantom{\int_A^B}\underbrace{
    \Delta^{\rm(PS)}(t_0,\mu^2;\args{a})}_{\text{unresolved}}\,
    O(\args{a})\right.\\
  &\quad+\,
    \sum_{\{\widetilde{\im\jm},\tilde{k}\}}
    \sum_{f_i=q,g}\frac{1}{16\pi^2}
    \int_{t_0}^{\mu^2}\done t\,\int_{z_{\rm min}}^{z_{\rm max}}\done z\,
    \int_0^{2\pi}\frac{\done\phi}{2\pi}\,J_{ij,k}(t,z,\phi)\;
    O(\rmap{\im\jm}{k}{\args{a}})\\
  &\qquad\quad\times\,
    \frac{1}{S_{ij}}\,\Bigg(\,
    \underbrace{\frac{8\pi\,\alpha_s}{t}\;\mc{K}_{ij,k}(t,z,\phi)\,
    \frac{\mc{L}(\rmap{\im\jm}{k}{f_i,t,z,\phi;\args{a}};t)}
         {\mc{L}(\args{a\,};t)}\;
    \Theta\rbr{\vphantom{\sum}Q_{\rm cut}
      -Q_{ij,k}(t,z,\phi)}}_{\text{resolved, PS domain}}\\
  &\qquad\qquad\qquad\quad+
  \underbrace{\frac{S(\rmap{\im\jm}{k}{\args{f\,}})}{S(\args{f\,})}\,
    \frac{\mr{R}_{ij,k}(\rmap{\im\jm}{k}{\args{a}})}{\mr{B}(\args{a})}\;
    \Theta\rbr{\vphantom{\sum}Q_{ij,k}(t,z,\phi)
     -Q_{\rm cut}}}_{\text{resolved, ME domain}}\,\Bigg)\;
    \Delta^{\rm(PS)}(t,\mu^2;\args{a})
  \left.\vphantom{\Bigg)_{\int}^{\int}}\right]\;.
  \end{split}
\end{equation}
There are three components to the differential cross section: The term 
describing unresolved emissions, which is generated 
in the standard parton-shower approach, and the resolved part, which is now split 
between the PS and the ME domain.  Within the ME domain, the matrix-element 
generator is directly invoked to define the real-emission configuration. 
This is possible due to the restricted phase space, removing all infrared 
divergent regions by applying the cut in $Q_{ij,k}$ and rendering the matrix
element finite.  In this case, the Sudakov form factor $\Delta^{\rm (PS)}$, 
which makes the matrix element exclusive, must be added explicitly.  It can 
either be calculated analytically, like in the original formulation 
of~\cite{Catani:2001cc,*Krauss:2002up}, or by utilising the shower itself to 
generate the correct probabilities.  This latter option is commonly referred 
to as the pseudoshower approach~\cite{Lonnblad:2001iq,Hoeche:2009rj}.

A complication arises if the phase-space separation criterion $Q_{ij,k}$ is 
different from the parton-shower evolution variable $t$.  This can imply 
the possibility of a shower emission $Q<Q_{\mr{cut}}$ being allowed ``between'' 
two branchings at $Q>Q_{\mr{cut}}$ in the parton-shower history of the 
matrix element.  In such cases, in order not to spoil the logarithmic accuracy 
of the parton shower, the existing branchings need to be embedded into the 
subsequent parton-shower evolution.  This leads to a truncated shower 
algorithm~\cite{Nason:2004rx,Hoeche:2009rj}.

If the expectation value $\abr{O}$ in Eq.~\eqref{eq:master_meps} is the total
cross section, i.e.\ if $O=1$, we can write
\begin{equation}\label{eq:meps_unitarity}
  \begin{split}
  &\sigma^{\rm (MEPS)}\,=\;\sum_{\args{f\,}}\int\done\Phi_B(\args{p})\,
  \mr{B}(\args{a})\left[\,\vphantom{\int_A^B}\underbrace{
    \Delta^{\rm(PS)}(t_0,\mu^2;\args{a})}_{\text{unresolved}}\right.\\
  &\quad+\,
    \sum_{\{\widetilde{\im\jm},\tilde{k}\}}
    \sum_{f_i=q,g}\frac{1}{16\pi^2}
    \int_{t_0}^{\mu^2}\done t\,\int_{z_{\rm min}}^{z_{\rm max}}\done z\,
    \int_0^{2\pi}\frac{\done\phi}{2\pi}\,J_{ij,k}(t,z,\phi)\;
    O(\rmap{\im\jm}{k}{\args{a}})\\
  &\qquad\quad\times\,
    \frac{1}{S_{ij}}\,\frac{8\pi\,\alpha_s}{t}\;\mc{K}_{ij,k}(t,z,\phi)\,
    \frac{\mc{L}(\rmap{\im\jm}{k}{f_i,t,z,\phi;\args{a}};t)}
         {\mc{L}(\args{a\,};t)}\;
  \Delta^{\rm(PS)}(t,\mu^2;\args{a})\\
  &\qquad\qquad\quad\times\,\Bigg(\,\underbrace{
    \Theta\rbr{\vphantom{\sum}Q_{\rm cut}
      -Q_{ij,k}(t,z,\phi)}}_{\text{resolved, PS domain}}
  +\underbrace{w(\rmap{\im\jm}{k}{f_i,t,z,\phi;\args{a}})\,
    \Theta\rbr{\vphantom{\sum}Q_{ij,k}(t,z,\phi)
     -Q_{\rm cut}}}_{\text{resolved, ME domain}}\,\Bigg)\;
  \left.\vphantom{\Bigg)_{\int}^{\int}}\right]\;.
  \end{split}
\end{equation}
where (cf.~\cite{Hoeche:2010pf})
\begin{equation}\label{eq:me_correction}
  w(\args{a})\,=\;\sbr{\,\dst\sum_{\{m,n\}}\sum_{l\neq m,n}
    \frac{S(\bmap{mn}{l}{\args{f\,}})}{S(\args{f\,})}\,
    \frac{\mc{B}(\bmap{mn}{l}{\args{a}})}{\mc{R}(\args{a})}\;
    \frac{8\pi\,\alpha_s}{2\,p_mp_n}\,
    \mc{K}_{mn,l}(\args{a})\,}^{-1}\;.
\end{equation}
If $w(\args{a})$ equals one, i.e.\ if the parton shower approximation
is equal to the real-radiation matrix element, the $t$-integral can be performed 
easily and the square bracket in Eq.~\eqref{eq:meps_unitarity} equals one. The 
total cross section is therefore identical to the leading order result, as
a direct consequence of the unitarity constraint for the parton shower. 
The more common configuration will however be, that $w(\args{a})\ne 1$. 
In this case, the square bracket in Eq.~\eqref{eq:meps_unitarity} is different
from one and the total cross section is not equal to the leading order result.
The origin of the mismatch is a difference in the emission rate of the 
parton shower compared to the ratio $R/B$ of matrix elements. While the former 
is exponentiated into the Sudakov form factor, the latter appears in the 
differential real-radiation probability only. We refer to this effect as an 
``emission-rate difference'' in the following.

To investigate its consequences on arbitrary observables, 
Eq.~\eqref{eq:master_meps} can be expanded in powers of $\alpha_s$, 
resulting in 
\begin{equation}\label{eq:meps_expansion}
  \begin{split}
  &\abr{O}^{\rm (MEPS)}\,=\;\sum_{\args{f\,}}\int\done\Phi_B(\args{p})\;
    \mr{B}(\args{a})\;O(\args{a})\\
  &\quad+\,
    \sum_{\args{f\,}}
    \sum_{\{\widetilde{\im\jm},\tilde{k}\}}
    \sum_{f_i=q,g}\frac{1}{16\pi^2}\int\done\Phi_B(\args{p})
    \int_{t_0}^{\mu^2}\done t\,\int_{z_{\rm min}}^{z_{\rm max}}\done z\,
    \int_0^{2\pi}\frac{\done\phi}{2\pi}\,J_{ij,k}(t,z,\phi)\\
  &\qquad\quad\times\,
    \frac{8\pi\,\alpha_s}{t}\,\frac{1}{S_{ij}}\;\mr{B}(\args{a})\,\mc{K}_{ij,k}(t,z,\phi)\,
    \frac{\mc{L}(\rmap{\im\jm}{k}{f_i,t,z,\phi;\args{a}};t)}
         {\mc{L}(\args{a\,};t)}\;
    \Big[O(\rmap{\im\jm}{k}{\args{a}})-O(\args{a})\Big]\\
  &\quad+\,
    \sum_{\args{f\,}}
    \sum_{\{\widetilde{\im\jm},\tilde{k}\}}
    \sum_{f_i=q,g}\frac{1}{16\pi^2}\int\done\Phi_B(\args{p})
    \int_{t_0}^{\mu^2}\done t\,\int_{z_{\rm min}}^{z_{\rm max}}\done z\,
    \int_0^{2\pi}\frac{\done\phi}{2\pi}\,J_{ij,k}(t,z,\phi)\\
  &\qquad\quad\times\,
    \frac{8\pi\,\alpha_s}{t}\,\frac{1}{S_{ij}}\;\mr{B}(\args{a})\,\mc{K}_{ij,k}(t,z,\phi)\,
    \frac{\mc{L}(\rmap{\im\jm}{k}{f_i,t,z,\phi;\args{a}};t)}
         {\mc{L}(\args{a\,};t)}\;
    O(\rmap{\im\jm}{k}{\args{a}})\\
  &\qquad\quad\times\,
    \Theta\rbr{\vphantom{\sum}Q_{ij,k}(t,z,\phi)-Q_{\rm cut}}\,
    \Big[w(\rmap{\im\jm}{k}{f_i,t,z,\phi;\args{a}})-1\Big]\\
  &\quad+\;\;\order(\alpha_s^2)\vphantom{\int_A^B}\;\;.
  \end{split}
\end{equation}
While the first and second terms represent the usual parton-shower prediction, 
the third term encodes the emission-rate differences effected by matrix element 
corrections in the region of well separated partons.
In the ME+PS approach they enter at $\order(\alpha_s)$.

Emission-rate differences seem to be an undesirable side-effect of the ME+PS 
method at first. However, given Eq.~\eqref{eq:meps_expansion}, they can serve 
as an indicator for the relevance of higher-order real-emission corrections. 
We will elaborate on this fact in some more detail in Sec.~\ref{Sec::Results}.

\section{Merging \texorpdfstring{\POWHEG}{POWHEG} and ME+PS - The \texorpdfstring{\MENLOPS}{MENLOPS} approach}
\label{Sec::MENLOPS}

In this section, the two master equations for the 
\POWHEG (Eq.~\eqref{eq:master_powheg}) and ME+PS (Eq.~\eqref{eq:master_meps}) 
approaches are combined into one single expression,
defining the \MENLOPS approach.  The aim of this combination algorithm is 
to simultaneously have NLO accuracy in the cross section, leading logarithmic 
accuracy as implemented in the parton shower and hard higher-order emissions 
corrected using tree-level matrix elements.

Our method of choice is to simply replace the unresolved and the PS resolved
part in Eq.~\eqref{eq:master_meps} with the respective 
\POWHEG expression.  This essentially amounts to the replacement of the 
parton-shower no-emission probability with the corresponding \POWHEG result, 
$\Delta^{\rm(PS)}\to \Delta^{\rm(ME)}$ and a substitution of the leading-order 
weight $\mr{B}$ by $\bar{\mr{B}}$, like in the \POWHEG method itself.

The ME part of the cross section is then generated separately, starting from
real-emission matrix elements, as described in Sec.~\ref{Sec::meps}.  This immediately
implies that it will not automatically benefit from a \POWHEG implementation 
regarding the local $K$-factor $\bar{\mr{B}}/\mr{B}$, and it is therefore
necessary to supply this $K$-factor explicitly.  There is no a-priori 
definition of a Born-level parton configuration in this 
context, because the ME event is defined in terms of a real-emission
configuration. One rather has to identify a branching history 
$\args{a}\to\bmap{\im\jm}{k}{\args{a}}$ such that $\bar{\mr{B}}/\mr{B}$ 
can be computed depending on $\bmap{\im\jm}{k}{\args{a}}$. The definition is 
achieved by clustering the real-emission configuration using an algorithm which 
is similar to a sequential recombination jet scheme and which determines 
the node to be clustered according to the related branching probability in
the parton shower. For more details on this technique we refer the reader 
to~\cite{Hoeche:2009rj}. 

Implementing these ideas, the master formula for the first emission in
\MENLOPS is obtained as
\begin{equation}\label{eq:master_menlops}
  \begin{split}
  &\abr{O}^{\rm (MENLOPS)}\,=\;\sum_{\args{f\,}}\int\done\Phi_B(\args{p})\,
  \bar{\mr{B}}(\args{a})\left[\,\vphantom{\int_A^B}\underbrace{
    \Delta^{\rm(ME)}(t_0,\mu^2;\args{a})}_{\text{unresolved}}\,
    O(\args{a})\right.\\
  &\qquad+\,
    \sum_{\{\widetilde{\im\jm},\tilde{k}\}}
    \sum_{f_i=q,g}\frac{1}{16\pi^2}\,
    \int_{t_0}^{\mu^2}\done t\,\int_{z_{\rm min}}^{z_{\rm max}}\done z\,
    \int_0^{2\pi}\frac{\done\phi}{2\pi}\,J_{ij,k}(t,z,\phi)\;
    O(\rmap{\im\jm}{k}{\args{a}})\\
  &\hspace*{20mm}\times\,
    \frac{1}{S_{ij}}\,\frac{S(\rmap{\im\jm}{k}{\args{f\,}})}{S(\args{f\,})}\,
    \frac{\mr{R}_{ij,k}(\rmap{\im\jm}{k}{\args{a}})}{\mr{B}(\args{a})}
    \,\Bigg(\,
          \underbrace{\Delta^{\rm(ME)}(t,\mu^2;\args{a})\,
                      \Theta\rbr{\vphantom{\sum}Q_{\rm cut}-Q_{ij,k}}}_{\text{resolved, PS domain}}\\
  &\hspace*{80mm}\;+
  \underbrace{\Delta^{\rm(PS)}(t,\mu^2;\args{a})\,\Theta\rbr{\vphantom{\sum}Q_{ij,k}
    -Q_{\rm cut}}}_{\text{resolved, ME domain}}\,\Bigg)\,
    \left.\vphantom{\Bigg)_{\int}^{\int}}\right]\;.
  \end{split}
\end{equation}
Note that the arguments of $Q_{ij,k}$ have been suppressed for ease of notation.
The resolved-ME part of the expression in square brackets exhibits an additional factor
\begin{equation}\label{eq:sud_correction_menlops}
  \begin{split}
  &\frac{\Delta^{\rm(PS)}(t,\mu^2;\args{a})}{\Delta^{\rm(ME)}(t,\mu^2;\args{a})}
    = 1+\sum_{\{\widetilde{\imath\jmath},\tilde{k}\}}\sum_{f_i=q,g}
      \frac{1}{16\pi^2}\int_{t}^{\mu^2}{\rm d}\bar{t}\int_{z_{\rm min}}^{z_{\rm max}}{\rm d}z
      \int_0^{2\pi}\frac{{\rm d}\phi}{2\pi}\,J_{ij,k}(\bar{t},z,\phi)\,\\&\qquad\times\frac{1}{S_{ij}}\,\left[
      \frac{S(r_{\widetilde{\imath\jmath},\tilde{k}}(\{\vec{f}\}))}{S(\{\vec{f}\})}\,
      \frac{{\rm R}_{ij,k}(r_{\widetilde{\imath\jmath},\tilde{k}}(\{\vec{a}\}))}{{\rm B}(\{\vec{a}\})}
      -\frac{8\pi\alpha_s}{\bar{t}}\,\mathcal{K}_{ij,k}(\bar{t},z,\phi)\,
      \frac{\mathcal{L}(r_{\widetilde{\imath\jmath},\tilde{k}}(\{\vec{a}\});\bar{t})}{
        \mathcal{L}(\{\vec{a}\};\bar{t})}\,\right]+\mathcal{O}(\alpha_s^2)
  \end{split}
\end{equation}
compared to the \POWHEG master formula. This makes the emission-rate difference
in the \MENLOPS method explicit. However, the expectation value of $O$ 
is still determined correct to $\order(\alpha_s)$, as can be seen by explicitly
expanding Eq.~\eqref{eq:master_menlops} in powers of $\alpha_s$:
\begin{equation}\label{eq:menlops_expansion}
  \begin{split}
&\abr{O}^{\rm (MENLOPS)}
\,=\;\sum_{\args{f\,}}\int\done\Phi_B(\args{p})\,
    \left(\mr{B}+\tilde{\mr{V}}+\mr{I}\right)(\args{a})\;O(\args{a})\\
  &\qquad+\sum_{\args{f\,}}\!
    \sum_{\{\widetilde{\im\jm},\tilde{k}\}}\!
    \sum_{f_i=q,g}\!\!
    \int\!\done\Phi_B(\args{p})\,\done\Phi_{R|B}^{ij,k}\!
  \sbr{\vphantom{\int}\,\mr{R}_{ij,k}(\rmap{\im\jm}{k}{\args{a}})\,\Theta(t_0\!-t)
    -\mr{S}_{ij,k}(\rmap{\im\jm}{k}{\args{a}})\,}O(\args{a})\\
  &\qquad+\sum_{\args{f\,}}\!
    \sum_{\{\widetilde{\im\jm},\tilde{k}\}}\!
    \sum_{f_i=q,g}\!\!
    \int\!\done\Phi_B(\args{p})\,\done\Phi_{R|B}^{ij,k}\;
    \mr{R}_{ij,k}(\rmap{\im\jm}{k}{\args{a}})\,
    \Theta(t-t_0)\,\Theta\rbr{\vphantom{\sum}Q_{\rm cut}-Q_{ij,k}}\;
    O(\rmap{\im\jm}{k}{\args{a}})\\
  &\qquad+\sum_{\args{f\,}}\!
    \sum_{\{\widetilde{\im\jm},\tilde{k}\}}\!
    \sum_{f_i=q,g}\!\!
    \int\!\done\Phi_B(\args{p})\,\done\Phi_{R|B}^{ij,k}\;
    \mr{R}_{ij,k}(\rmap{\im\jm}{k}{\args{a}})\,
    \Theta(t-t_0)\,\Theta\rbr{Q_{ij,k}-\vphantom{\sum}Q_{\rm cut}}\;
    O(\rmap{\im\jm}{k}{\args{a}})\\
  &\qquad+\,\order(\alpha_s^2) \vphantom{\sum_{f_i}}\;.
  \end{split}
\end{equation}
Comparing this with Eq.~\eqref{eq:powheg_expansion}, we find that
the $\order(\alpha_s)$ accuracy of the \MENLOPS method is identical to 
what is obtained from \POWHEG. The potential mismatch between exact 
higher-order tree-level matrix elements and their respective parton-shower 
approximation, leading to a difference between $\Delta^{\rm(ME)}$ and
$\Delta^{\rm(PS)}$, contributes terms of $\order(\alpha_s^2)$ or higher
as long as $\Theta\rbr{Q_{ij,k}-\vphantom{\sum}Q_{\rm cut}}$ enforces
$t>t_0$. The precise value of $Q_{\rm cut}$ must therefore be chosen
such that this constraint is satisfied.

Apart from being of order $\alpha_s$, the term in the square brackets 
of Eq.~\eqref{eq:sud_correction_menlops} should be rather small in practice, 
as potential differences between matrix-element and parton-shower expressions 
merely lie in subleading logarithmic and power corrections.
This corresponds to saying that the mismatch between \POWHEG and \MENLOPS results 
is at most of order $\alpha_s^2\log\left(\mu^2/Q_{\rm cut}^2\right)$ if the parton shower has 
LL accuracy, and of order $\alpha_s^2$ if it has NLL accuracy.

Generating a second emission using the ME+PS method supplemented 
with the above metioned local $K$-factor of course introduces additional 
emission-rate differences, as described by Eq.~\eqref{eq:meps_expansion}. 
However, because such terms are of $\order(\alpha_s^2)$, they do not spoil 
the next-to-leading order accuracy of the method.

At this point we would like to stress that in their publication Hamilton and
Nason arrived at the same ideas~\cite{Hamilton:2010wh}.

\section{Results}
\label{Sec::Results}

This section collects results obtained with an implementation of the algorithm 
described in the previous sections in the \Sherpa framework.  It aims at 
detailing the improved description of data collected in various collider 
experiments and at quantifying some of the systematic uncertainties inherent 
to the \MENLOPS method, in particular those related to the merging of the
multijet tree-level contributions. Note again, that the \MENLOPS approach
is designed to merge the next-to-leading order accurate description of a given 
core interaction (like for example $e^+e^-\!\to\! q\bar{q}$) through the \POWHEG
method with higher-order tree-level contributions (like $e^+e^-\!\to\! q\bar{q}gg$)
described in the ME+PS approach. Since the total cross section is essentially defined
by the \POWHEG expression of the core process in question, uncertainties like
those related to the choice of scales are encoded mostly there. They have
been discussed in our parallel publication~\cite{Hoeche:2010pf}, while 
uncertainties related to the ME+PS method were discussed for example 
in~\cite{Hoeche:2009rj,Carli:2010cg}.

However, a comparison with results of the ME+PS and \POWHEG techniques alone
is extremely useful to assess the quality of the approach and the improvements
related to it.  The precise setup of \Sherpa for this comparison, including in 
particular a parton shower based on Catani-Seymour subtraction terms~\cite{Schumann:2007mg}
and an automated implementation of the Catani-Seymour subtraction method~\cite{Gleisberg:2007md}
in the matrix-element generator \Amegic~\cite{Krauss:2001iv} was described 
in detail in our parallel publication~\cite{Hoeche:2010pf}.  Throughout the
paper, we use the CTEQ6.6 parton distribution functions~\cite{Nadolsky:2008zw},
(and, correspondingly, the $\overline{\rm MS}$ subtraction scheme) with 
$\alpha_S(m_Z)=0.118$ and running at two-loop. If not stated otherwise, hadronisation 
is not accounted for. Multiple parton interactions are not included in the simulation.
We use the Rivet program package~\cite{Buckley:2008vh,*Buckley:2010ar} and 
the HZTool library~\cite{Waugh:2006ip} for analyses and comparison with data.

\subsection*{Merging Systematics}

\begin{table}[p]
  \begin{center}
  \begin{tabular}{|c|c|c|c|c|c|c|c|c|}
  \hline
  \multirow{3}{*}{$\log_{10} y_\text{cut}$} & \multicolumn{8}{c|}{$N_{\mr{max}}$} \\ \cline{2-9}
                 & \multicolumn{2}{c|}{0}   & \multicolumn{2}{c|}{1}
                                                & \multicolumn{2}{c|}{2}
                                                    & \multicolumn{2}{c|}{3}
    \\ \cline{2-9}
                 & LO+PS & \hspace*{-0.5mm}\POWHEG\hspace*{-1mm}
                                            & ME+PS & \MENLOPS
                                                & ME+PS & \MENLOPS
                                                    & ME+PS & \MENLOPS
      \\ \hline
           -1.25 & \multirow{3}{*}{28.37(1)}
                         & \multirow{3}{*}{29.43(1)}
                                            & 28.09(2) & 29.60(2)
                                                & 28.11(2) & 29.63(2)
                                                    & 28.09(2) & 29.63(2)
  \\ \cline{1-1}\cline{4-9}
           -1.75 &       &                  & 27.46(2) & 29.47(3)
                                                & 27.56(3) & 29.43(3)
                                                    & 27.48(3) & 29.43(3)
  \\ \cline{1-1}\cline{4-9}
           -2.25 &       &                  & 27.11(3) & 29.38(3)
                                                & 26.90(3) & 29.18(4)
                                                    & 26.93(4) & 29.17(4)
  \\ \hline
  \end{tabular}
  \end{center}
  \captionsetup{width=\textwidth}
  \caption{Dependence of the inclusive $e^+e^-\to jets$ cross section in nb on the number of 
           extra jets generated in both the ME+PS and the \MENLOPS method. For 
           $N_\mr{max}=0$ these reduce to the LO+PS and \POWHEG methods, 
           respectively.
           \label{tab:eexsec}}
\end{table}

\begin{table}[p]
  \begin{center}
  \begin{tabular}{|c|c|c|c|c|c|c|c|c|}
  \hline
  \multirow{3}{*}{$\bar{Q}_\text{cut}$} & \multicolumn{8}{c|}{$N_{\mr{max}}$} \\ \cline{2-9}
                 & \multicolumn{2}{c|}{0}   & \multicolumn{2}{c|}{1}
                                                & \multicolumn{2}{c|}{2}
                                                    & \multicolumn{2}{c|}{3}    \\ \cline{2-9}
                 & LO+PS & \POWHEG          & ME+PS & \MENLOPS
                                                & ME+PS & \MENLOPS
                                                    & ME+PS & \MENLOPS      \\ \hline
          3 GeV  & \multirow{3}{*}{365.7(4)}
                         & \multirow{3}{*}{322.4(6)}
                                            & 380.5(6) & 333.5(9)\hphantom{1.}
                                                & 388.4(6) & 339.3(1.0)
                                                    & 396.6(9) & 344.4(1.2) \\ \cline{1-1}\cline{4-9}
          5 GeV  &       &                  & 380.0(6) & 330.6(9)\hphantom{1.}
                                                & 387.0(6) & 339.2(1.1)
                                                    & 393.2(8) & 342.8(1.2) \\ \cline{1-1}\cline{4-9}
          9 GeV  &       &                  & 381.8(7) & 330.4(1.0)
                                                & 388.1(8) & 337.9(1.2)
                                                    & 394.2(8) & 344.0(1.2) \\ \hline
  \end{tabular}
  \end{center}
  \captionsetup{width=\textwidth}
  \caption{Dependence of the DIS production cross section in nb on the number of 
           extra jets generated in both the ME+PS and the \MENLOPS method. For 
           $N_\mr{max}=0$ these reduce to the LO+PS and \POWHEG methods, 
           respectively. The precise definition of $\bar{Q}_\text{cut}$ is 
           given in \cite{Carli:2010cg}.
           \label{tab:disxsec}}
\end{table}

\begin{table}[p]
  \begin{center}
  \begin{tabular}{|c|c|c|c|c|c|c|c|c|}
  \hline
  \multirow{3}{*}{$Q_\text{cut}$} & \multicolumn{8}{c|}{$N_{\mr{max}}$} \\ \cline{2-9}
                 & \multicolumn{2}{c|}{0}   & \multicolumn{2}{c|}{1}
                                                & \multicolumn{2}{c|}{2}
                                                    & \multicolumn{2}{c|}{3}    \\ \cline{2-9}
                 & LO+PS & \POWHEG          & ME+PS & \MENLOPS
                                                & ME+PS & \MENLOPS
                                                    & ME+PS & \MENLOPS      \\ \hline
          15 GeV & \multirow{3}{*}{1993(1)}
                         & \multirow{3}{*}{2423(1)}
                                            & 2114(6) & 2549(9)
                                                & 2169(6) & 2587(9)
                                                    & 2159(7) & 2599(10) \\ \cline{1-1}\cline{4-9}
          20 GeV &       &                  & 2103(4) & 2516(6)
                                                & 2137(5) & 2548(9)
                                                    & 2135(5) & 2548(9)\hphantom{2} \\ \cline{1-1}\cline{4-9}
          40 GeV &       &                  & 2092(3) & 2477(9)
                                                & 2104(3) & 2485(7)
                                                    & 2101(3) & 2482(7)\hphantom{2} \\ \hline
  \end{tabular}
  \end{center}
  \captionsetup{width=\textwidth}
  \caption{Dependence of the $W$ production cross section in pb on the number of 
           extra jets generated in both the ME+PS and the \MENLOPS method. For 
           $N_\mr{max}=0$ these reduce to the LO+PS and \POWHEG methods, 
           respectively.
           \label{tab:wxsec}}
\end{table}

\begin{table}[p]
  \begin{center}
  \begin{tabular}{|c|c|c|c|c|c|c|c|c|}
  \hline
  \multirow{3}{*}{$Q_\text{cut}$} & \multicolumn{8}{c|}{$N_{\mr{max}}$} \\ \cline{2-9}
                 & \multicolumn{2}{c|}{0}   & \multicolumn{2}{c|}{1}
                                                & \multicolumn{2}{c|}{2}
                                                    & \multicolumn{2}{c|}{3}    \\ \cline{2-9}
                 & LO+PS & \hspace*{-0.5mm}\POWHEG\hspace*{-1.5mm}
                                            & ME+PS & \MENLOPS
                                                & ME+PS & \MENLOPS
                                                    & ME+PS & \MENLOPS      \\ \hline
          15 GeV & \multirow{3}{*}{394.7(1)}
                         & \multirow{3}{*}{477.8(1)}
                                            & 417.7(8) & 489.3(8)
                                                & 425.9(9) & 502.6(1.3)
                                                    & 429.4(1.1) & 503.6(1.3) \\ \cline{1-1}\cline{4-9}
          20 GeV &       &                  & 416.8(7) & 487.2(9)
                                                & 424.3(8) & 496.4(1.0)
                                                    & 423.6(8)\hphantom{1.} & 496.1(1.0) \\ \cline{1-1}\cline{4-9}
          40 GeV &       &                  & 417.1(4) & 486.6(6)
                                                & 419.8(5) & 489.1(6)\hphantom{1.}
                                                    & 420.6(5)\hphantom{1.} & 489.1(6)\hphantom{1.} \\ \hline
  \end{tabular}
  \end{center}
  \captionsetup{width=\textwidth}
  \caption{Dependence of the $Z$ production cross section in pb on the number of 
           extra jets generated in both the ME+PS and the \MENLOPS method. For 
           $N_\mr{max}=0$ these reduce to the LO+PS and \POWHEG methods, 
           respectively.
           \label{tab:zxsec}}
\end{table}

\begin{table}[p]
  \begin{center}
  \begin{tabular}{|c|c|c|c|c|c|c|c|c|}
  \hline
  \multirow{3}{*}{$Q_\text{cut}$} & \multicolumn{8}{c|}{$N_{\mr{max}}$} \\ \cline{2-9}
                 & \multicolumn{2}{c|}{0}   & \multicolumn{2}{c|}{1}
                                                & \multicolumn{2}{c|}{2}
                                                    & \multicolumn{2}{c|}{3}    \\ \cline{2-9}
                 & LO+PS & \POWHEG         & ME+PS & \MENLOPS
                                                & ME+PS & \MENLOPS
                                                    & ME+PS & \MENLOPS      \\ \hline
          15 GeV & \multirow{3}{*}{1.063(1)}
                         & \multirow{3}{*}{2.425(1)}
                                            & 1.217(4) & 2.698(11)
                                                & 1.297(6) & 2.846(15)
                                                    & 1.353(11) & 2.937(27) \\ \cline{1-1}\cline{4-9}
          20 GeV &       &                   & 1.195(3) & 2.627(9)\hphantom{0}
                                                & 1.255(5) & 2.762(13)
                                                    & 1.270(7)\hphantom{0} & 2.769(19)\\ \cline{1-1}\cline{4-9}
          40 GeV &       &                   & 1.177(2) & 2.488(6)\hphantom{0}
                                                & 1.215(3) & 2.571(10)
                                                    & 1.212(4)\hphantom{0} & 2.598(13) \\ \hline
  \end{tabular}
  \end{center}
  \captionsetup{width=\textwidth}
  \caption{Dependence of the Higgs production cross section in pb in gluon fusion on 
           the number of extra jets generated in both the ME+PS and the 
           \MENLOPS method. For $N_\mr{max}=0$ these reduce to the LO+PS and 
           \POWHEG methods, respectively.
           \label{tab:hxsec}}
\end{table}

\begin{table}[p]
  \begin{center}
  \begin{tabular}{|c|c|c|c|c|c|c|c|c|}
  \hline
  \multirow{3}{*}{$Q_\text{cut}$} & \multicolumn{8}{c|}{$N_{\mr{max}}$} \\ \cline{2-9}
                 & \multicolumn{2}{c|}{0}   & \multicolumn{2}{c|}{1}
                                                & \multicolumn{2}{c|}{2}
                                                    & \multicolumn{2}{c|}{3}    \\ \cline{2-9}
                 & LO+PS & \POWHEG         & ME+PS & \MENLOPS
                                                & ME+PS & \MENLOPS
                                                    & ME+PS & \MENLOPS      \\ \hline
          20 GeV & \multirow{3}{*}{0.949(1)}
                         & \multirow{3}{*}{1.333(1)}
                                             & 1.305(4) & 1.369(5)
                                                & 1.391(6) & 1.469(7)
                                                    & 1.405(7) & 1.470(9) \\ \cline{1-1}\cline{4-9}
          40 GeV &       &                   & 1.277(3) & 1.379(5)
                                                & 1.361(4) & 1.459(6)
                                                    & 1.378(6) & 1.465(6)\\ \cline{1-1}\cline{4-9}
          80 GeV &       &                   & 1.227(2) & 1.405(5)
                                                & 1.285(3) & 1.451(5)
                                                    & 1.298(5) & 1.460(5) \\ \hline
  \end{tabular}
  \end{center}
  \captionsetup{width=\textwidth}
  \caption{Dependence of the $W^+W^-$-production cross section in pb on the number of 
           extra jets generated in both the ME+PS and the \MENLOPS method. For 
           $N_\mr{max}=0$ these reduce to the LO+PS and \POWHEG methods, 
           respectively.
           \label{tab:wwxsec}}
\end{table}

\begin{figure}[p]
  \begin{center}
  \includegraphics[width=0.4\textwidth]{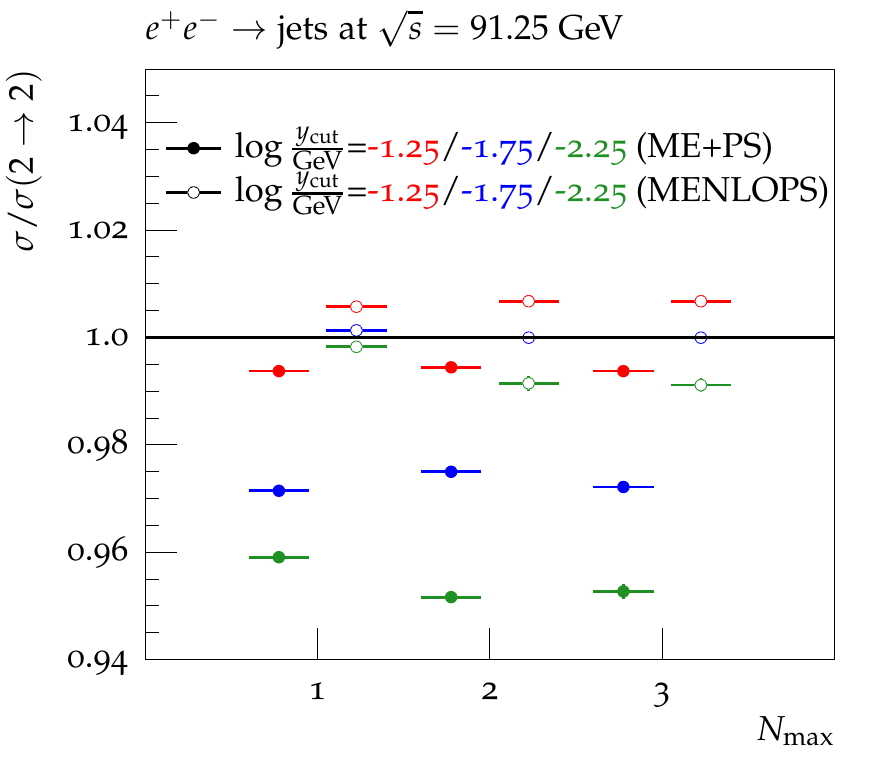}
  \hspace*{0.05\textwidth}
  \includegraphics[width=0.4\textwidth]{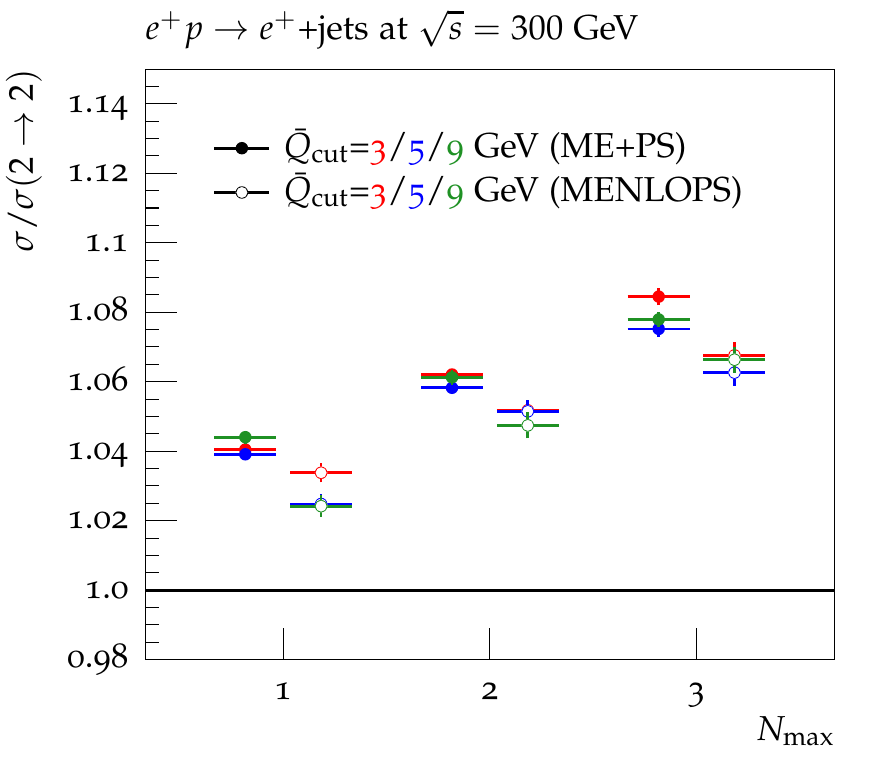}
  \vspace*{2mm}\\
  \includegraphics[width=0.4\textwidth]{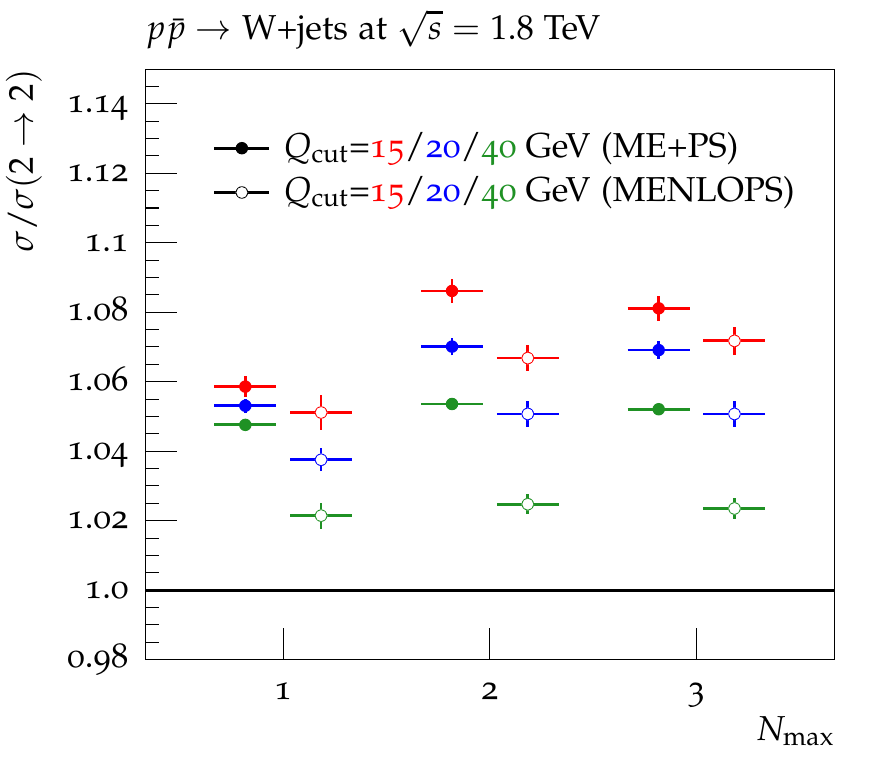}
  \hspace*{0.05\textwidth}
  \includegraphics[width=0.4\textwidth]{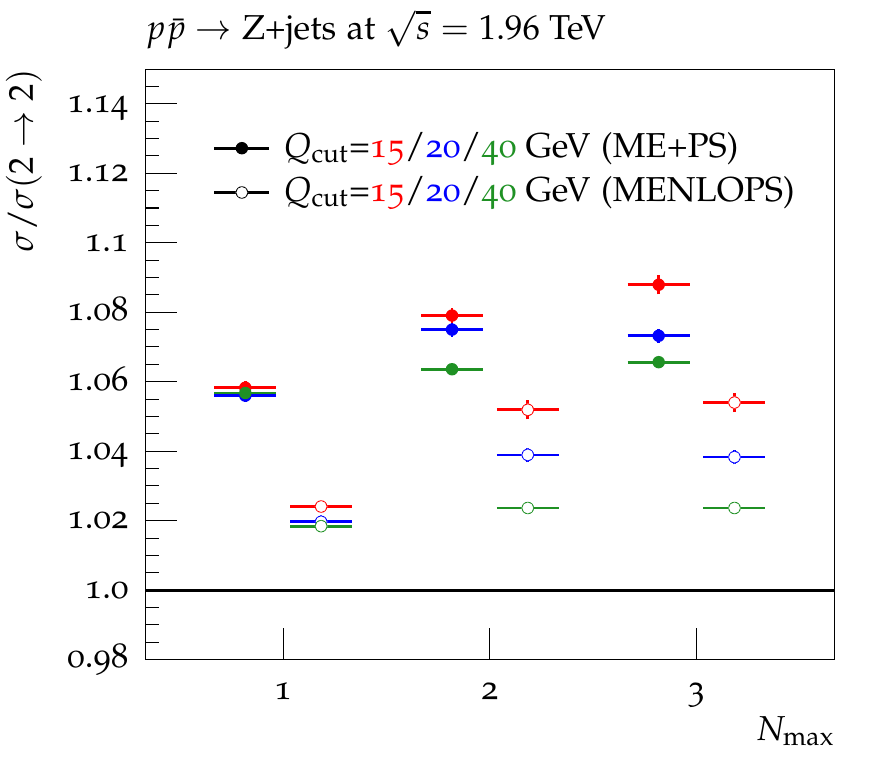}
  \vspace*{2mm}\\
  \includegraphics[width=0.4\textwidth]{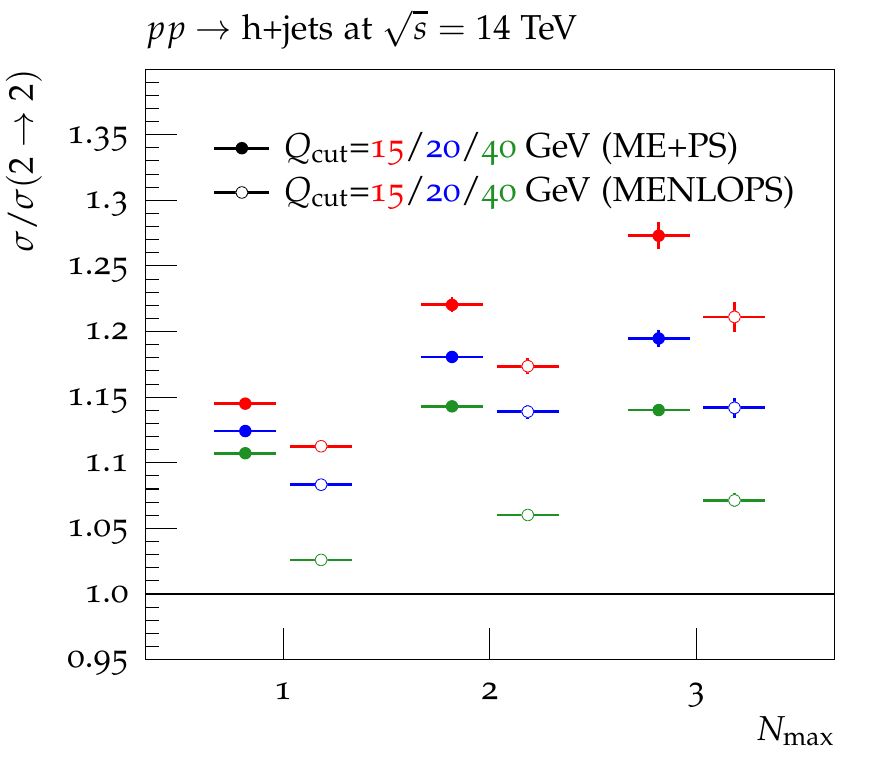}
  \hspace*{0.05\textwidth}
  \includegraphics[width=0.4\textwidth]{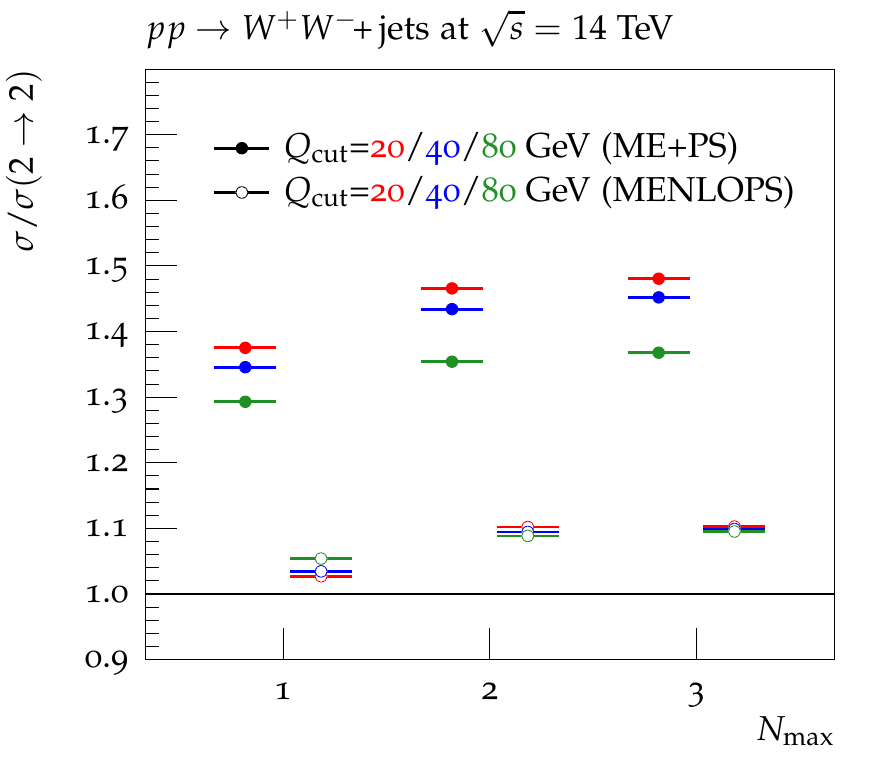}
  \end{center}
  \captionsetup{width=0.85\textwidth}\caption{
  Visualisation of the emission-rate differences induced by the ME+PS and 
  \MENLOPS master formulas, Eqs.~\eqref{eq:master_menlops} and 
  \eqref{eq:master_meps}, respectively.
  \label{fig:xsfigs}}
\end{figure}

\begin{figure}[p]
  \begin{center}
  \includegraphics[width=0.4\textwidth]{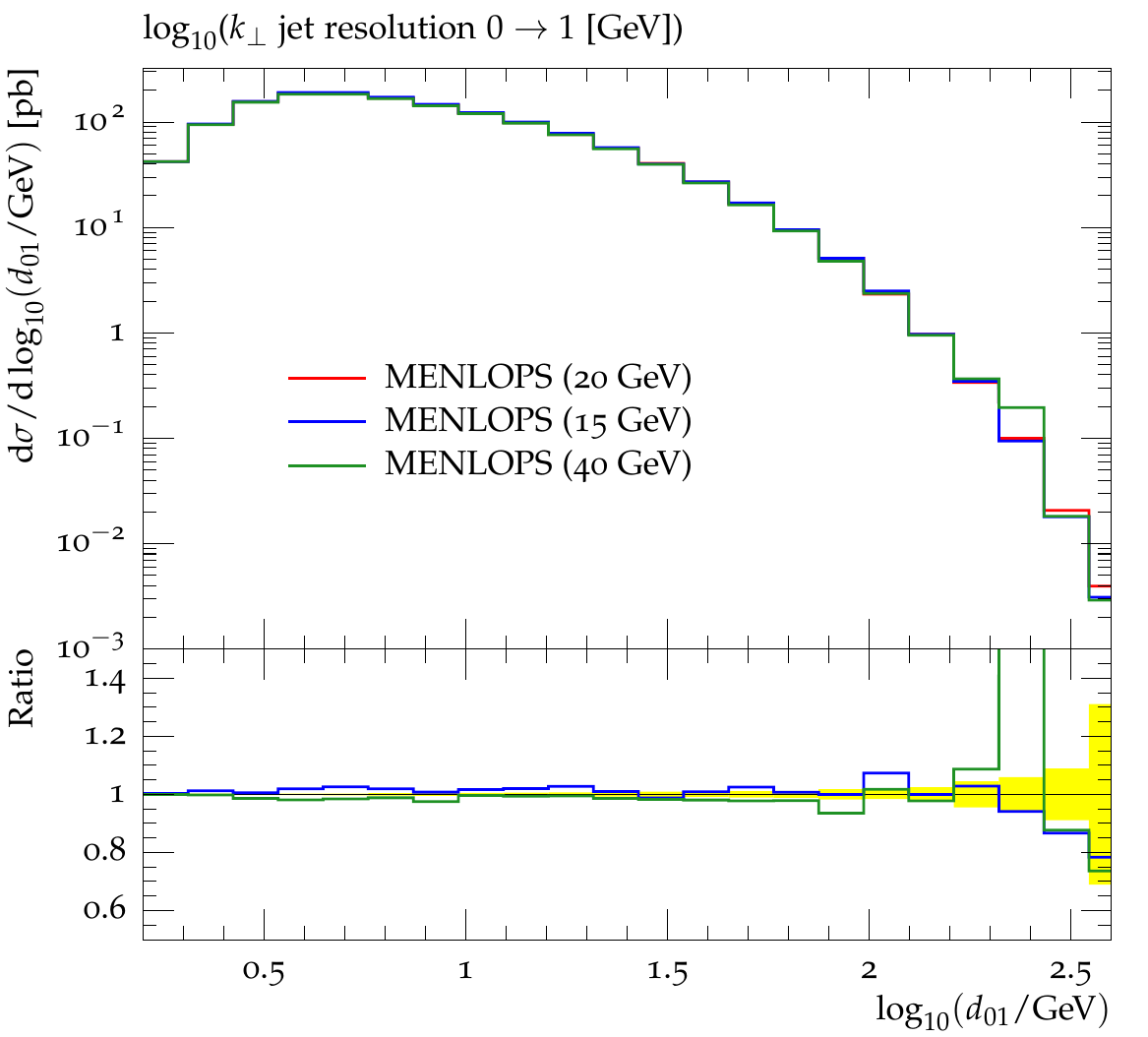}
  \hspace*{0.05\textwidth}
  \includegraphics[width=0.4\textwidth]{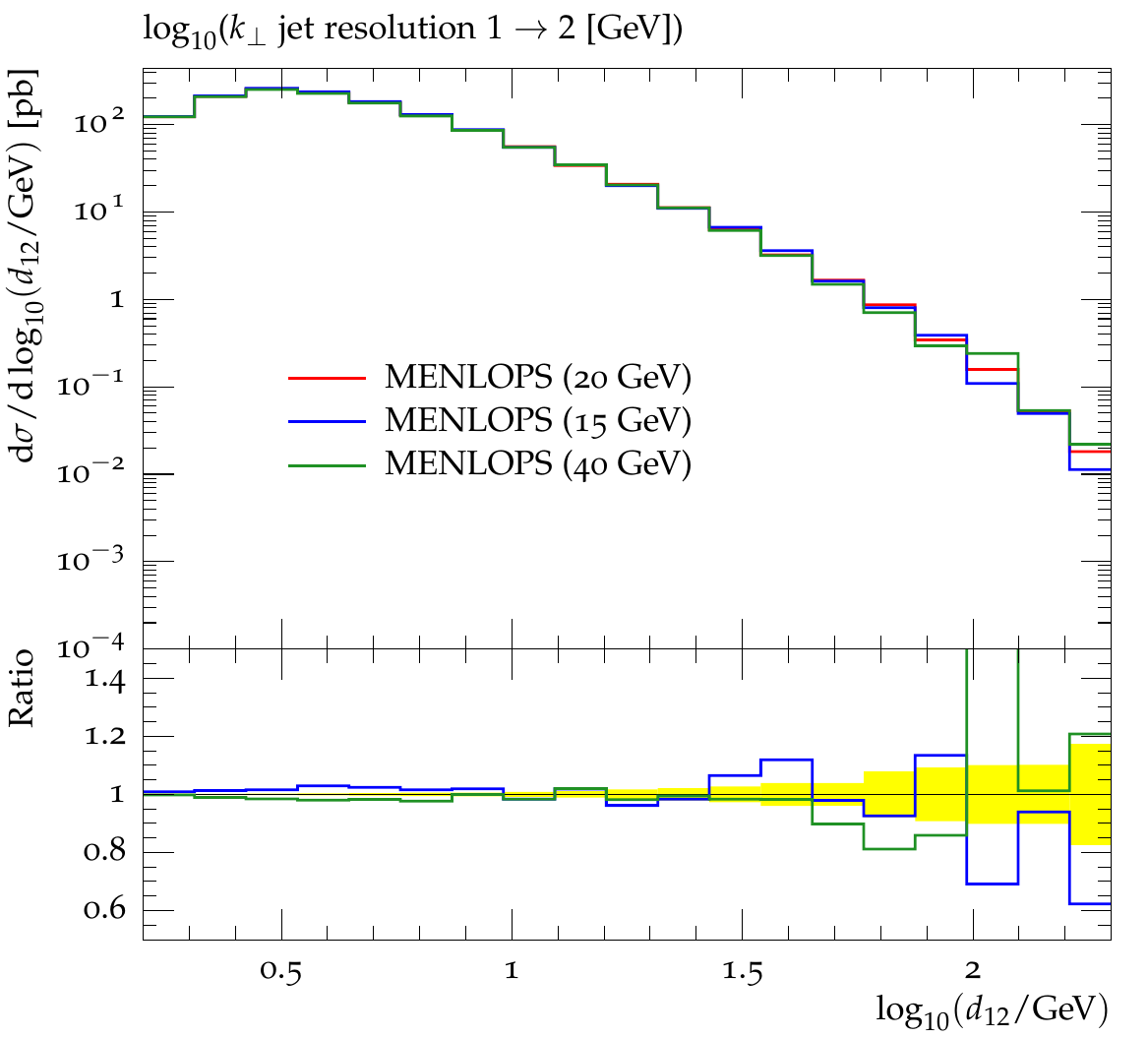}
  \vspace*{2mm}\\
  \includegraphics[width=0.4\textwidth]{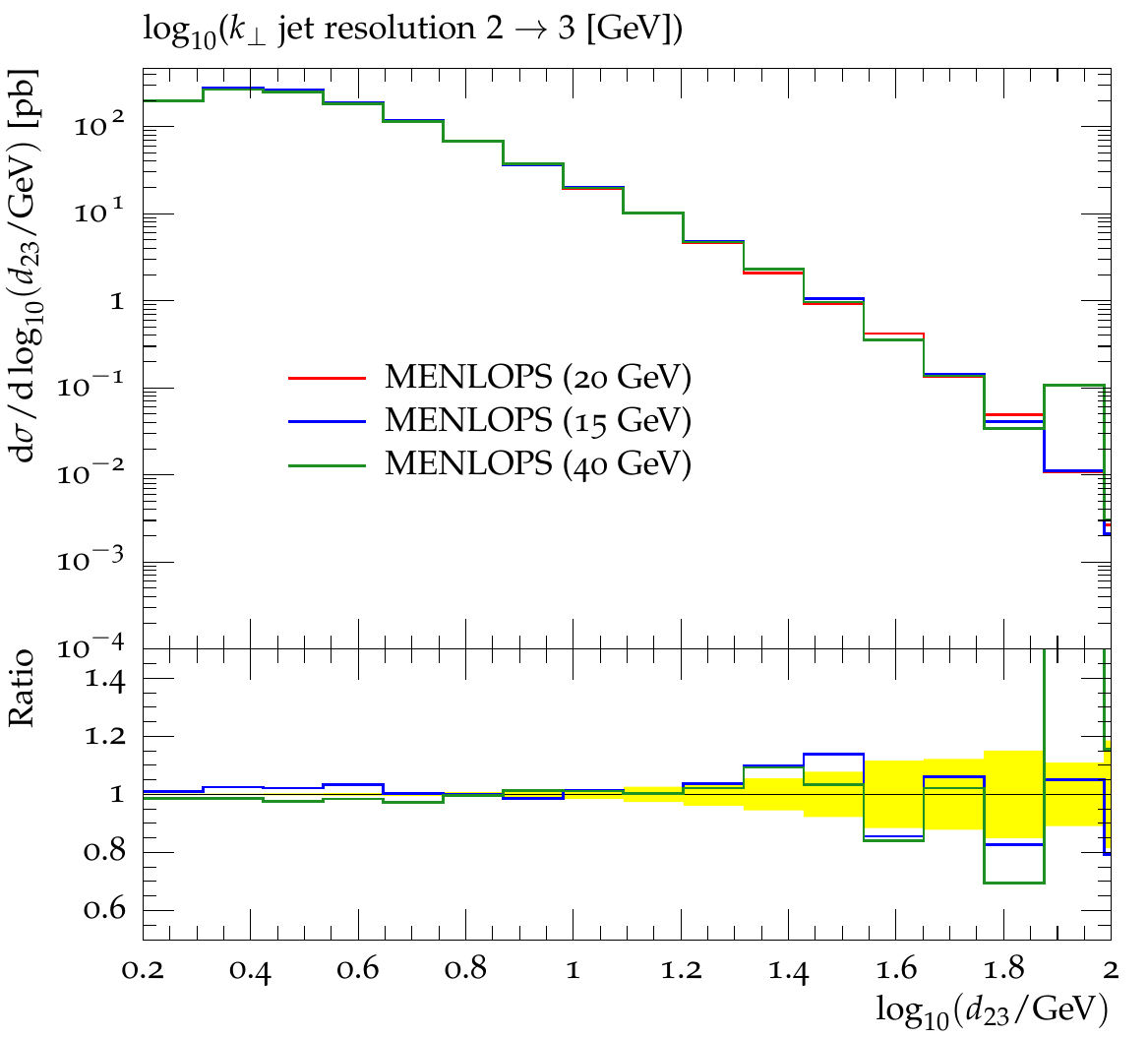}
  \hspace*{0.05\textwidth}
  \includegraphics[width=0.4\textwidth]{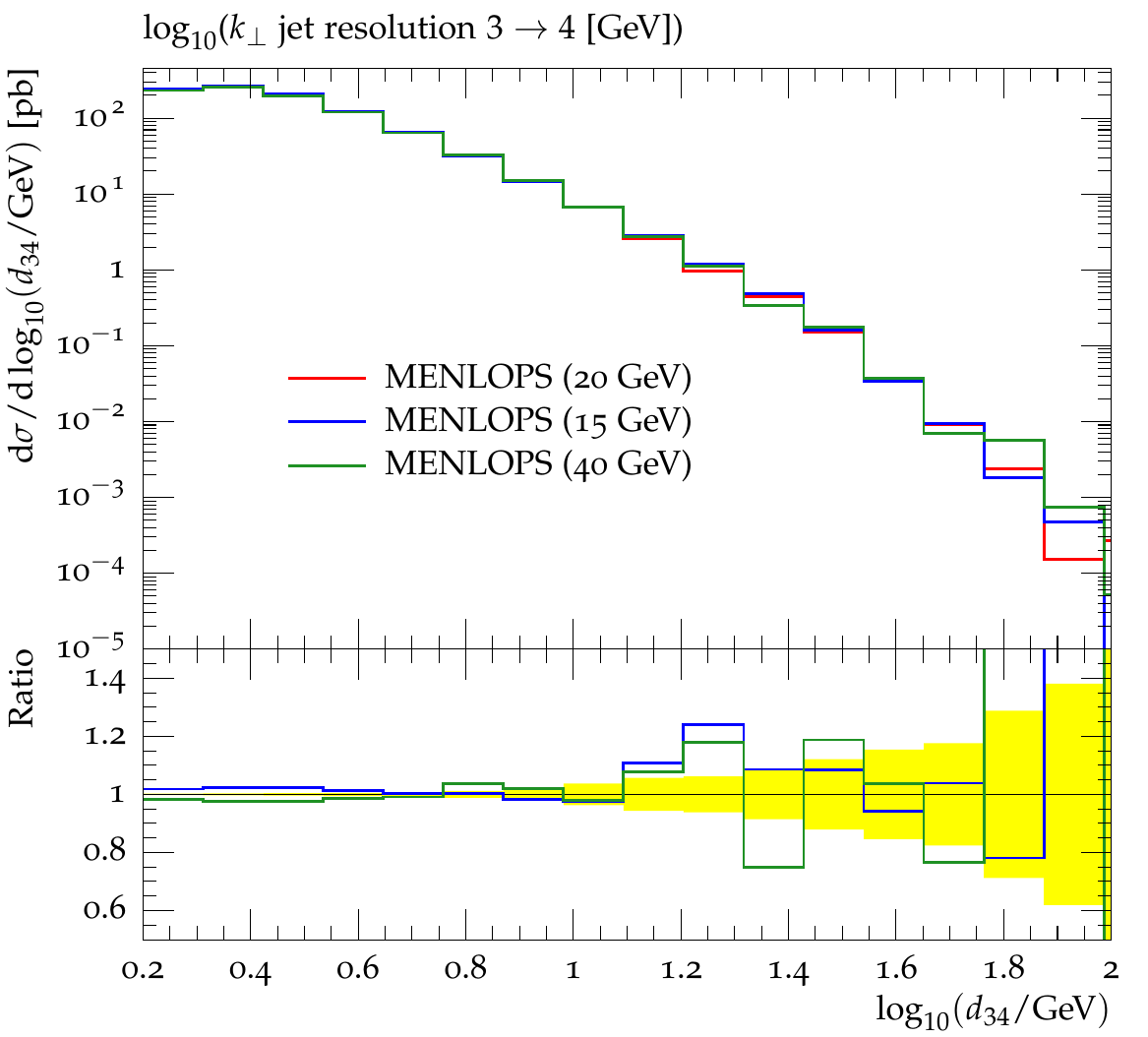}
  \end{center}
  \captionsetup{width=0.85\textwidth}\caption{
  Differential jet rates $d_{n\,n+1}$ for three different merging cuts, $Q_\mr{cut}$,
  in Drell-Yan lepton-pair production at the Tevatron at $\sqrt{s}=1.96$ TeV.
  \label{fig:tev:jetrates_syst}}
\end{figure}

As pointed out in the previous section, the ME+PS approach violates the 
unitarity of the parton-shower simulation.  This discrepancy is directly 
inherited by the \MENLOPS method.  The extent of this effect depends 
entirely on the quality of the parton-shower algorithm, as can be seen in 
Eq.~\eqref{eq:sud_correction_menlops}: If the parton-shower approximation to the 
real-emission matrix element is good, the correction factor, 
Eq.~\eqref{eq:sud_correction_menlops}, is close to one.

We test the quality of our algorithms in the reactions  $e^+e^-\to$ hadrons,
deep-inelastic lepton-nucleon scattering, Drell-Yan lepton-pair production, 
$W$- and Higgs-boson production and $W^+W^-$-production by varying the phase-space 
separation cut, $Q_\mr{cut}$, and the maximum number of partons, $N_\mr{max}$, 
which is simulated with matrix elements in the \MENLOPS approach.  This is in close 
correspondence to the sanity checks of the ME+PS method which have been 
presented in~\cite{Hoeche:2009rj}. The respective results are summarised in 
Tabs.~\ref{tab:eexsec}-\ref{tab:wwxsec} and depicted in Fig.~\ref{fig:xsfigs}. 
It is interesting to note that differences in the total
cross-section are smallest for the \MENLOPS samples with $N_\mr{max}=1$,
and that they increase steadily for larger $N_\mr{max}$. This indicates that
the parton shower tends to underestimate the cross section of higher-order
tree-level contributions. We observe two important effects, which allow us 
to judge the quality of the \MENLOPS approach with respect to NLO accuracy: 
Firstly, for $N_\mr{max}=1$ the emission-rate differences never exceed the size 
of the NLO corrections.
Secondly, for any given value of $Q_{\rm cut}$ and $N_{\rm jet}$, the relative
difference between the cross sections from \MENLOPS and \POWHEG is always smaller 
than the one between the cross sections from ME+PS and LO+PS.
This is best seen in Fig.~\ref{fig:xsfigs}, and it gives some confidence that 
the \MENLOPS technique can help to improve perturbative QCD predictions 
from parton-shower Monte Carlo.

The above analysis can be seen from a different perspective as follows:
Usually, the biggest intrinsic uncertainty of the ME+PS approach stems from 
the freedom to choose the phase-space separation cut, $Q_\mr{cut}$, as 
explained and exemplified in a number of processes in~\cite{Hoeche:2009rj}. 
Since the \MENLOPS method relies on identical ideas to separate the 
real-emission phase space, it naturally inherits this source of uncertainty.
Deviations of \MENLOPS results from results with different values of 
$Q_\mr{cut}$ are to be expected.  However, their small size in a reasonable
range of $Q_\mr{cut}$ is a sign of the algorithm working well. The following rule 
of thumb can be applied: If the value of $Q_\mr{cut}$ is chosen too large, too much 
extra emission phase space is left to the \POWHEG simulation, typically 
leading to an underestimation of jet rates, since \POWHEG only simulates the 
first emission through matrix elements.  If, on the other hand, this value is 
too small, too much phase space is filled by matrix elements with large 
final-state multiplicity, which may lead to noticeable emission-rate differences. 
The value of $Q_\mr{cut}$ should therefore lie well between the parton-shower 
cutoff and the factorisation scale of the core process, with some margin on 
either side of this interval.

We exemplify the stability of our \MENLOPS implementation with respect to 
variations of $Q_\mr{cut}$ in Fig.~\ref{fig:tev:jetrates_syst}.  Due to their
similarity to $Q_{ij,k}$, the differential jet rates shown there are extremely 
sensitive to the details of the radiation pattern and thus to the accuracy
of the ME+PS implementation. They tend to expose even the slightest mismatch
between PS and ME subsamples, which then shows up as a kink in the distribution.
However, when varying $Q_\mr{cut}$ in a rather wide range, we observe no  
sizable discrepancies between the respective \MENLOPS predictions, which is 
a very encouraging result regarding the quality of the algorithm and its 
implementation in \Sherpa.

\subsection{\texorpdfstring{$e^+e^-\to$}{e+ e- to} jets}
\label{Sec:ee}

\begin{figure}[p]
  \begin{center}
  \includegraphics[width=0.45\textwidth]{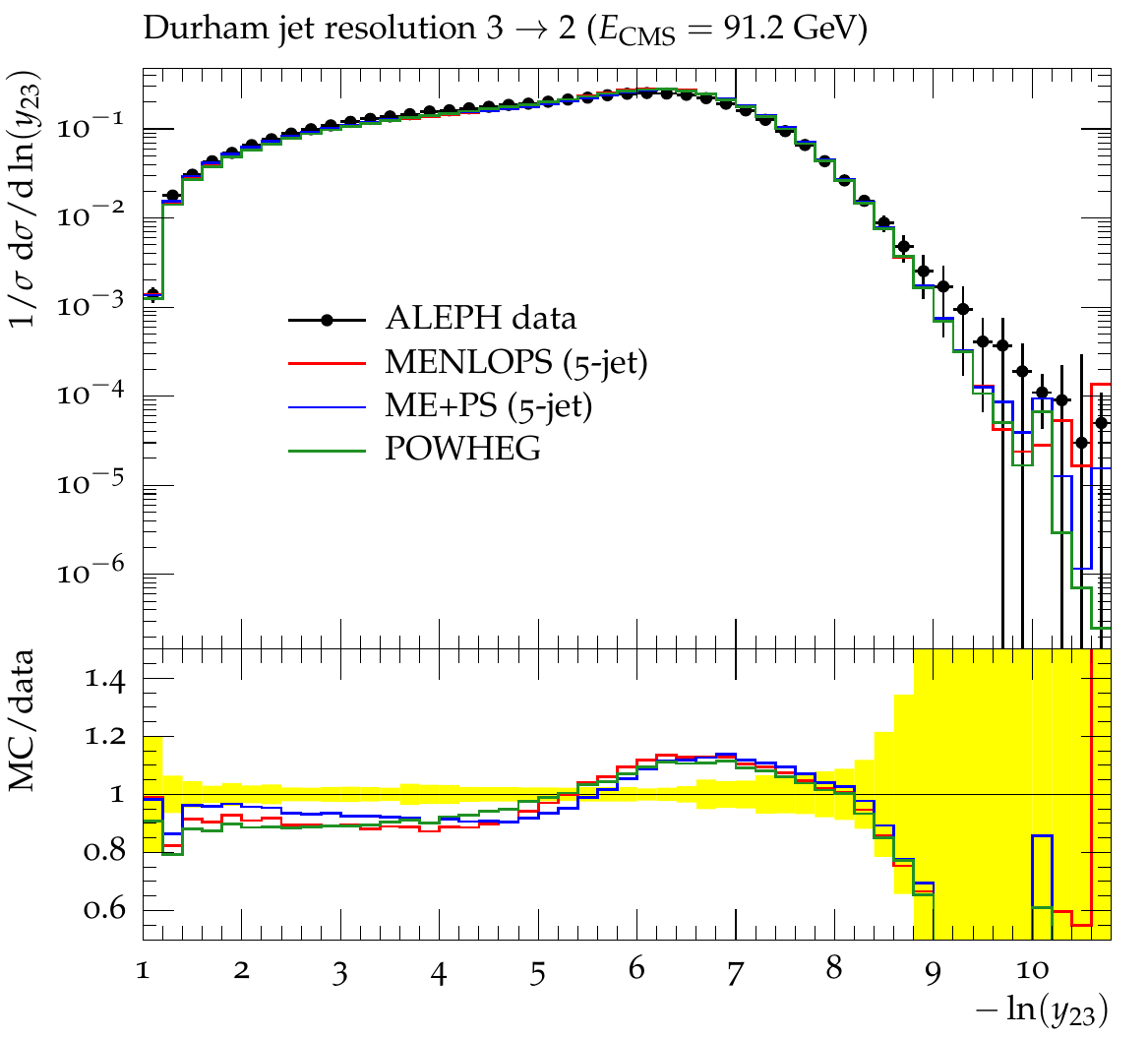}
  \hspace*{0.05\textwidth}
  \includegraphics[width=0.45\textwidth]{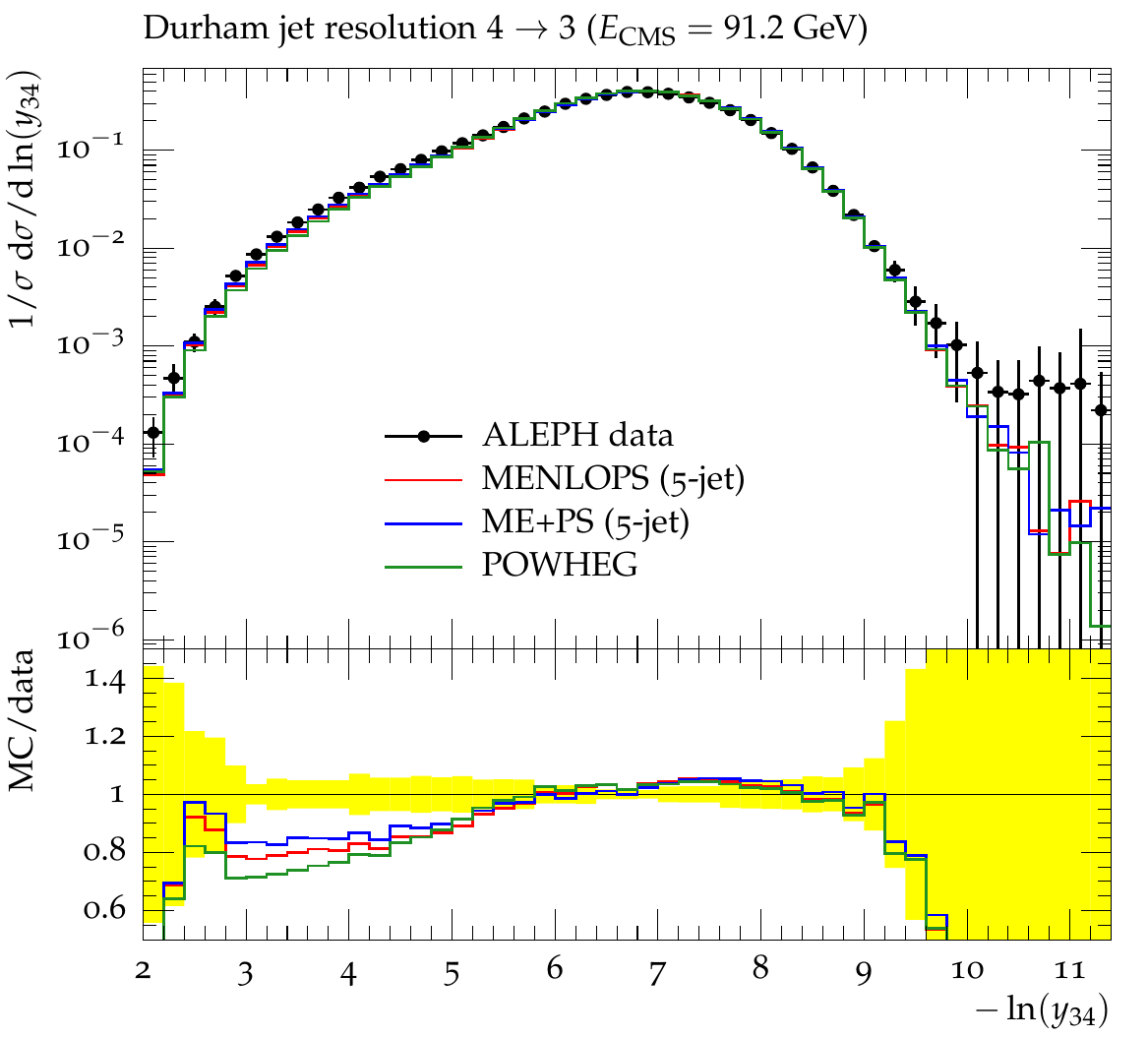}
  \vspace*{2mm}\\
  \includegraphics[width=0.45\textwidth]{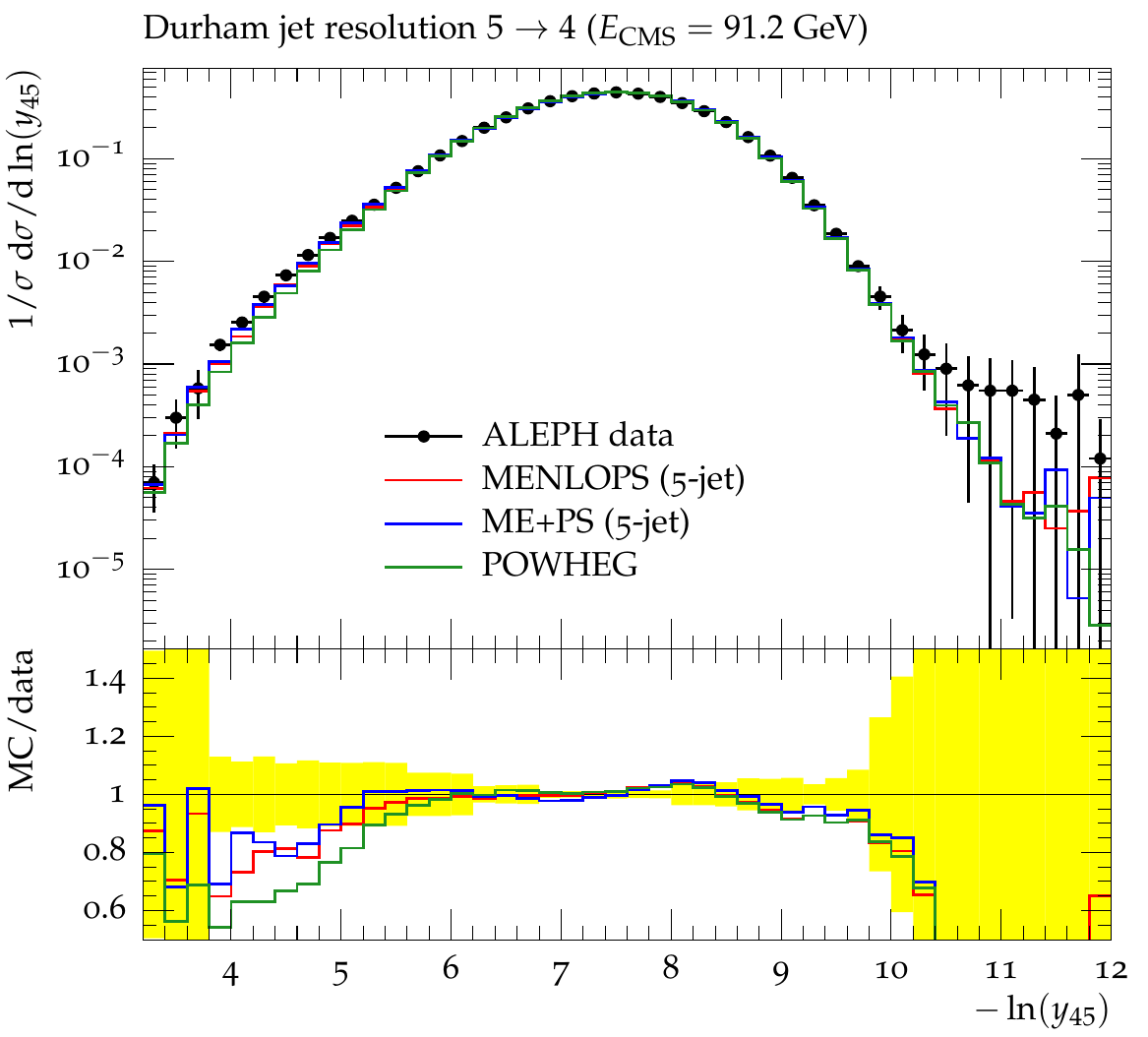}
  \hspace*{0.05\textwidth}
  \includegraphics[width=0.45\textwidth]{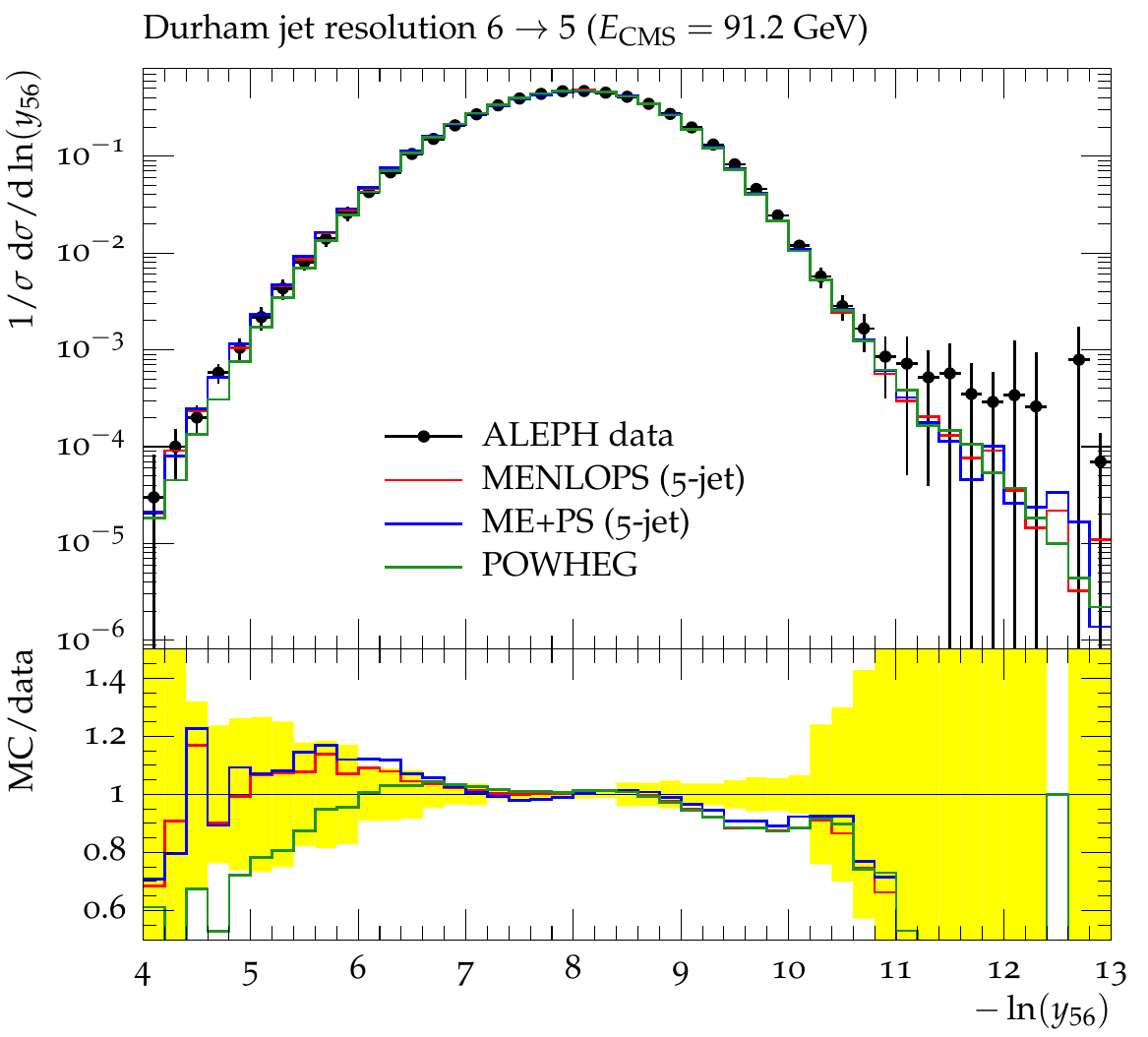}
  \end{center}
  \caption{
  Durham $d_{n\;n+1}$ jet resolutions at LEP compared to data taken 
  by the ALEPH experiment~\cite{Heister:2003aj}.
  \label{fig:lep:23jetrate_34jetrate}}
\end{figure}

\begin{figure}[p]
  \begin{center}
  \includegraphics[width=0.45\textwidth]{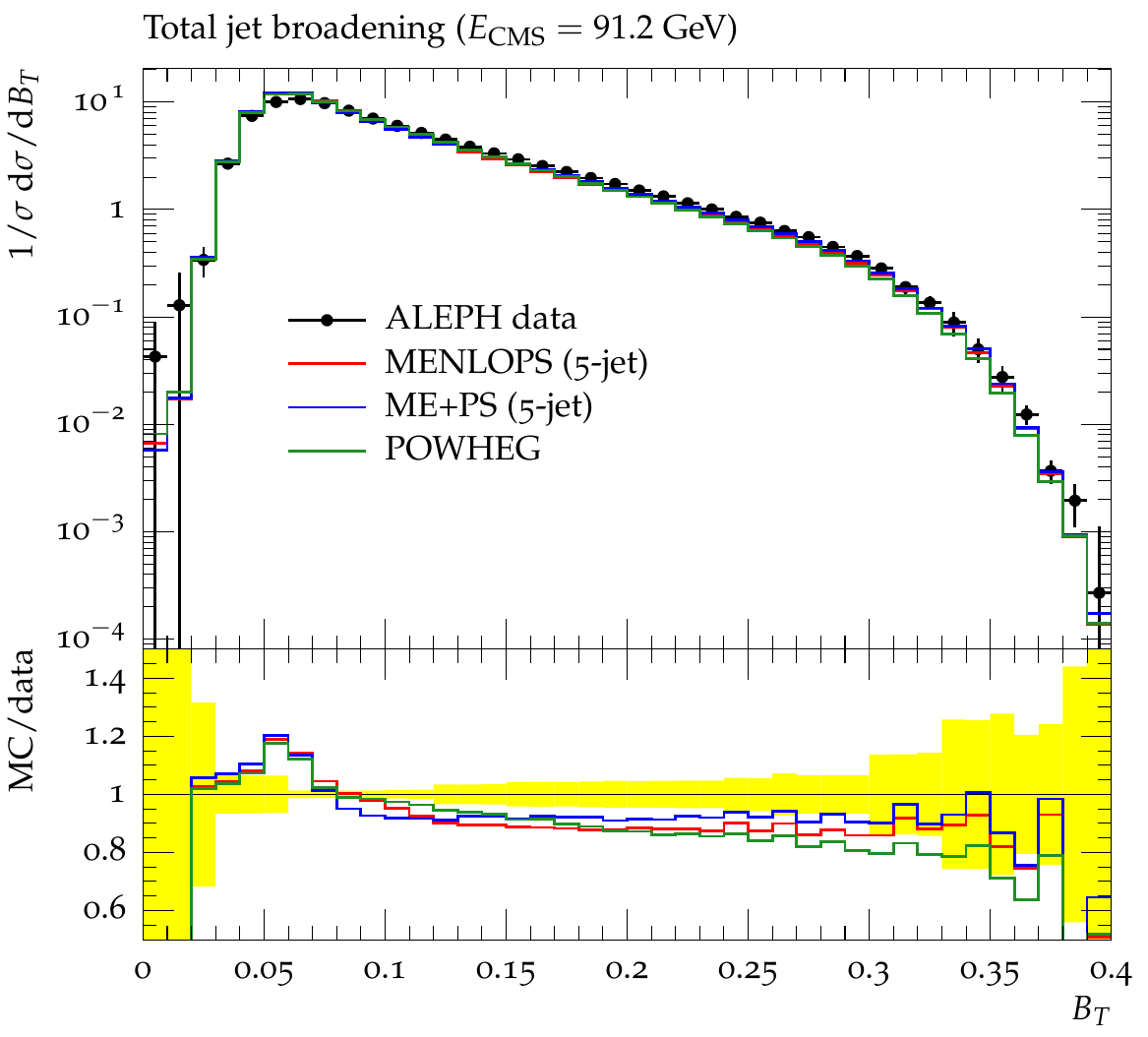}
  \hspace*{0.05\textwidth}
  \includegraphics[width=0.45\textwidth]{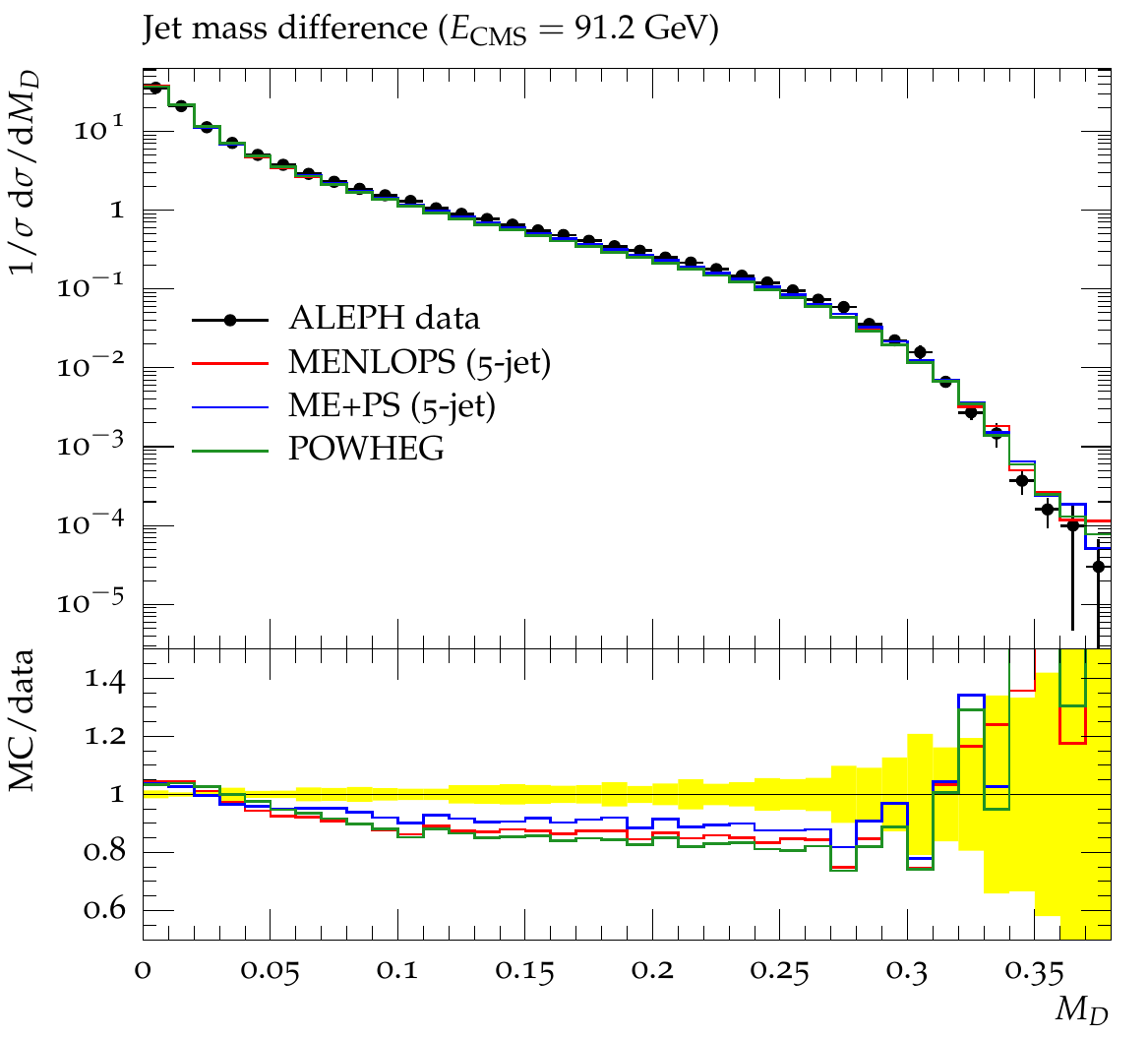}
  \end{center}
  \caption{
  Total jet broadening  and jet mass difference at LEP compared to data taken by the 
  ALEPH experiment~\cite{Heister:2003aj}.
  \label{fig:lep:jetbroadening_cparam}}
\end{figure}

\begin{figure}[p]
  \begin{center}
  \includegraphics[width=0.45\textwidth]{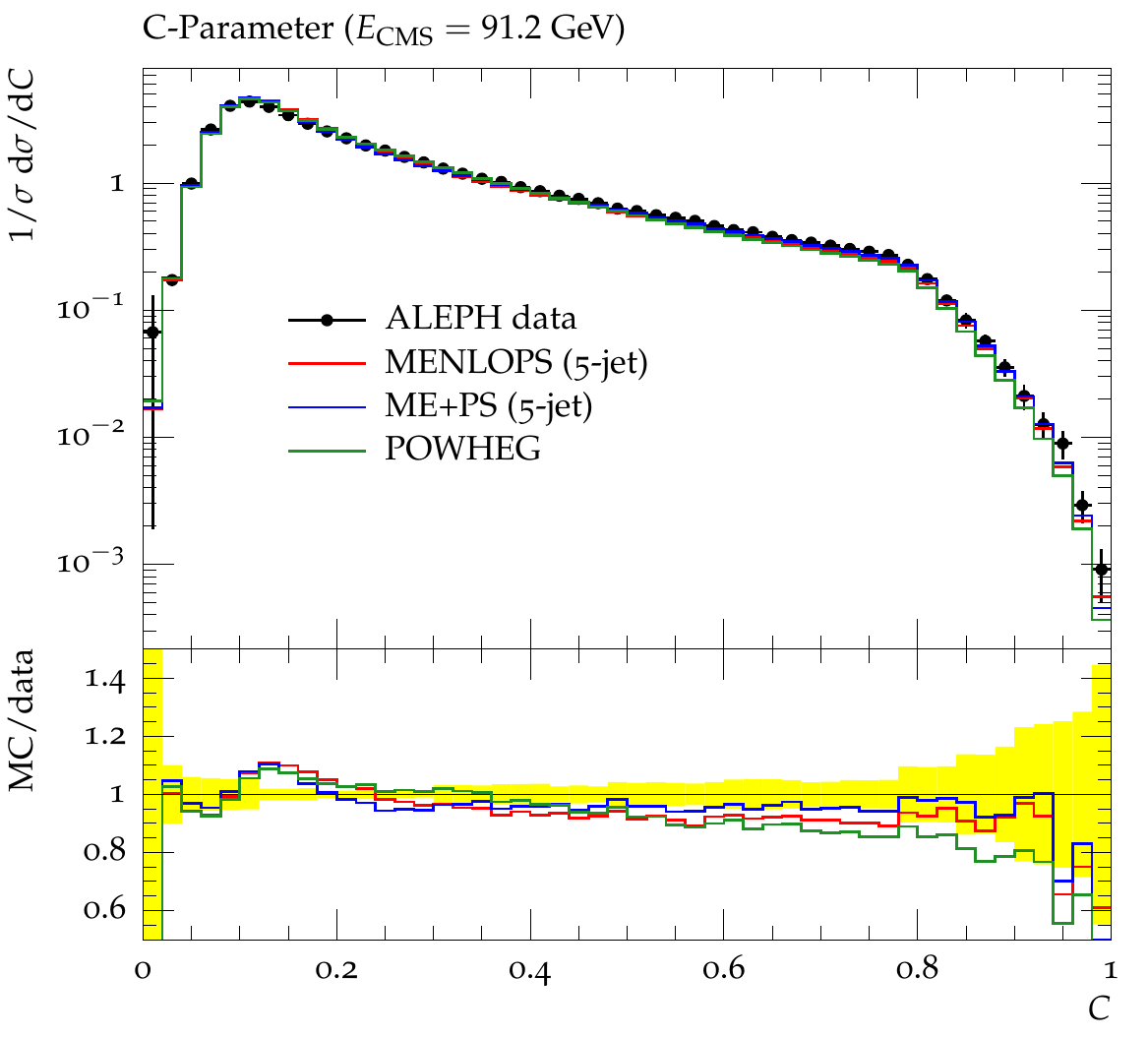}
  \hspace*{0.05\textwidth}
  \includegraphics[width=0.45\textwidth]{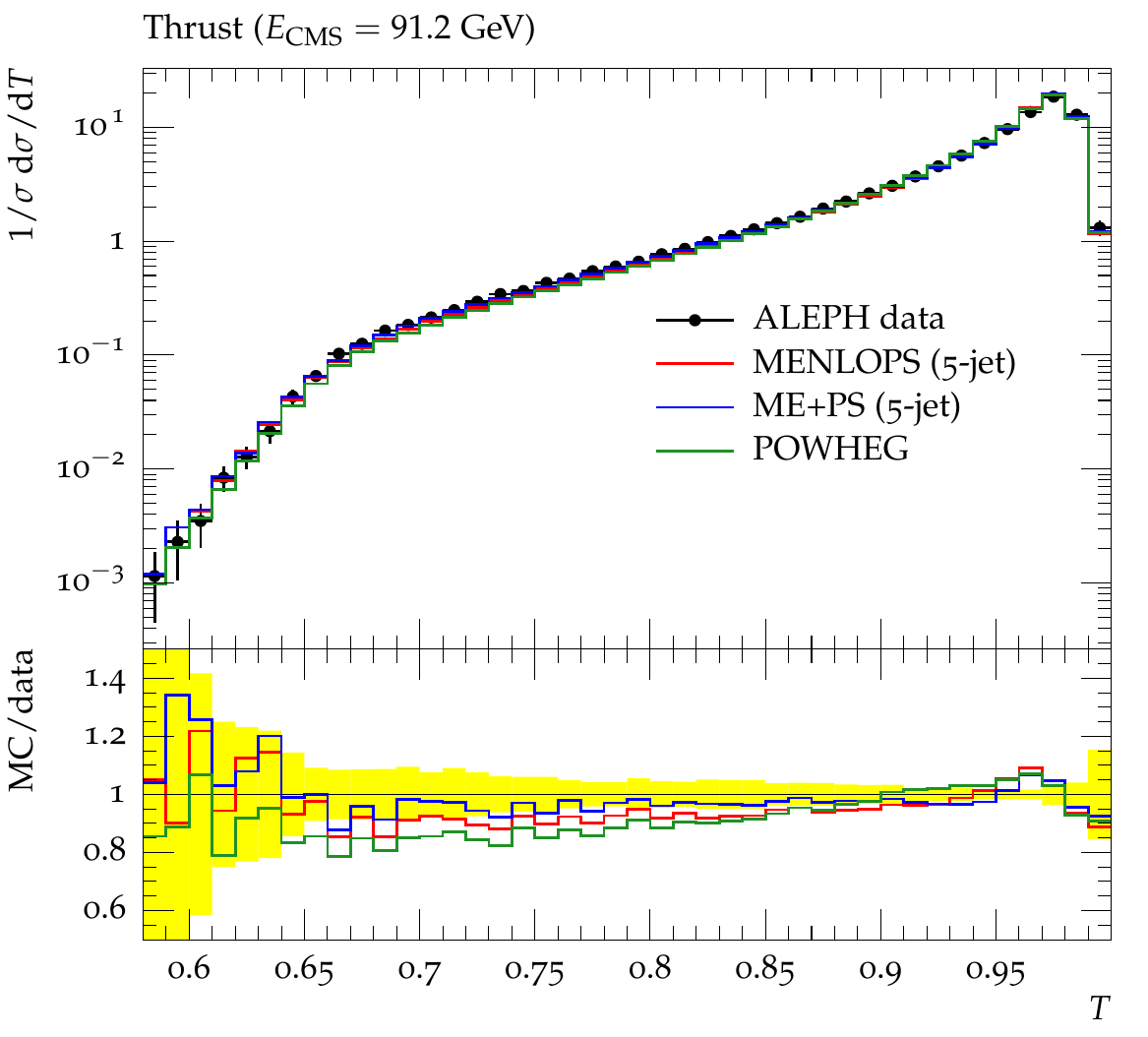}
  \end{center}
  \caption{
  C parameter and thrust distribution at LEP compared to data taken by 
  the ALEPH experiment~\cite{Heister:2003aj}.
  \label{fig:lep:jetmassdiff_thrust}}
\end{figure}

\begin{figure}[p]
  \begin{center}
  \includegraphics[width=0.45\textwidth]{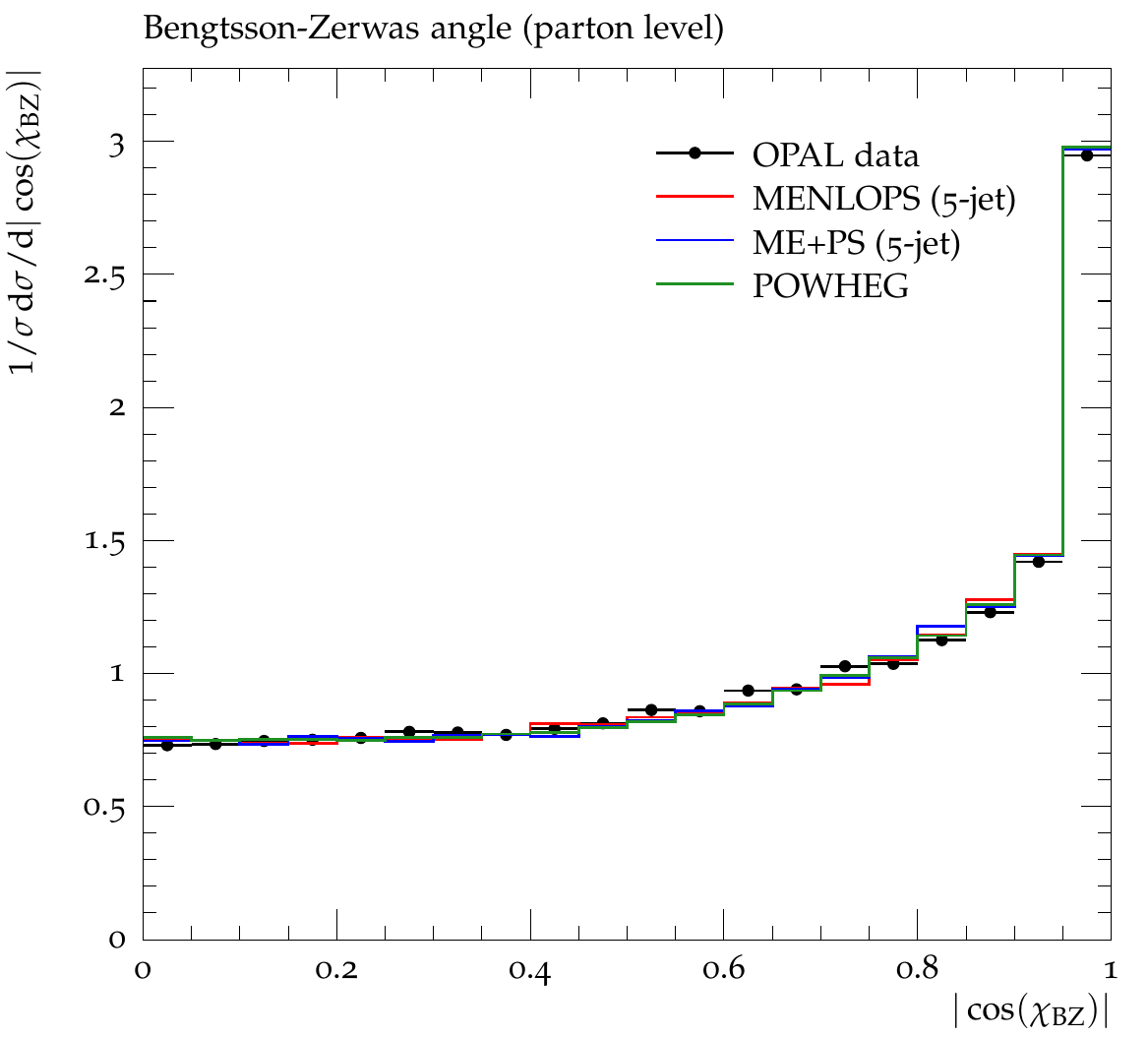}
  \hspace*{0.05\textwidth}
  \includegraphics[width=0.45\textwidth]{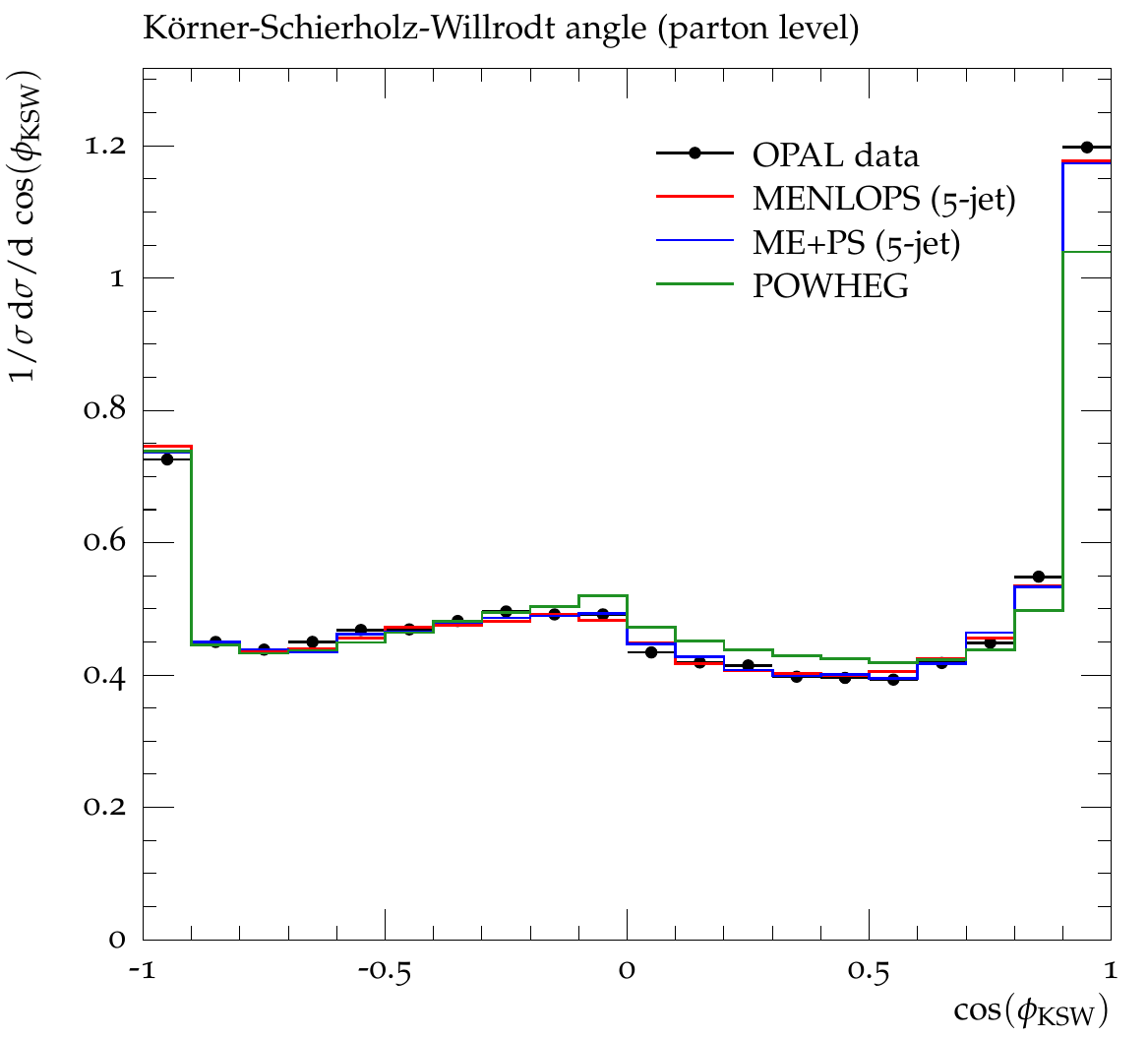}
  \vspace*{2mm}\\
  \includegraphics[width=0.45\textwidth]{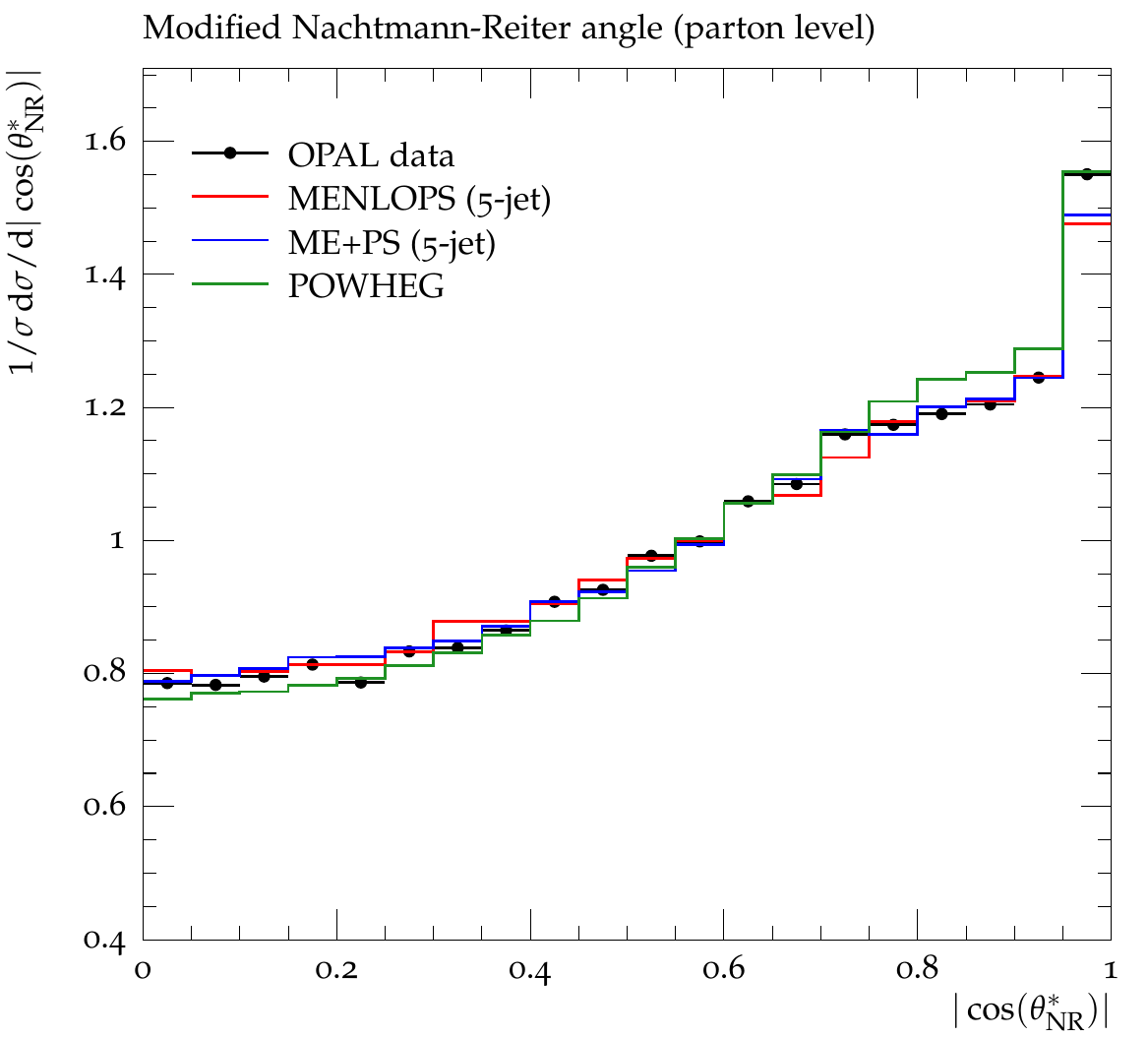}
  \hspace*{0.05\textwidth}
  \includegraphics[width=0.45\textwidth]{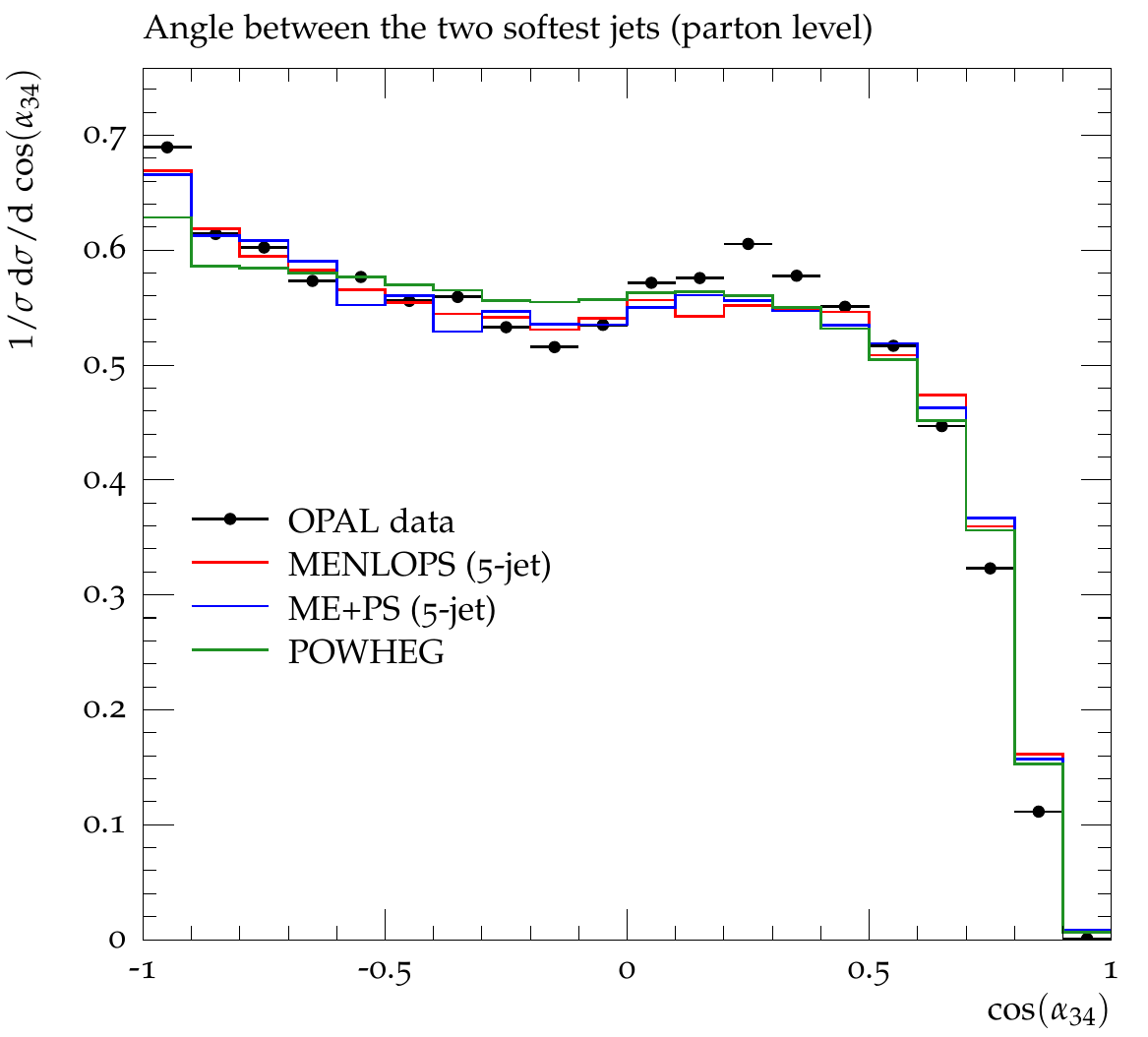}
  \end{center}
  \caption{
  Angles between the leading (in energy) four jets defined using the Durham 
  algorithm with $y_\mathrm{cut}=0.008$. Results at the parton level
  are compared to data from the OPAL experiment~\cite{Abbiendi:2001qn}.
  \label{fig:lep:4jetangles}}
\end{figure}

In this section we focus on electron-positron annihilation into hadrons at LEP 
energies ($\sqrt{s}=$91.25 GeV). The core process of the simulation is therefore
the reaction $e^+e^-\to q\bar{q}$. A full wealth of experimental data has been
provided by the LEP experiments, which allows to assess the quality of the 
\MENLOPS approach in this simplest realistic scenario.  Although the 
improvements discussed in this paper concern only the perturbative QCD part of
the Monte-Carlo simulation, our results account for hadronisation effects 
using the Lund model~\cite{Andersson:1983ia,*Andersson:1998tv,Sjostrand:1995iq,*Sjostrand:2006za}
to make them comparable to experimental data.  
Virtual matrix elements needed for the simulation were supplied by code 
provided by the \BlackHat collaboration~\cite{whitehat:2010aa,*Berger:2009zg,*Berger:2009ep,*Berger:2010vm}.

Figure~\ref{fig:lep:23jetrate_34jetrate} highlights the improvement in the 
description of jet data. In the hard-emission region the \MENLOPS results 
for the $2\to3$-, the $3\to4$- and the $4\to5$-jet rate are generally closer 
to the data than the \POWHEG ones, which hints at the success of the simulation. 
Deviations in the $5\to6$-jet rate are most likely due to the fact that
matrix elements for six-jet production are not included. Note that these 
distributions are normalised to the total cross section, such that no rate 
difference between the ME+PS and the \MENLOPS samples can be observed. 

Figures~\ref{fig:lep:jetbroadening_cparam} and~\ref{fig:lep:jetmassdiff_thrust}
show examples of event-shape variables, which are all very well described in the
hard-emission region by the \MENLOPS simulation. 
Several distributions for jet angular correlations in 4-jet production, that
have been important for the analysis of QCD and searches for physics beyond 
the Standard Model are investigated in Fig.~\ref{fig:lep:4jetangles}.
The good fit to those data proves that correlations amongst the final-state 
partons are correctly implemented by the higher-order matrix elements.


\subsection{Deep-inelastic lepton-nucleon scattering}
\label{Sec:DIS}

\begin{figure}[p]
  \begin{center}
  \includegraphics[width=0.6\textwidth]{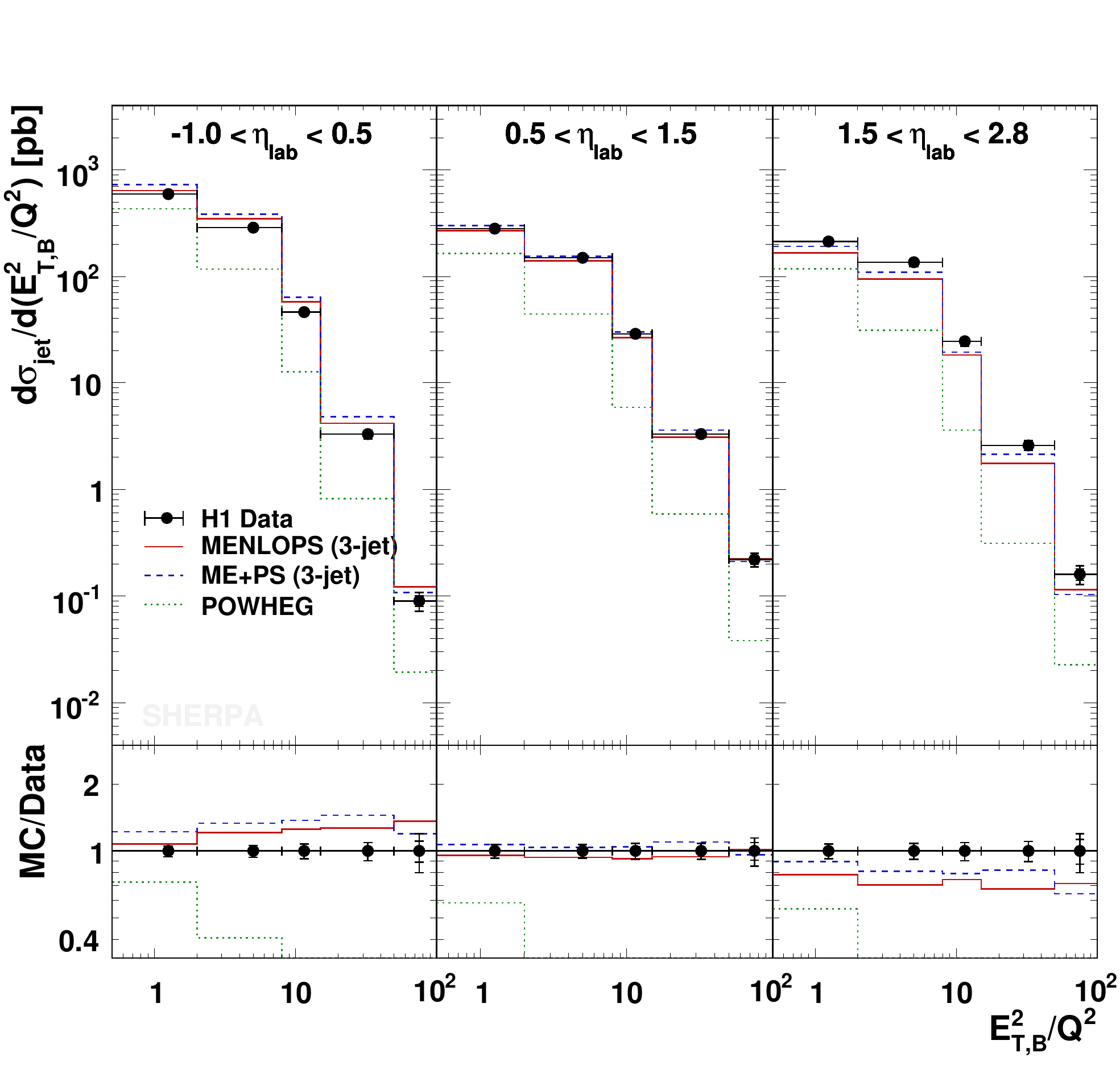}\hfill
  \includegraphics[width=0.39\textwidth]{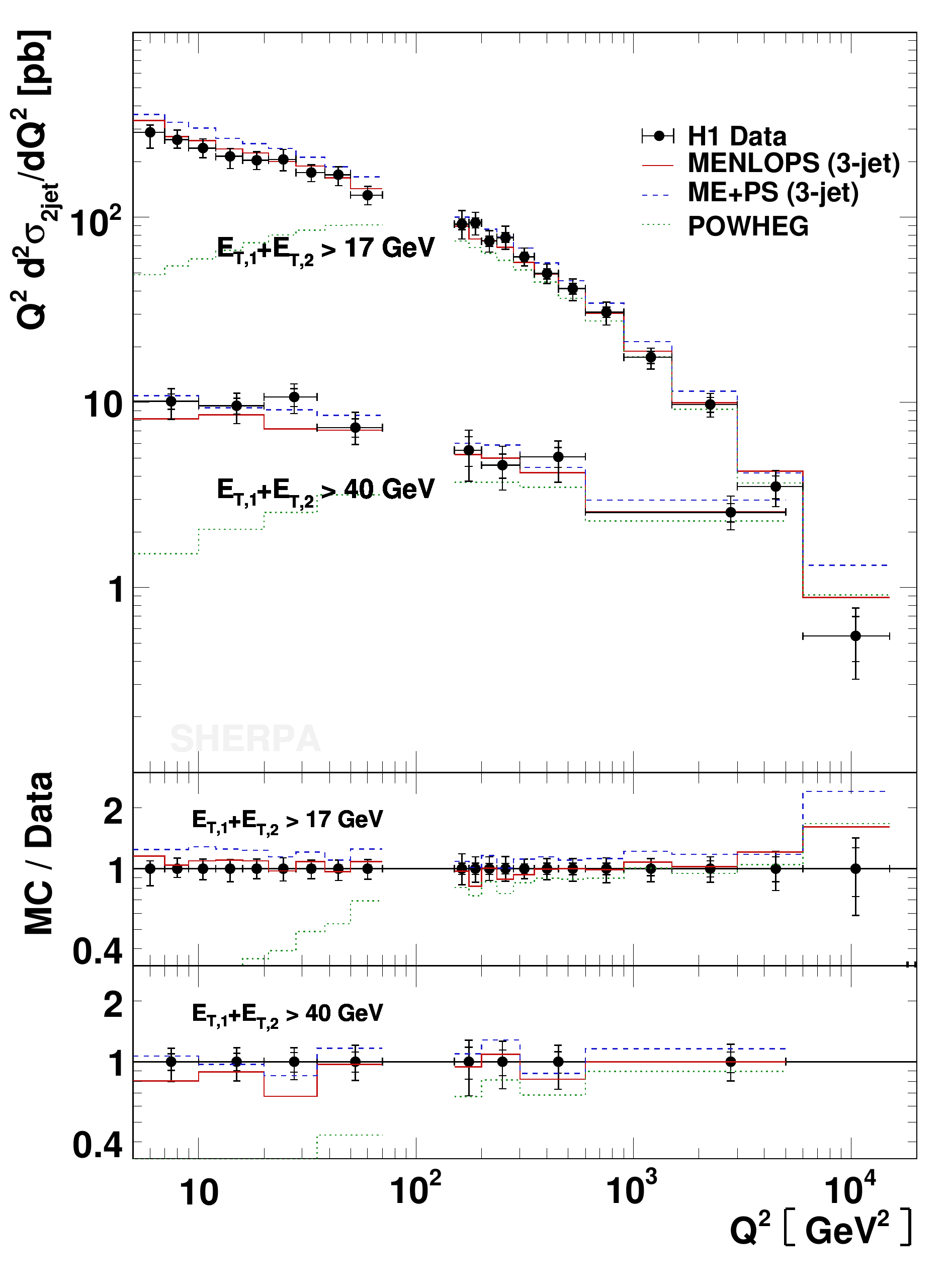}
  \end{center}
  \caption{
  Left: The inclusive jet cross section as a function of $E_{T,B}^2/Q^2$ in bins
  of $\eta_{lab}$, compared to data from the H1 collaboration~\protect\cite{Adloff:2002ew}.
  $E_{T,B}^2$ is the jet transverse energy in the Breit frame, while $\eta_{lab}$
  denotes the jet rapidity in the laboratory frame.
  Right: The dijet cross section as a function of $Q^2$ in bins of $E_{T,1}+E_{T,2}$,
  compared to data from the H1 collaboration~\protect\cite{Adloff:2000tq}.
  \label{fig:dis:q2}}
\end{figure}

\begin{figure}[p]
  \begin{center}
  \includegraphics[width=0.475\textwidth]{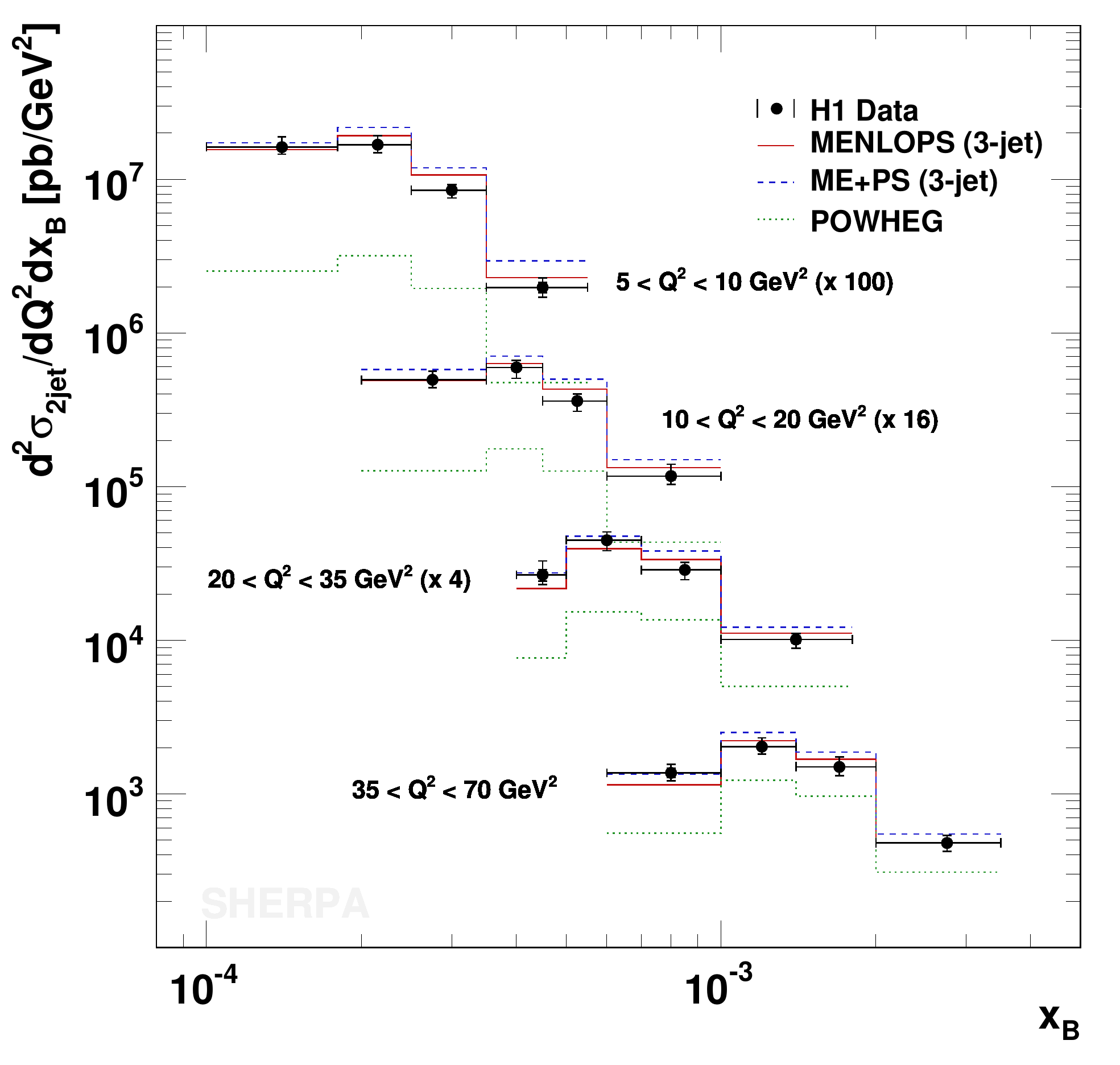}\hfill
  \includegraphics[width=0.475\textwidth]{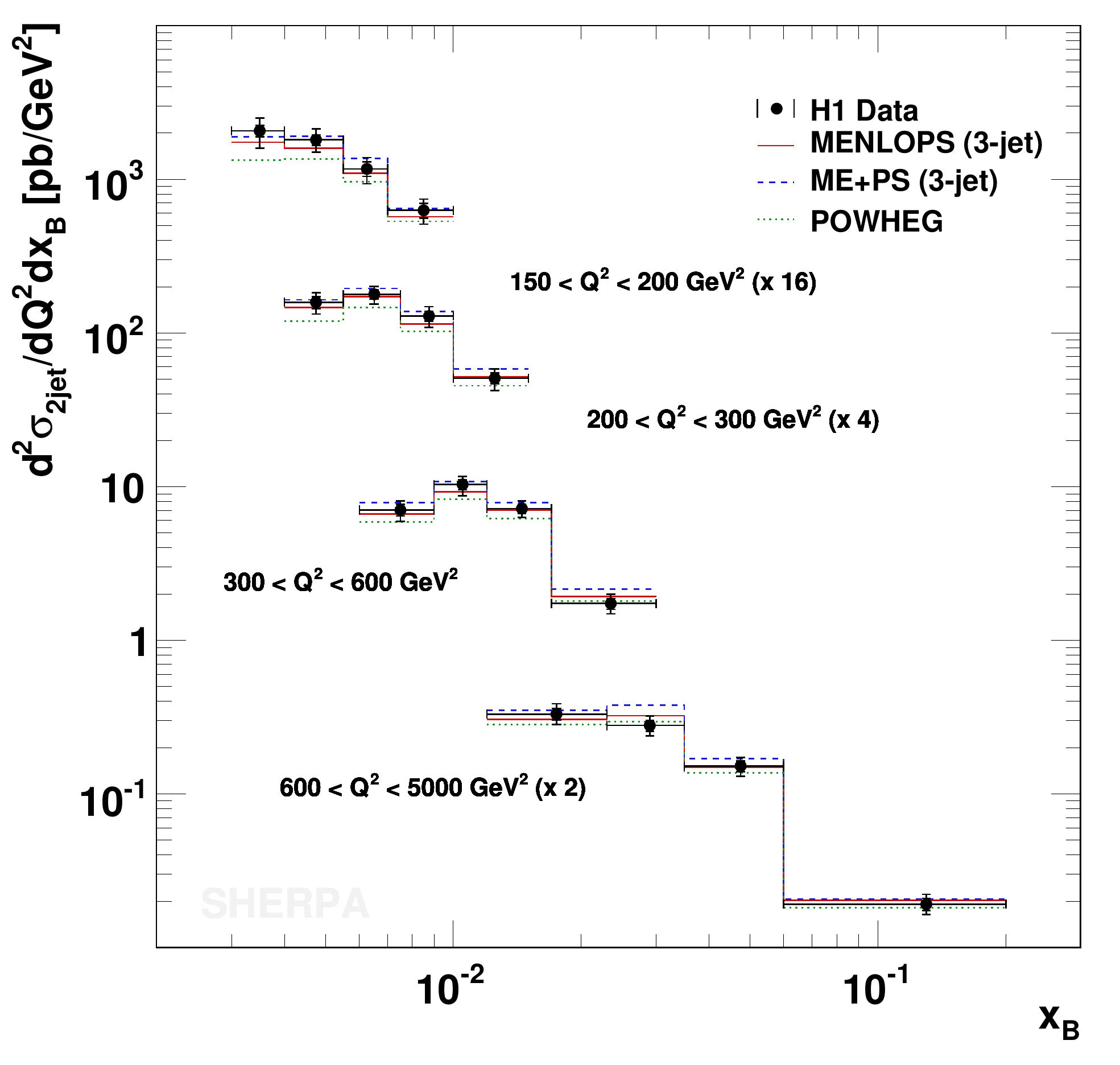}
  \end{center}
  \caption{
  The dijet cross section as a function of the Bj{\o}rken variable $x_{B}$,
  compared to data from the H1 collaboration~\protect\cite{Adloff:2000tq}.
  \label{fig:dis:xb_twojet}}
\end{figure}

\begin{figure}[p]
  \begin{center}
  \includegraphics[width=0.495\textwidth]{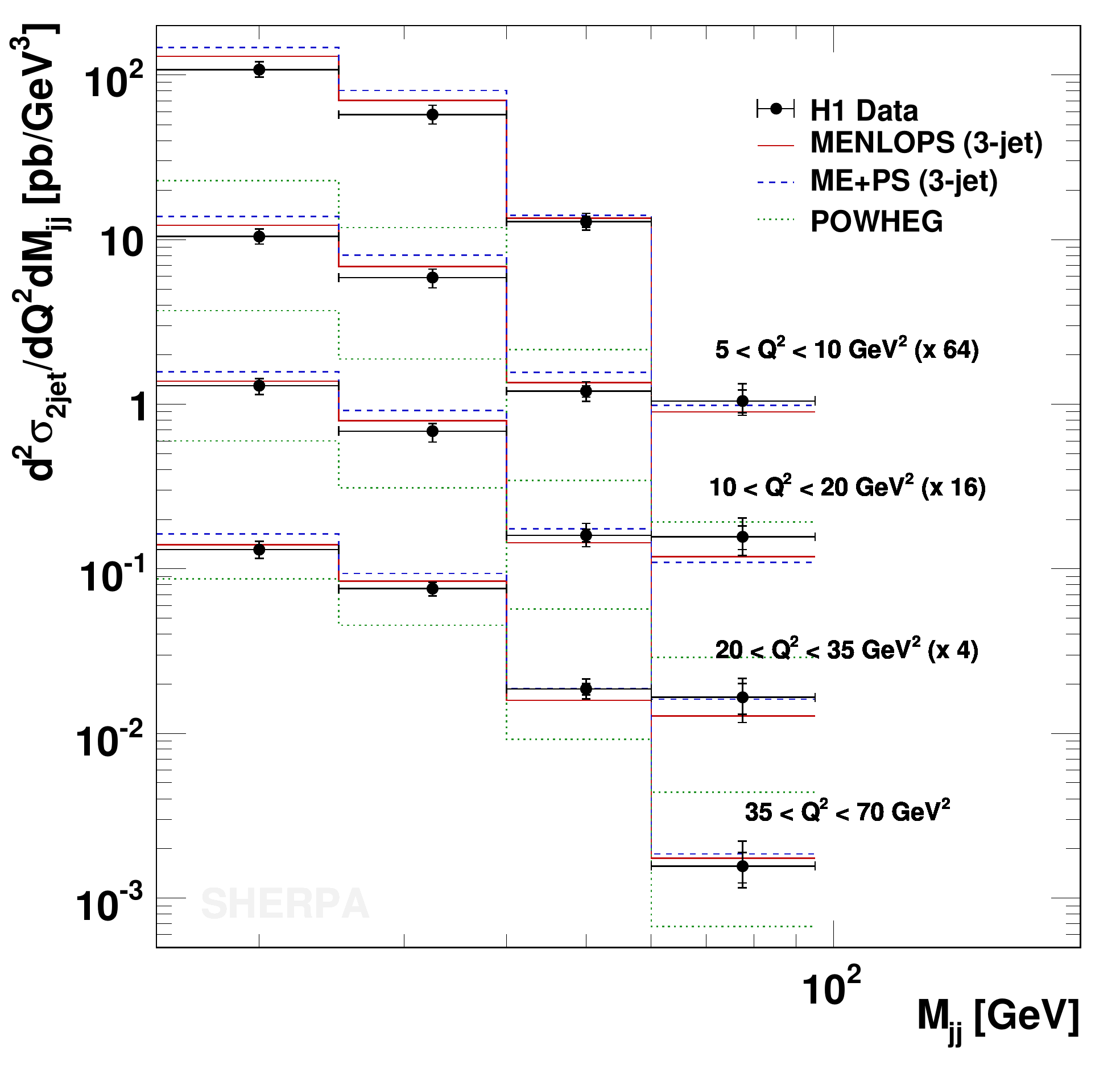}\hfill
  \includegraphics[width=0.495\textwidth]{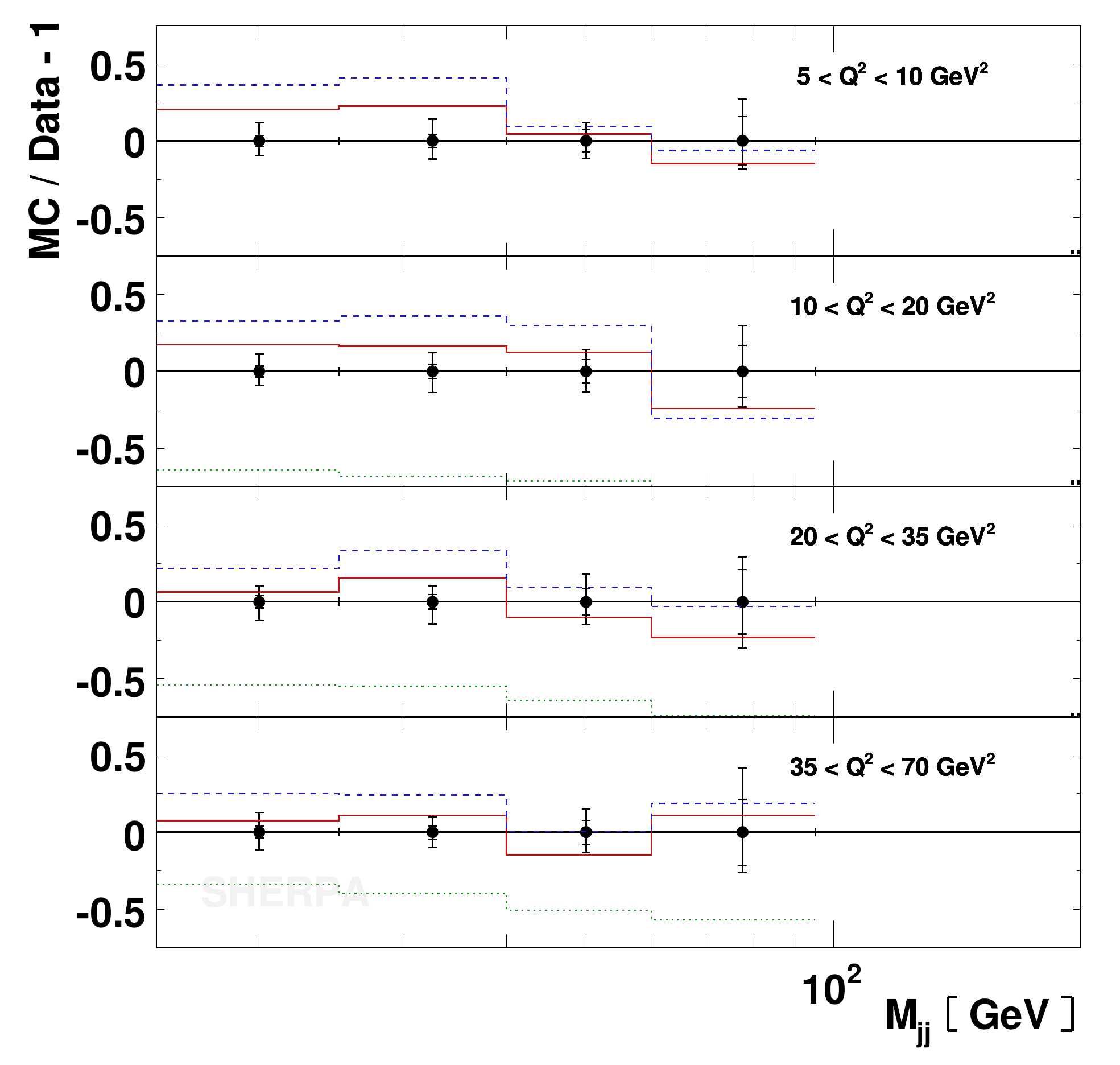}
  \vspace*{2mm}\\
  \includegraphics[width=0.495\textwidth]{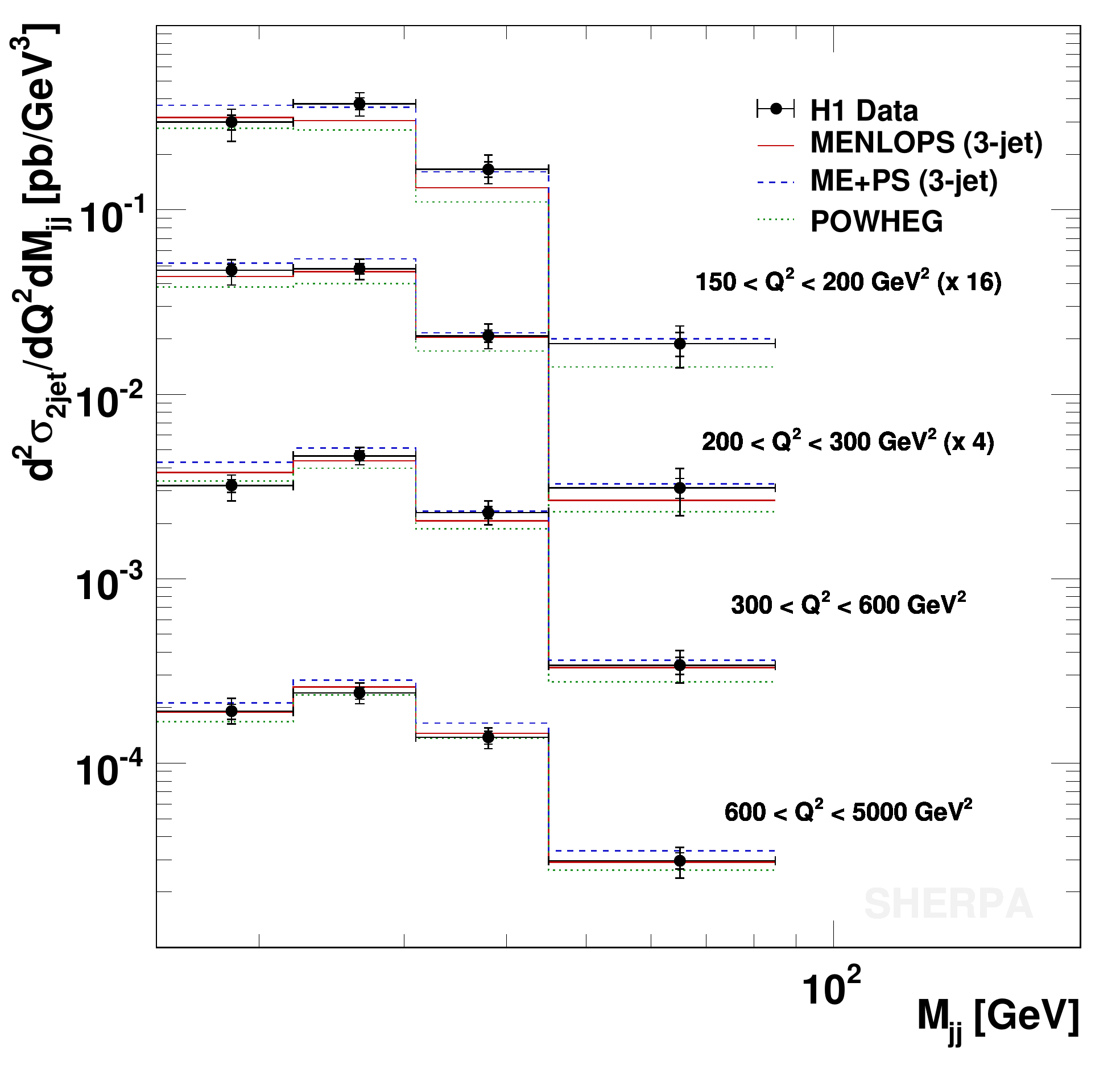}\hfill
  \includegraphics[width=0.495\textwidth]{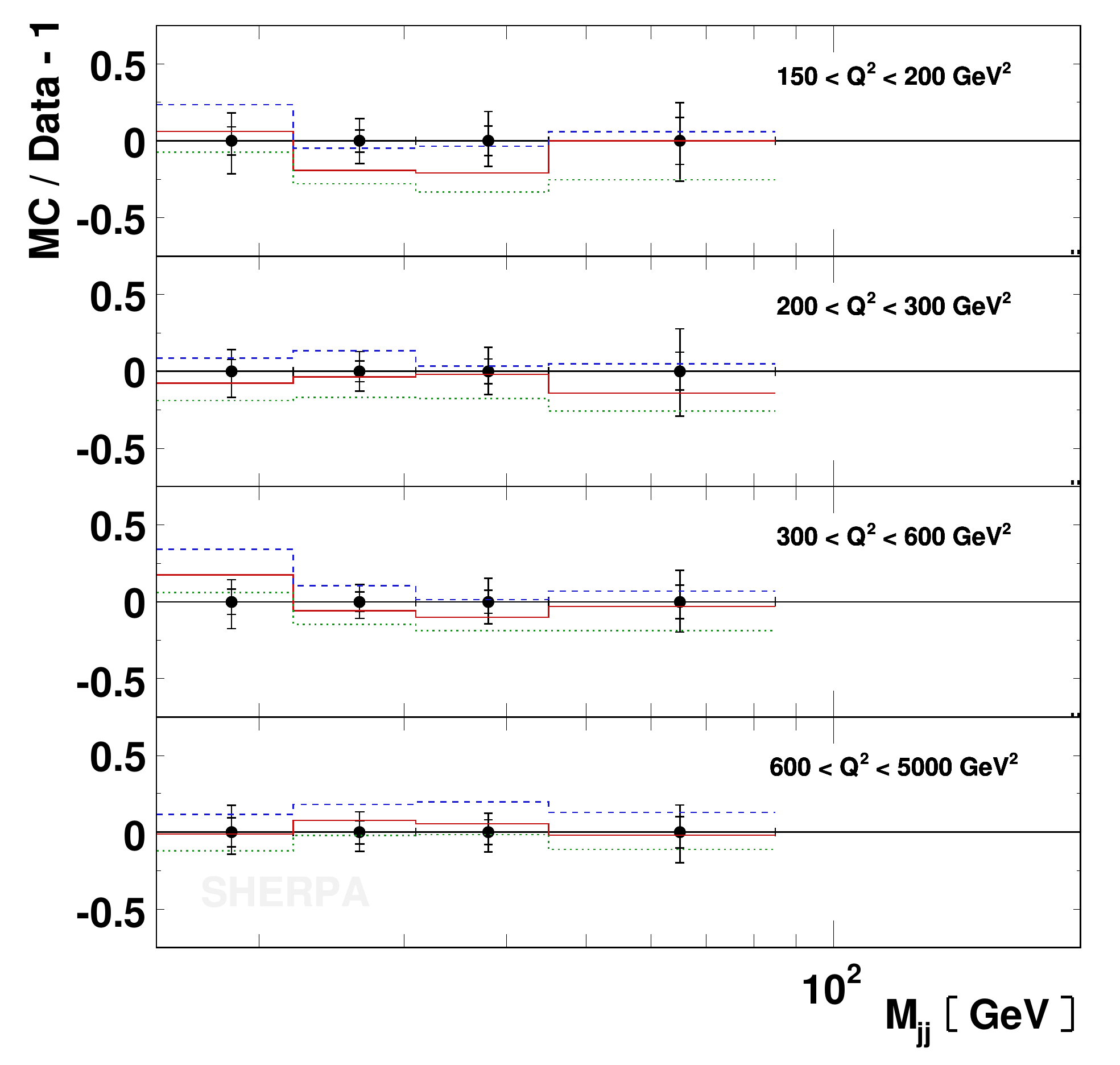}
  \end{center}
  \caption{
  The dijet cross section as a function of the dijet mass $m_{jj}$,
  compared to data from the H1 collaboration~\protect\cite{Adloff:2000tq}.
  \label{fig:dis:mjj_twojet}}
\end{figure}

Deep-inelastic scattering (DIS) is one of the best understood processes in 
perturbative QCD.  However, it has been an obstacle for a very long time to 
properly simulate hadronic final states in DIS using general-purpose Monte 
Carlo based on collinear factorisation.  Only recently, a consistent approach 
was presented~\cite{Carli:2010cg}, that allows to describe jet data throughout 
the experimentally accessible range of $Q^2$, the negative virtuality of the 
exchanged virtual $\gamma^*\!/Z$-boson.  It is absolutely mandatory 
for this method that a large number of final-state partons can be described 
by hard matrix elements in order to lift the severe restrictions on the 
real-emission phase space of the parton shower, which are imposed by the 
factorisation theorem.

In all our simulations we use the core process $e^+q\to e^+q$.
We present results for two analyses.  The first is the measurement of 
inclusive jet production in~\cite{Adloff:2002ew}, which covers different 
ranges of jet-pseudorapidity in the laboratory frame, $\eta_{lab}$, in 
the low-$Q^2$ domain $5<Q^2<100\,\rm GeV^2$.  Jets are defined using the 
inclusive $k_T$-algorithm~\cite{Ellis:1993tq,*Catani:1993hr} and are 
constrained to $E_{T,B}>5\,\rm GeV$ and the pseudorapidity range 
$-1<\eta_{lab}<2.8$, where $E_{T,B}$ is the jet transverse energy in the Breit 
frame.  The second analysis corresponds to the measurement of dijet 
production in~\cite{Adloff:2000tq}, which covered a wider range of $Q^2$ and 
produced many doubly differential jet spectra. The acceptance region
is $5<Q^2<15000\,\rm GeV^2$ and $-1<\eta_{lab}<2.5$. 
Jet transverse energies are subject to the cuts $E_{T,B\,1,2}>5\,\rm GeV$ and 
$E_{T,B\,1}+E_{T,B\,2}>17\,\rm GeV$.  The latter requirement is introduced to 
avoid $E_{T,B\,1}\approx E_{T,B\,2}$, which is the region of the phase space 
where next-to-leading order corrections are unstable due to implicit 
restrictions on soft emissions~\cite{Frixione:1997ks}.

As outlined in~\cite{Carli:2010cg}, a crucial observable is given by the 
inclusive jet cross section, differential with respect to $E_{T,B}^2/Q^2$.  
For $E_{T,B}^2/Q^2>1$ it probes a part of the phase space where leading order 
Monte-Carlo models without the inclusion of low-$x$ effects are bound to fail 
in their description of jet spectra.  Another very good observable to 
validate the proper Monte-Carlo simulation is the dijet cross section as a 
function of $Q^2$. While still a relatively inclusive quantity, it is 
an important indicator for the correct simultaneous implementation of 
inclusive DIS and the additional production of hard QCD radiation.  The 
high quality of the \MENLOPS prediction for the two above observables is 
confirmed in Fig.~\ref{fig:dis:q2}.  Discrepancies in the description of the 
$E_{T,B}^2/Q^2$-spectrum in the forward region can be attributed to the fact that 
the simulation is limited to three additional partons in the hard matrix 
elements. This restriction is imposed by the usage of the matrix-element 
generator \Amegic~\cite{Krauss:2001iv}.  Figures~\ref{fig:dis:xb_twojet} 
and~\ref{fig:dis:mjj_twojet} exemplify again that the \MENLOPS simulation 
correctly predicts multijet differential distributions in all 
regions of the phase space, while the \POWHEG approach fails in the low-$Q^2$ 
domain.

\clearpage

\subsection{Drell-Yan lepton-pair production}
\label{Sec:DY}

\begin{figure}[p]
  \begin{center}
  \includegraphics[width=0.5\textwidth]{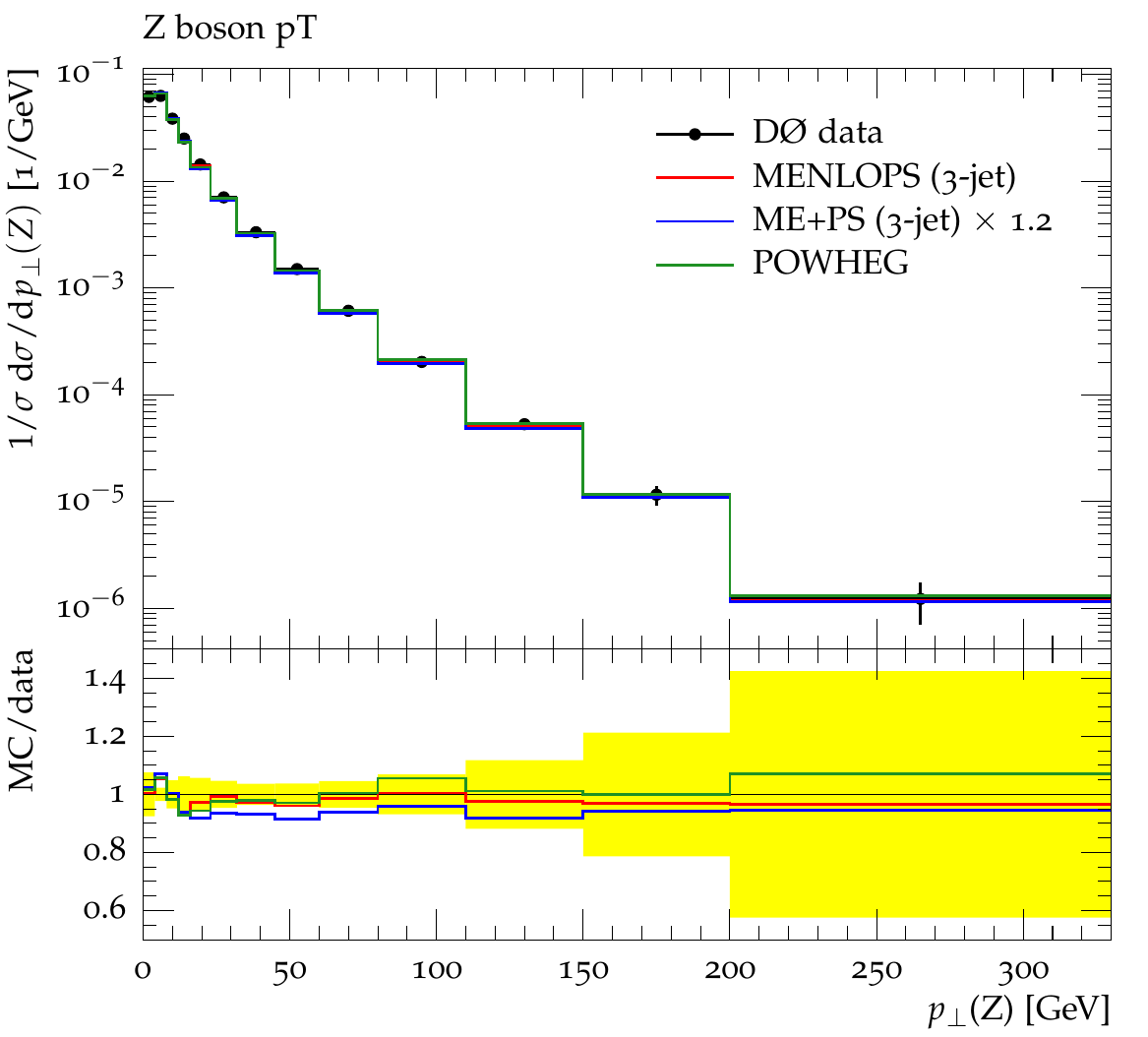}\nolinebreak
  \includegraphics[width=0.5\textwidth]{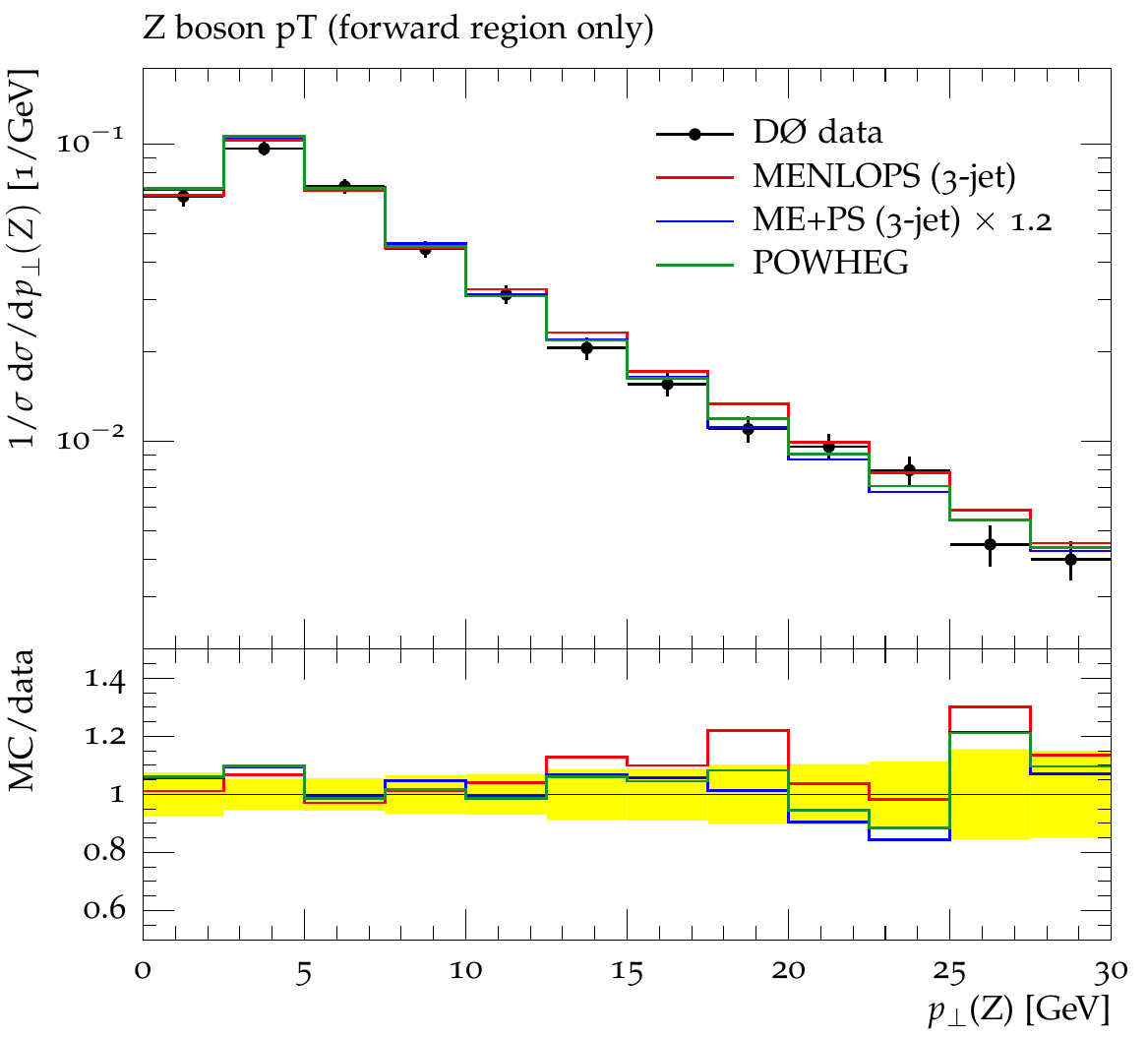}
  \end{center}
  \caption{
  The transverse momentum of the reconstructed $Z$ boson in Drell-Yan lepton-pair
  production at the Tevatron at $\sqrt{s}=1.96$ TeV. Experimental data stem
  from the D\O\ experiment~\cite{Abazov:2010kn,Abazov:2007nt} and are described in
  the text.
  \label{fig:tev:ZpT_Zy}}
\end{figure}

\begin{figure}[p]
  \begin{center}
  \includegraphics[width=0.5\textwidth]{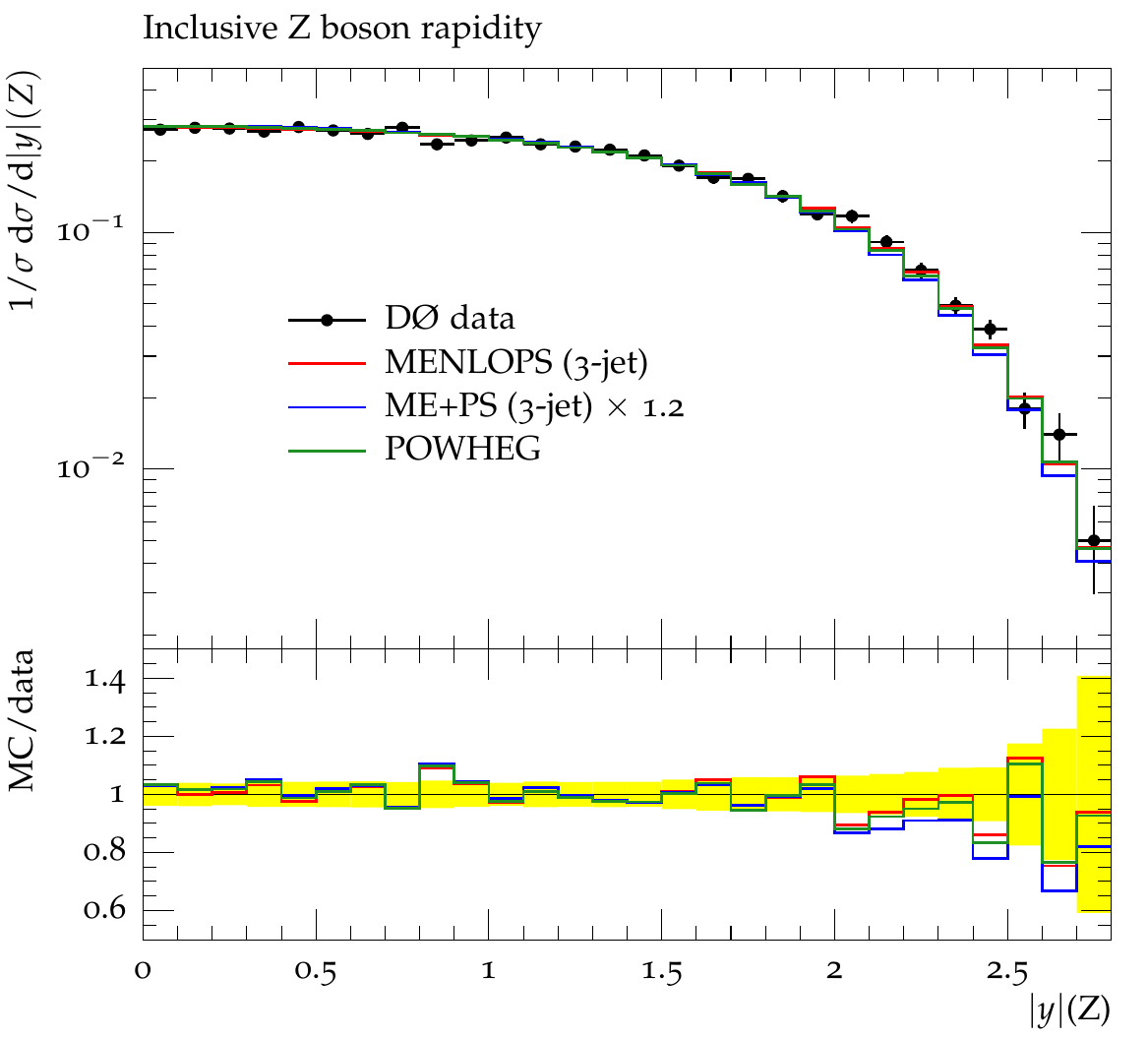}\nolinebreak
  \includegraphics[width=0.5\textwidth]{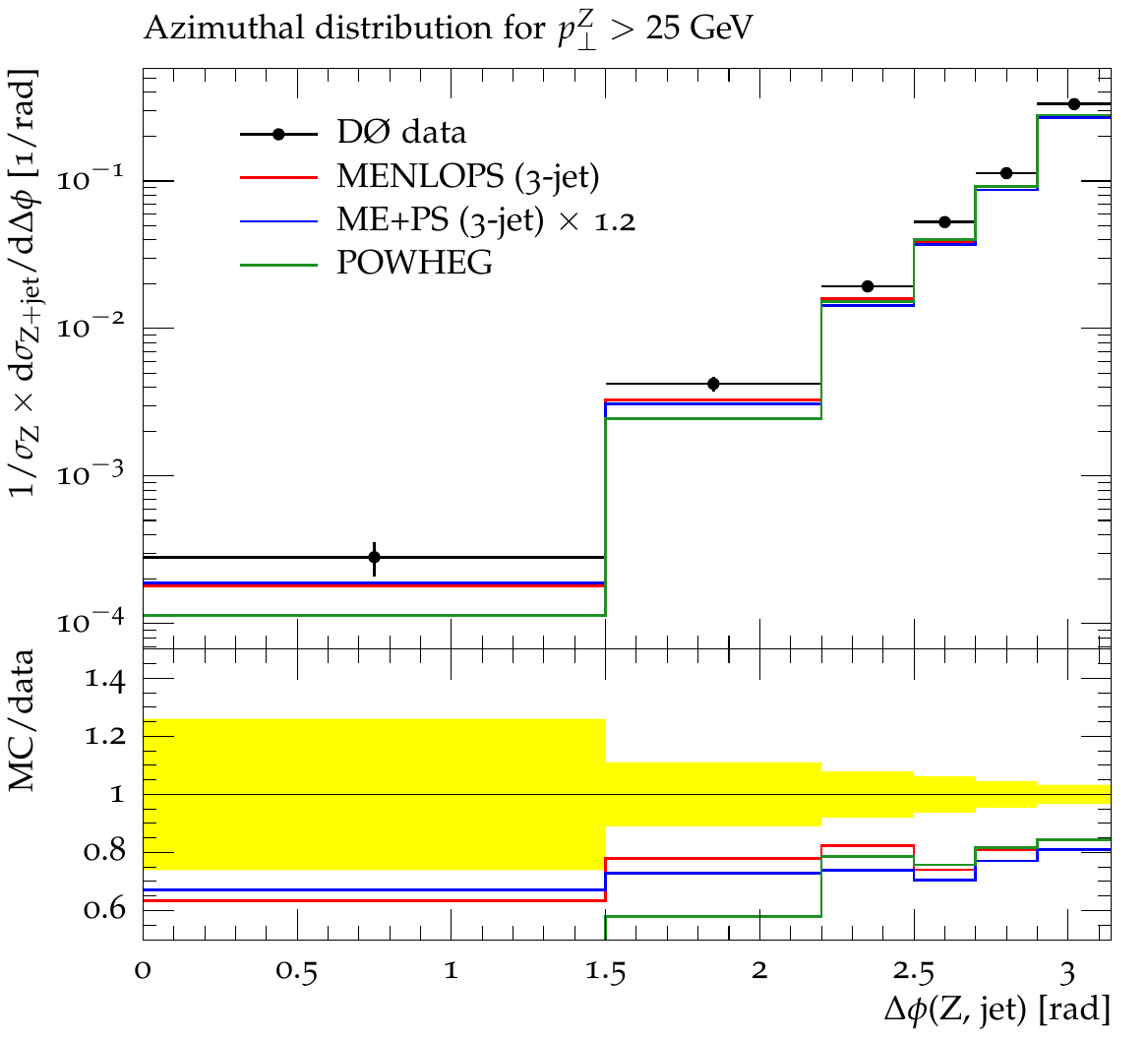}
  \end{center}
  \caption{
  Rapidity of the reconstructed $Z$ boson~\cite{Abazov:2007jy} (left)
  and azimuthal separation of the boson and the leading
  jet~\cite{Abazov:2009pp} (right)
  in Drell-Yan lepton-pair production at the Tevatron at $\sqrt{s}=1.96$ TeV.
  \label{fig:tev:Zy_dphi}}
\end{figure}

\begin{figure}[p]
  \begin{center}
  \includegraphics[width=0.5\textwidth]{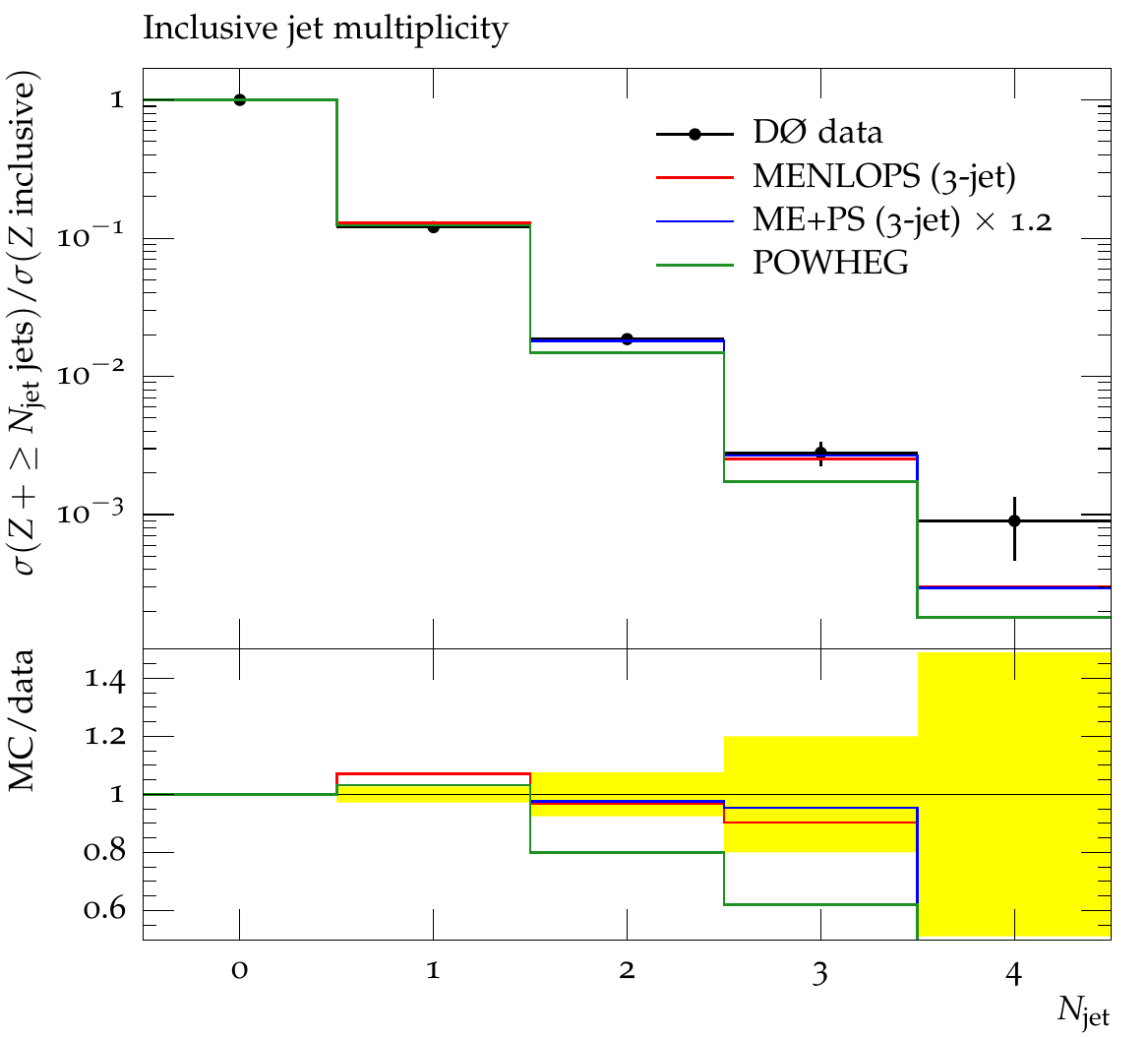}\nolinebreak
  \includegraphics[width=0.5\textwidth]{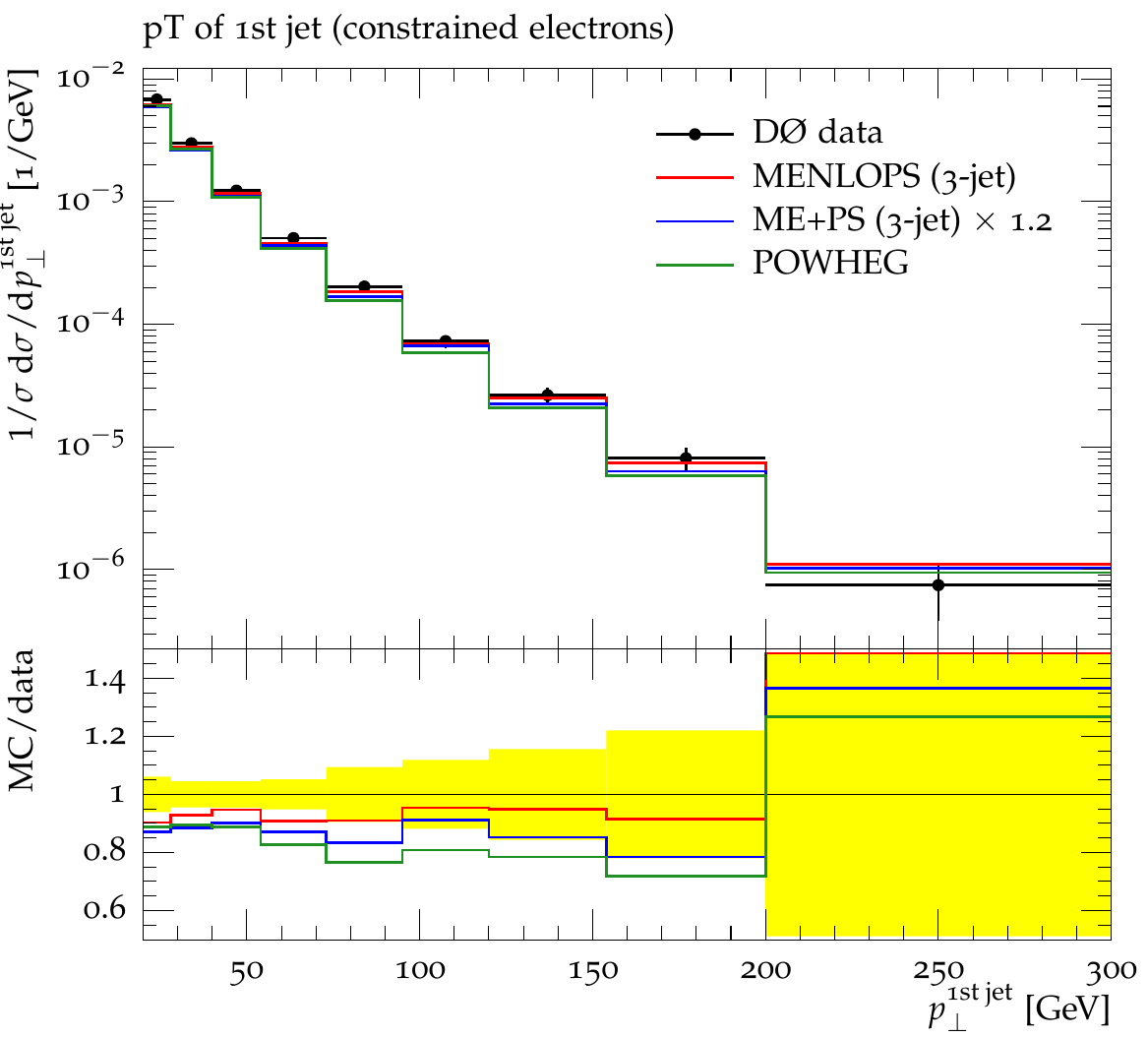}
  \end{center}
  \caption{
  Inclusive jet multiplicity~\cite{Abazov:2006gs} (left) and
  transverse momentum of the leading jet~\cite{Abazov:2009av} (right)
  in $Z+$jets events at the Tevatron at $\sqrt{s}=1.96$ TeV.
  \label{fig:tev:zjetmulti_zjet1}}
\end{figure}

\begin{figure}[p]
  \begin{center}
  \includegraphics[width=0.5\textwidth]{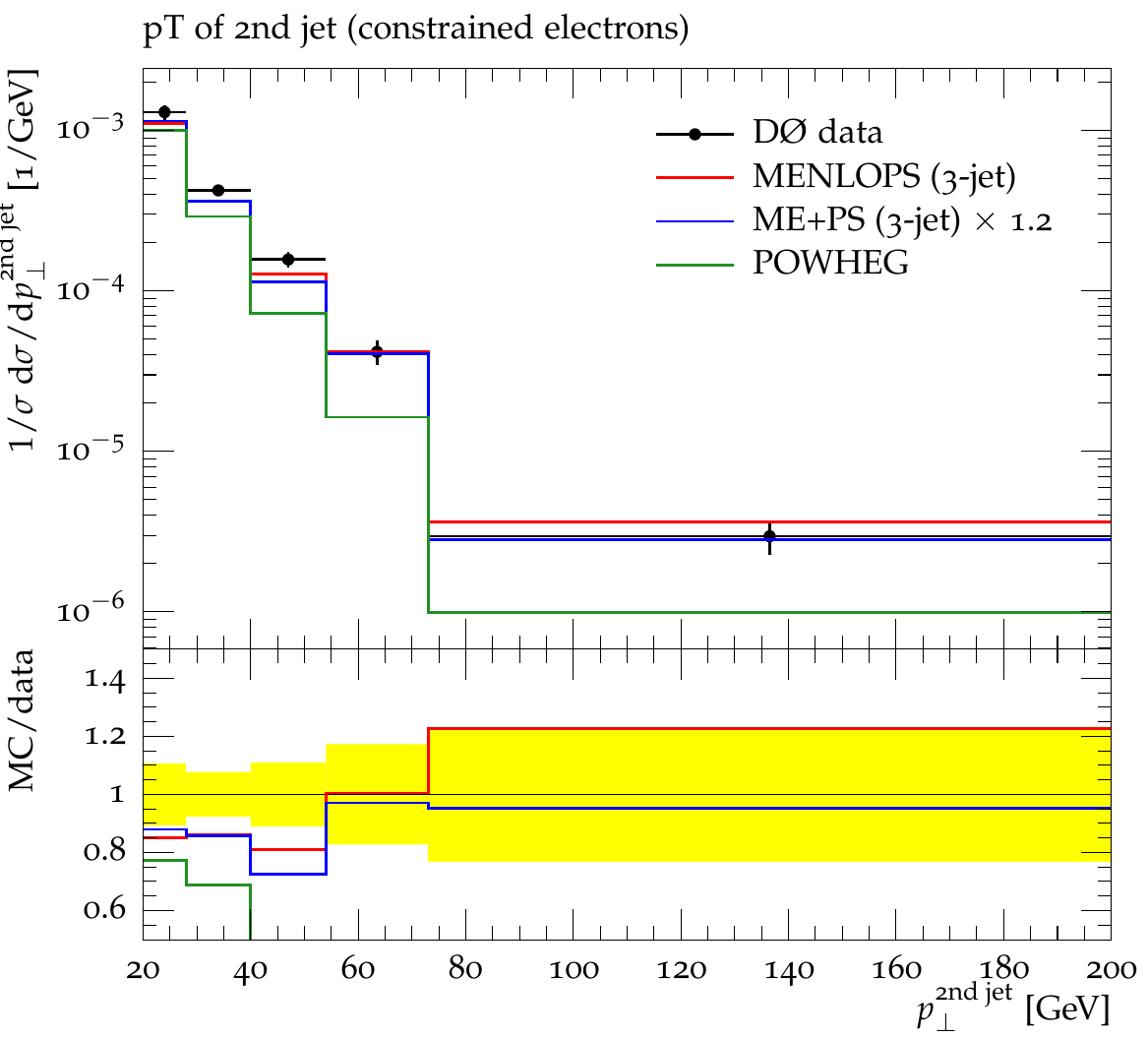}\nolinebreak
  \includegraphics[width=0.5\textwidth]{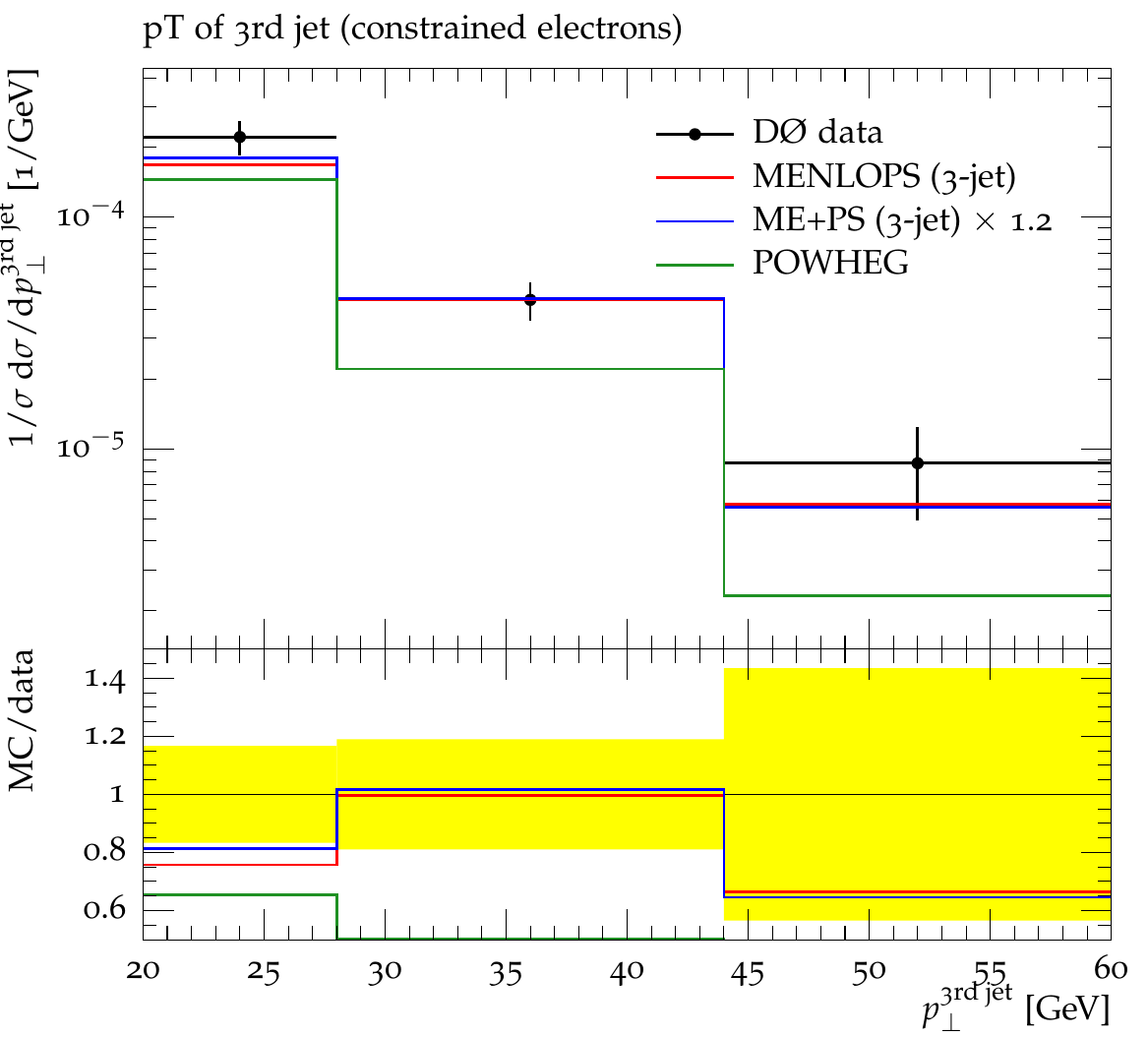}
  \end{center}
  \caption{
  Transverse momentum of the second and third jet~\cite{Abazov:2009av}
  in $Z+$jets events at the Tevatron at $\sqrt{s}=1.96$ TeV.
  \label{fig:tev:zjet2_zjet3}}
\end{figure}

Results for lepton-pair production through the Drell-Yan process are compared 
to data from the Tevatron at $\sqrt{s}=1.96$ TeV 
in Figs.~\ref{fig:tev:ZpT_Zy}-\ref{fig:tev:zjet2_zjet3}, using the core process 
$q\bar{q}\to \ell\ell$, where $\ell=e,\mu$. The invariant mass of the
lepton pair was restricted to be within $66<m_{\ell\ell}/\mathrm{GeV}<116$
in the simulation. The \MENLOPS and ME+PS samples use tree-level matrix elements
up to $Z+3$ jets with a merging cut of $Q_\text{cut}=20$~GeV.
Virtual matrix elements are provided by 
\BlackHat~\cite{whitehat:2010aa,*Berger:2009zg,*Berger:2009ep,*Berger:2010vm}.
The $Z\to \ell^+\ell^-$ decay is corrected for QED next-to-leading
order and soft-resummation effects using the Yennie-Frautschi-Suura (YFS)
approach~\cite{Schonherr:2008av}.

The Tevatron experiments provide a wealth of measurements sensitive to QCD
corrections in Drell-Yan production. Fig.~\ref{fig:tev:ZpT_Zy} shows the
transverse momentum distribution of the lepton pair in two different analyses
from the \DO experiment.  The left hand plot displays a very recent analysis
using the $Z\to \mu\mu$ channel~\cite{Abazov:2010kn} to measure the
$Z$-$p_\perp$ distribution normalised to the inclusive cross section. It
requires muons with $p_\perp>15$~GeV in a mass window of
$65<m_{\mu\mu}/\mathrm{GeV}<115$ and with $|\eta|<1.7$. The muon signal is 
corrected to the particle level including photons clustered in a cone 
of radius $R=0.2$ around each lepton.
The plot on the right hand side stems from an analysis in the electron
channel~\cite{Abazov:2007nt} which uses Monte-Carlo models to correct the
leptons for all acceptances including the pseudorapidity range and minimal
transverse momentum. Here we display the peak region of the transverse momentum
of forward $Z$ bosons with $|y_Z|>2$.  The agreement between all three 
approaches and the measurement is outstanding. In the bins at $p_\perp<10$~GeV
non-perturbative effects like the intrinsic transverse momentum of partons in
a proton might play a role. Related Monte-Carlo models in \Sherpa could be tuned 
to reach an even better agreement. Still, the Monte-Carlo prediction lies within 
the experimental error band over the full range.

Two more measurements from the \DO experiment are displayed in
Fig.~\ref{fig:tev:Zy_dphi}. The pseudorapidity of the $Z$
boson~\cite{Abazov:2007jy} was measured in the electron channel requiring
electrons with $p_\perp>15$~GeV in the mass window
$71<m_\text{ee}/\mathrm{GeV}<111$.  Again, all three Monte-Carlo approaches
agree very well with the experimental data. The right hand plot shows the
azimuthal correlation between the $Z$ boson and the leading
jet~\cite{Abazov:2009pp}.  This is a measurement in the muon channel
with the same selection cuts as described above.  The distribution has been
normalised using the inclusive $Z$ cross section and the comparison shows
that the three approaches underestimate the total rate for $Z+$jet production
with respect to inclusive $Z$ production by approximately 10\%.  This might
hint at the need for NLO accuracy also in the $Z+$jet process.  It is remarkable
though that the inclusion of higher-order tree-level matrix elements
significantly improves the shape of the distribution with respect to the
\POWHEG{} simulation.

The observables presented so far are mainly sensitive to the correct
description of the leading jet. For that reason even the \POWHEG{} approach is
well capable of providing sufficient accuracy in their prediction.
We now proceed to observables sensitive to higher-order corrections.

Figure~\ref{fig:tev:zjetmulti_zjet1} (left) shows the inclusive jet
multiplicity~\cite{Abazov:2006gs} for jets constructed using the D\O\ improved
legacy cone algorithm~\cite{Blazey:2000qt} with a cone radius of $R=0.5$ and
$p_\perp>20$~GeV. Jets were required to lie in $|\eta|<2.5$ and to be separated
from the leptons by $\Delta R(\ell,\text{jet})>0.4$.
While \POWHEG{} agrees with the data for the $N_\text{jet}=1$ bin it fails to
predict the rate of events with more than one jet. The \MENLOPS{} and ME+PS
predictions impressively demonstrate the effect of higher-order
corrections provided by tree-level matrix elements up to the third jet. They
agree with the measurement within the error bands for $N_\text{jet}=2,3$ and
as expected fail to predict the correct four-jet rate because no matrix-element
corrections have been applied at that multiplicity.

Transverse momentum spectra of the three leading jets accompanying the $Z$ boson
were measured by D\O\ in~\cite{Abazov:2009av}. The distributions in
Fig.~\ref{fig:tev:zjetmulti_zjet1} (right) and \ref{fig:tev:zjet2_zjet3} are
normalised to the inclusive cross section for $Z$ production and the jets have
been constructed using the same settings as in the multiplicity measurement.
Both \MENLOPS and ME+PS deliver a very good description of these spectra while
\POWHEG fails to describe the rate and shape for the second and third jet.

\clearpage

\subsection{\texorpdfstring{$W+$}{W+}jets Production}
\label{Sec:Wjets}

\begin{figure}[p]
  \begin{center}
  \includegraphics[width=0.45\textwidth]{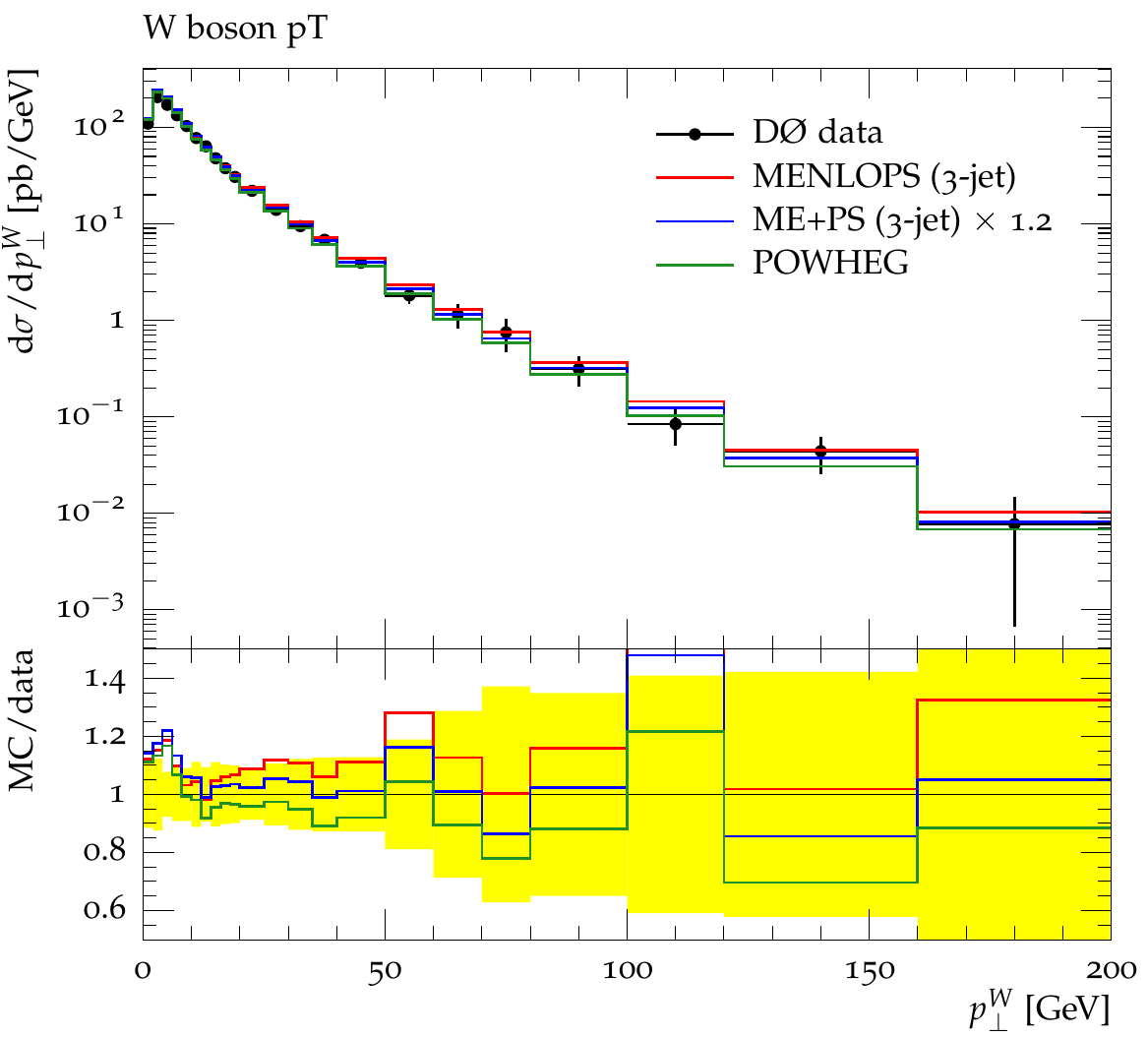}
  \hspace*{0.05\textwidth}
  \includegraphics[width=0.45\textwidth]{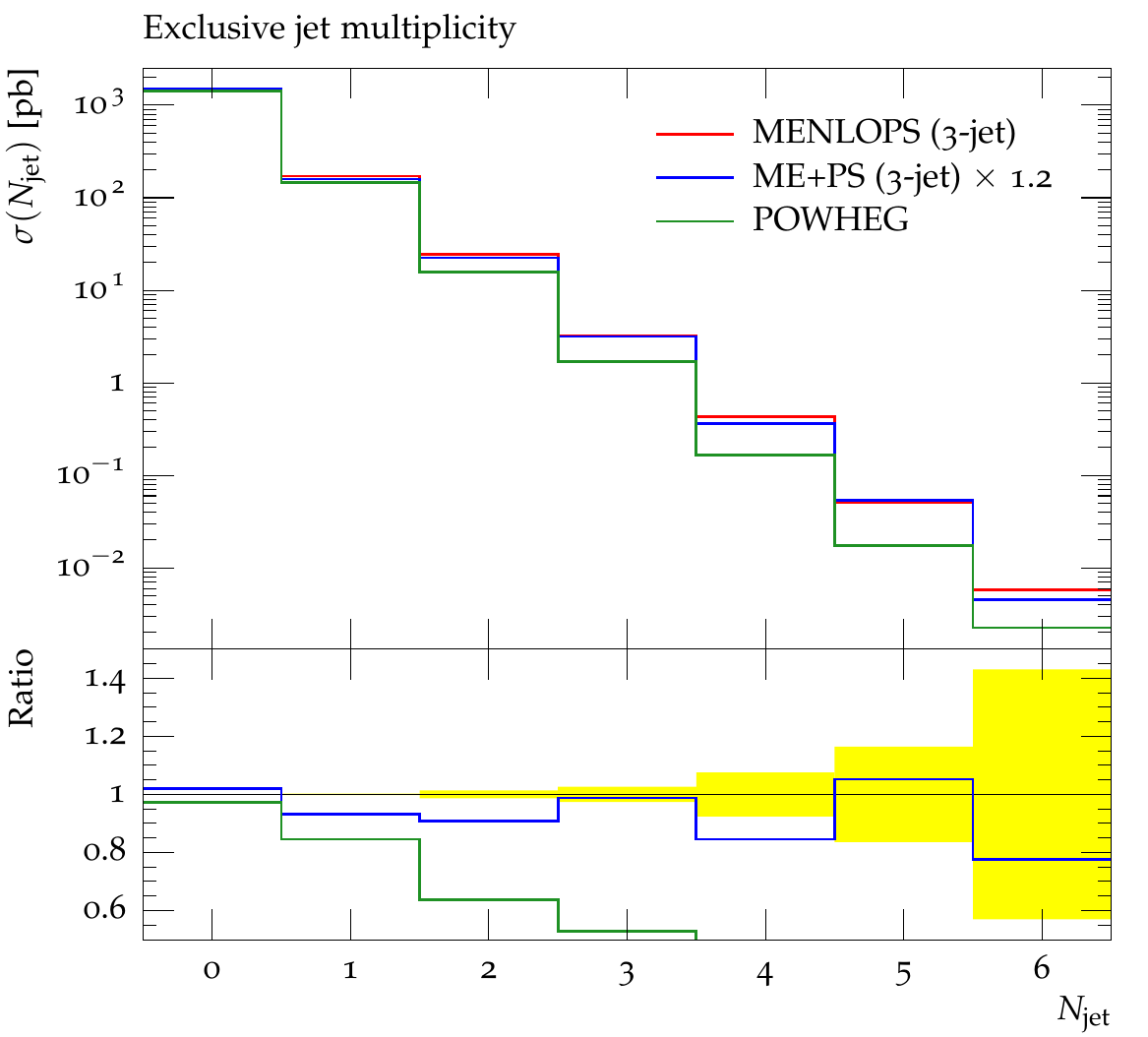}
  \end{center}
  \caption{
  Transverse momentum of the $W$, compared to data taken by the \protect\DO 
  collaboration~\cite{Abbott:2000xv}, and the exclusive jet multiplicity in 
  inclusive $W$ production at the Tevatron at $\sqrt{S}=1.8$ TeV.
  \label{fig:wjets:hpt_jetsdr12}}
\end{figure}

\begin{figure}[p]
  \begin{center}
  \includegraphics[width=0.45\textwidth]{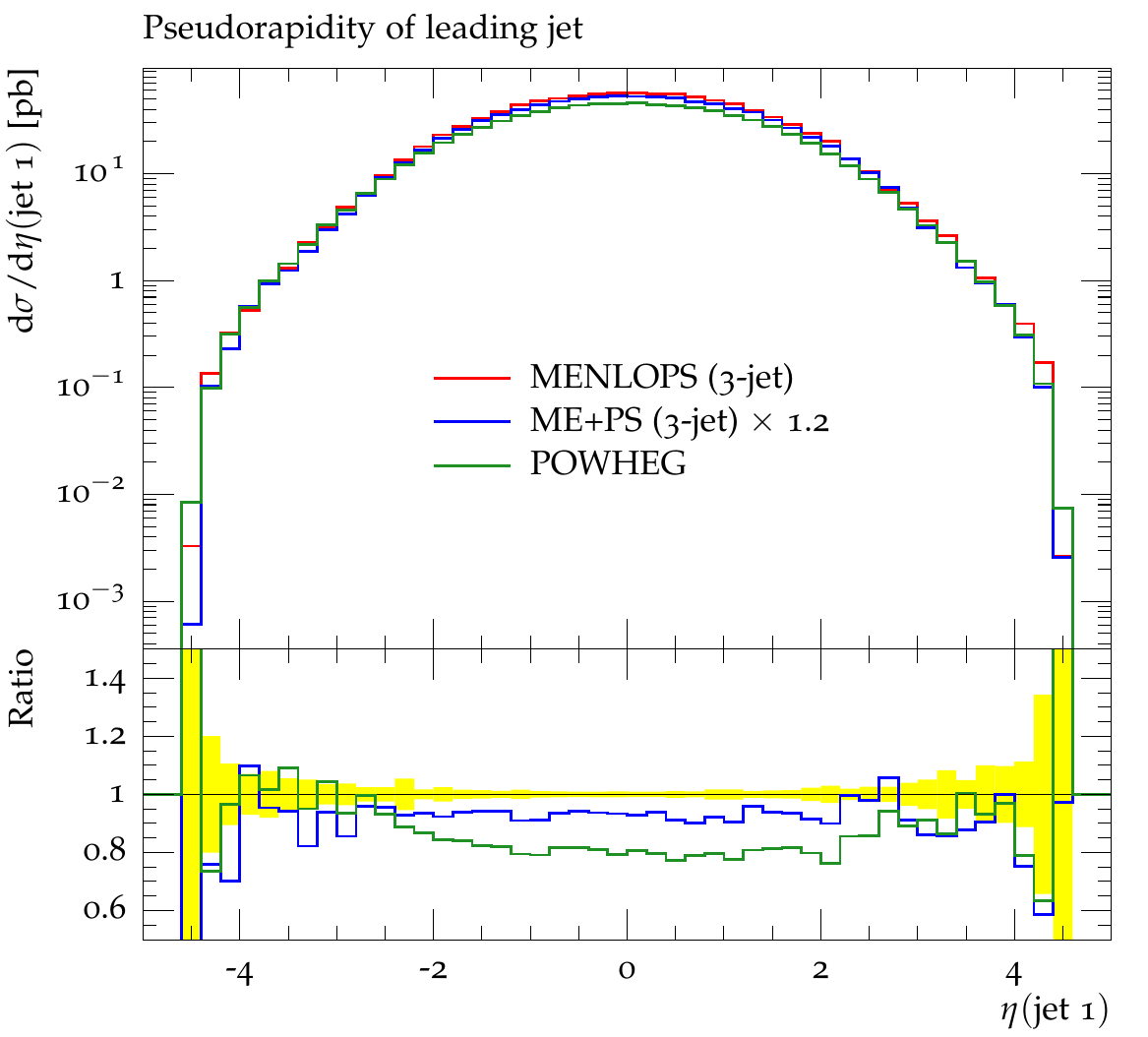}
  \hspace*{0.05\textwidth}
  \includegraphics[width=0.45\textwidth]{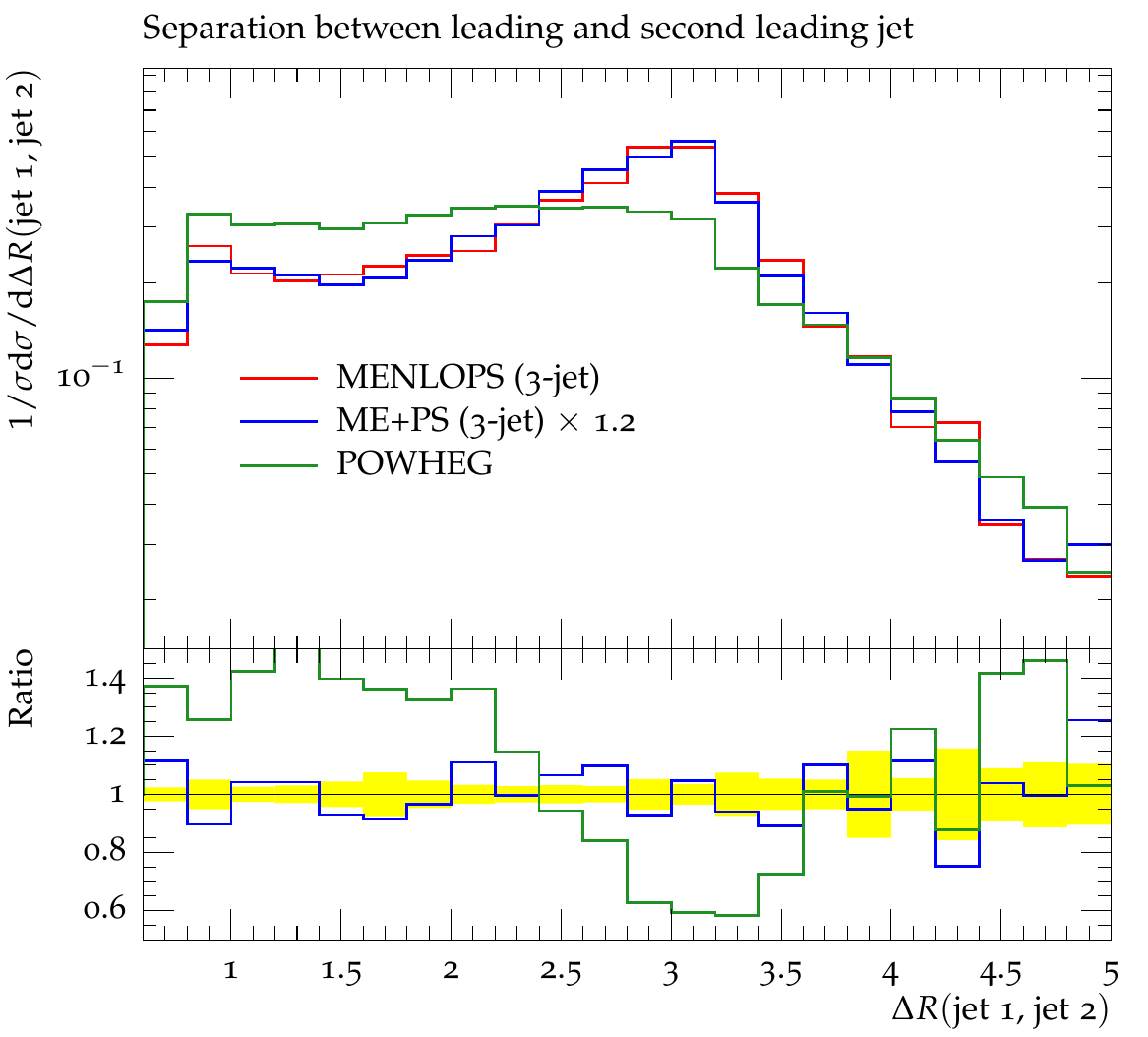}
  \end{center}
  \caption{
  Pseudorapidity of the hardest jet and angular separation of the first two hardest 
  jets in inclusive $W$ production at the Tevatron at $\sqrt{S}=1.8$~TeV.
  \label{fig:wjets:jets}}
\end{figure}

\begin{figure}[p]
  \begin{center}
  \includegraphics[width=0.45\textwidth]{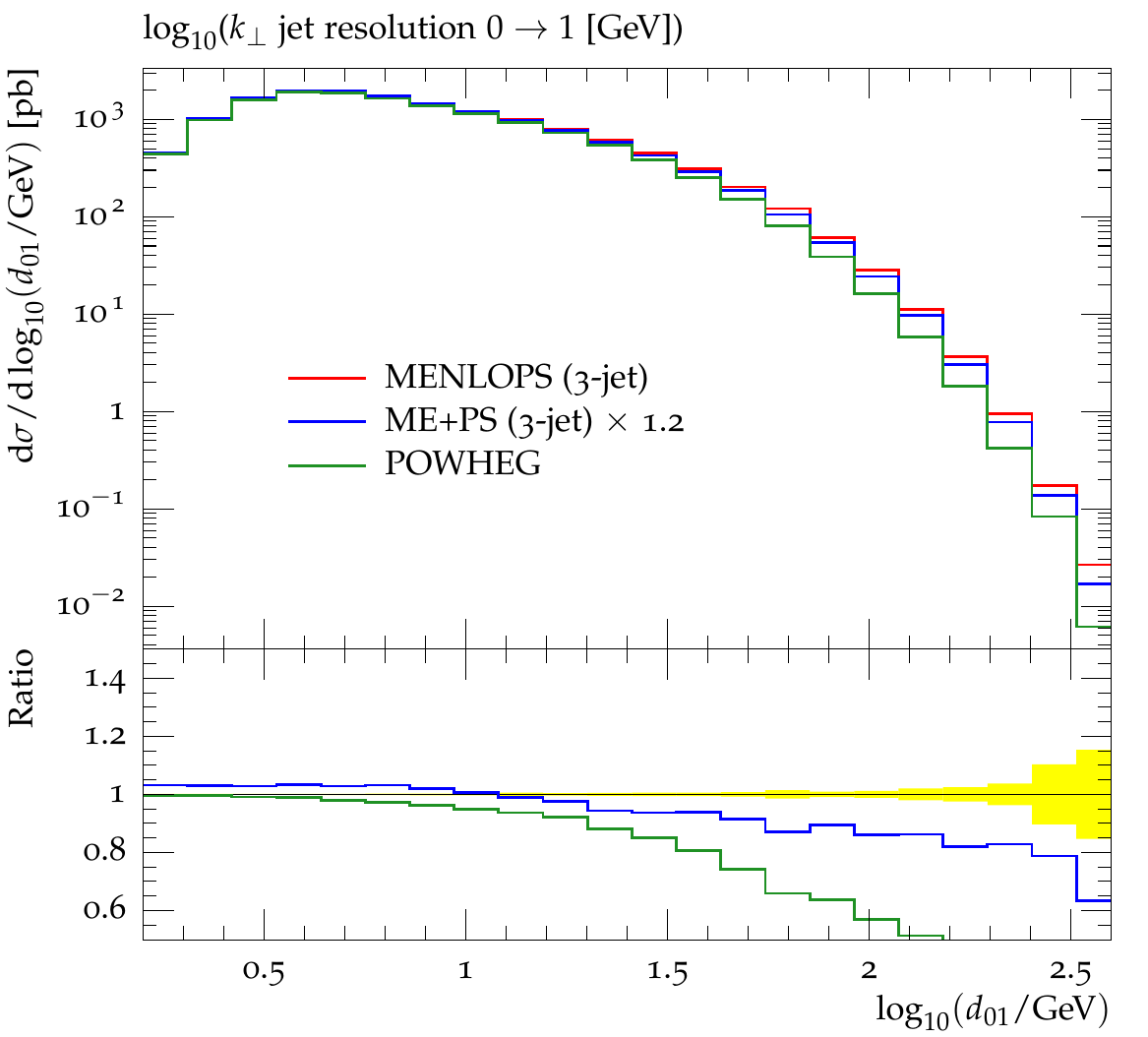}
  \hspace*{0.05\textwidth}
  \includegraphics[width=0.45\textwidth]{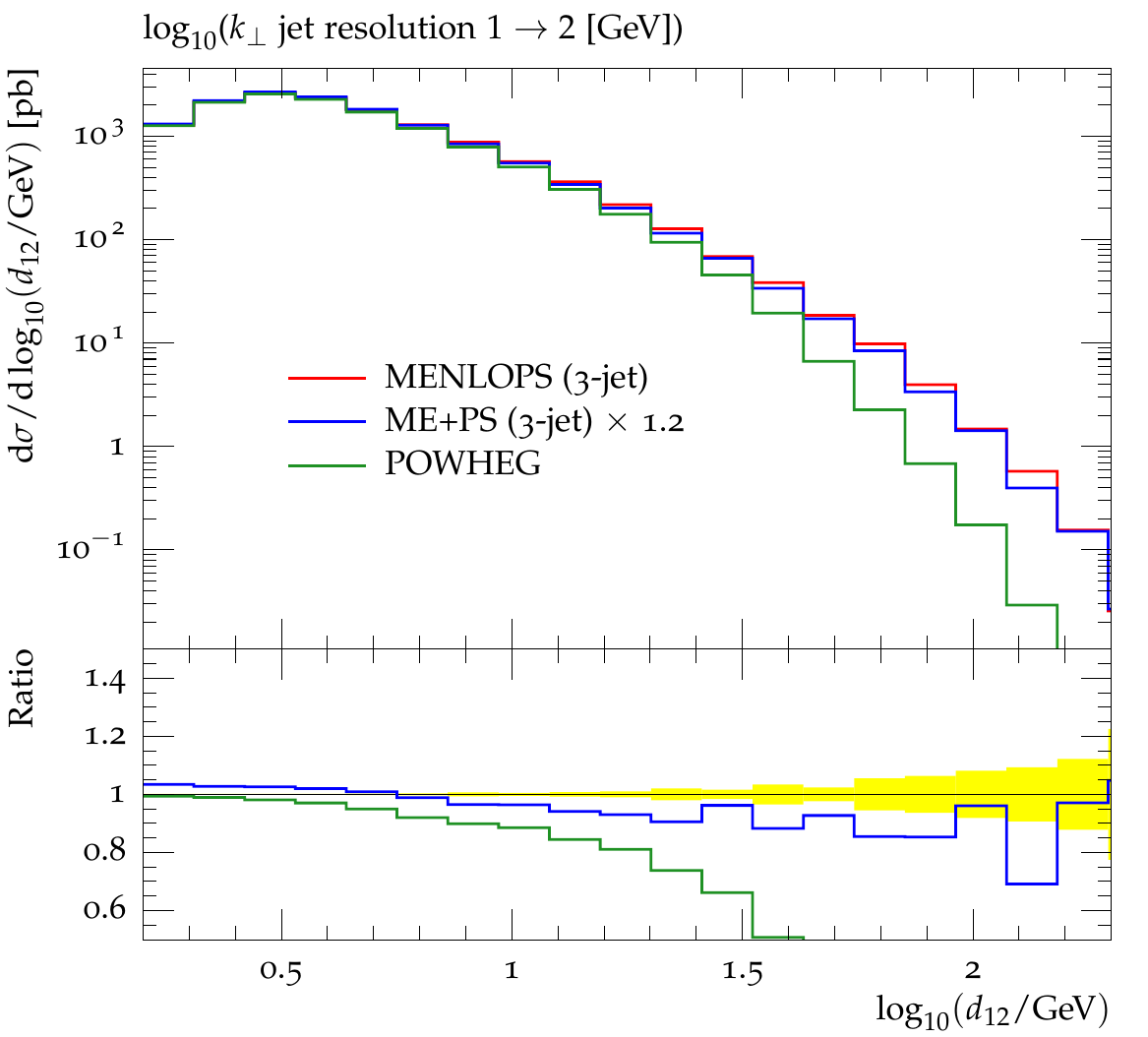}
  \vspace*{2mm}\\
  \includegraphics[width=0.45\textwidth]{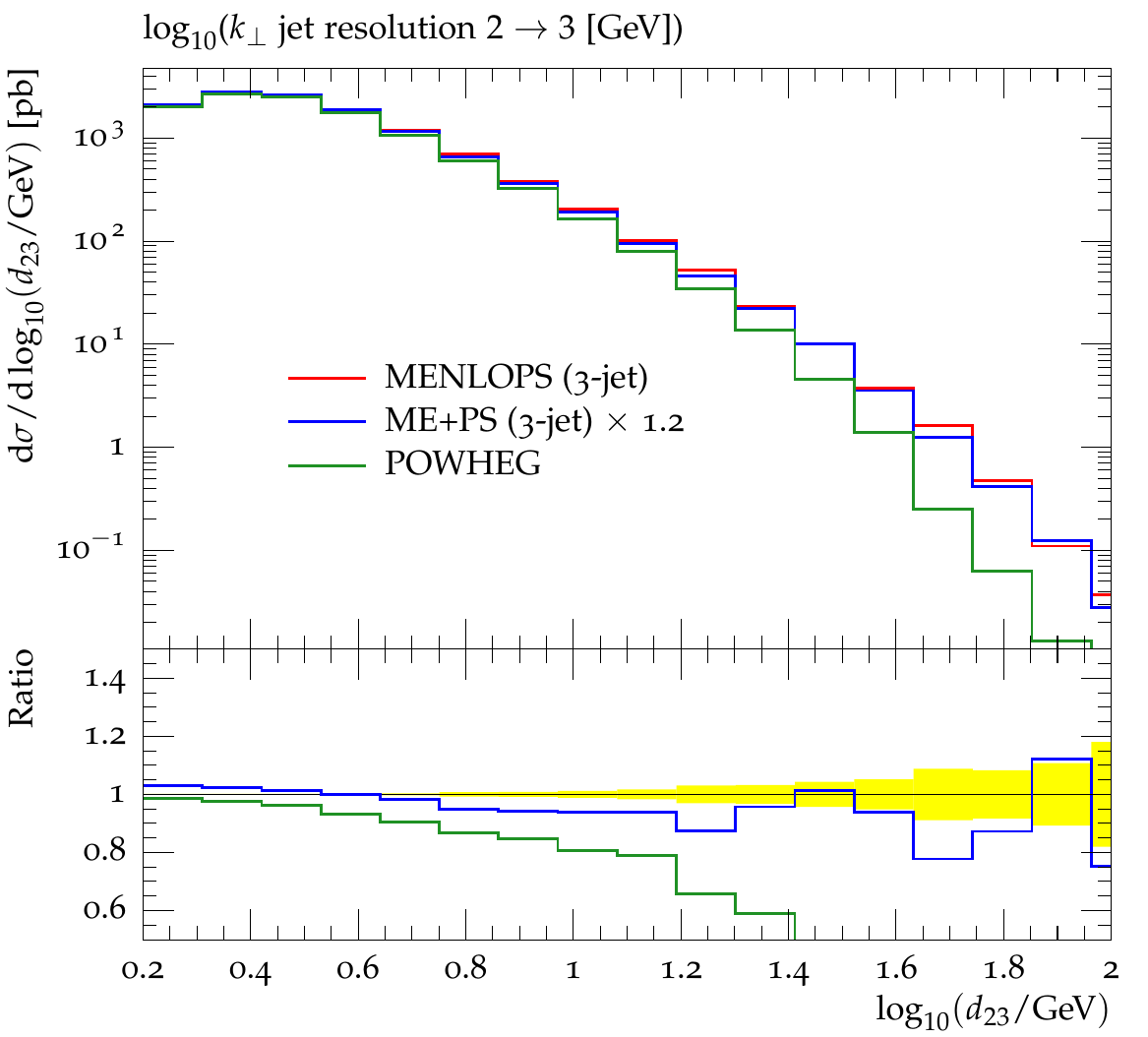}
  \hspace*{0.05\textwidth}
  \includegraphics[width=0.45\textwidth]{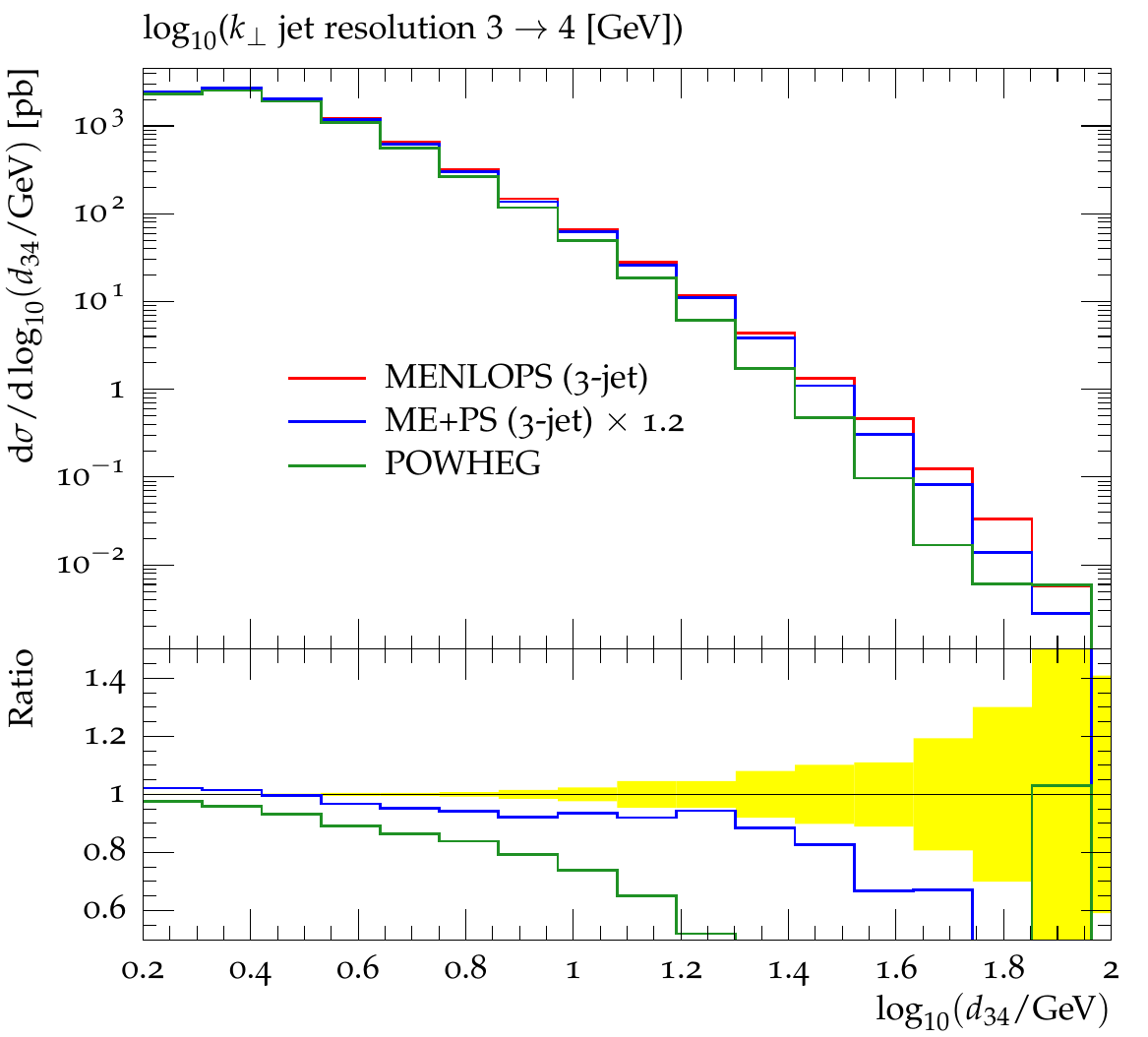}
  \end{center}
  \caption{
  Differential jet rates $d_{n\,n+1}$ in $W$ production
  at the Tevatron at $\sqrt{s}=1.8$ TeV.
  \label{fig:tev:jetrates_w}}
\end{figure}

In this section we focus on the production of $W$-bosons and their subsequent decay 
into an electron-neutrino pair at the Tevatron at $\sqrt{s}=1.8$ TeV. The core process
of the Monte-Carlo simulation is therefore $q\bar{q}'\to\ell\bar\nu$.
The separation criterion is set to $Q_\text{cut}=20$ GeV and up to three extra jets 
are taken into account. The electron-neutrino pair 
is required to have an invariant mass of $m_{e\nu}>10$ GeV. The $W\to e\nu$ decay
is corrected for QED next-to-leading order and soft-resummation effects using 
the YFS approach~\cite{Schonherr:2008av}. Virtual matrix elements are supplied
by \BlackHat~\cite{whitehat:2010aa,*Berger:2009zg,*Berger:2009ep,*Berger:2010vm}.

The left panel of Fig.~\ref{fig:wjets:hpt_jetsdr12} displays the transverse 
momentum of the $W$-boson as compared to data taken by the \DO 
collaboration~\cite{Abbott:2000xv}, while the right panel shows the exclusive 
jet multiplicity of $k_\perp$-clustered jets ($D\!=\!0.7$) with at least 
20 GeV.  Although the event sample generated using the \POWHEG technique only 
provides the best match to the central value of the data, all 
three event samples are well within the experimental uncertainties. On the 
other hand, already in the rate of single-jet events deviations between the 
\POWHEG sample and both the \MENLOPS and ME+PS samples are visible, with 
the latter two agreeing very well.  Similarly, the \POWHEG sample 
underestimates the amount of radiation into the central detector region, as 
exemplified in Fig.~\ref{fig:wjets:jets}. The right panel of this figure 
shows that, since the \POWHEG approach is capable of modelling the second 
hardest emission using the soft-collinear approximation of the parton shower 
only, its description of the angular separation of the the first two hardest 
jets is missing prominent features originating in the wide angle region. 
These features are of course present in the approaches having fixed-order 
matrix elements at their disposal.

Figure \ref{fig:tev:jetrates_w} shows the differential jet rates $d_{01}$, 
$d_{12}$, $d_{23}$ and $d_{34}$ using the above $k_\perp$-algorithm. While the 
first three of them, for the matrix-element merged samples, are described by 
matrix element to matrix element transitions, only the softer part of 
$d_{01}$ is described by such a transition for the \POWHEG sample.  The harder 
part of the $d_{01}$ receives corrections by matrix elements of higher jet 
multiplicity which are clustered into a single hard jet first.  Of course, 
these corrections are missing in the \POWHEG sample.  Furthermore, $d_{12}$ 
is described by a matrix element to parton shower transition only in the 
\POWHEG sample.  Hence, it strongly underestimates the amount of hard 
wide-angle radiation. Similarly, both $d_{23}$ and $d_{34}$ are described by 
the parton shower only in the \POWHEG sample, showing the same behaviour. It 
is worth noting that both the \MENLOPS sample, implementing local $K$-factors, 
and the ME+PS, scaled by a global $K$-factor, agree within their respective 
statistical uncertainties over the whole range, indicating the well 
known fact of the approximate momentum independence of the virtual 
corrections to the leading order process.

\clearpage

\subsection{Higgs boson production}
\label{Sec:Hjets}

\begin{figure}[p]
  \vspace*{-10mm}
  \begin{center}
  \includegraphics[width=0.45\textwidth]{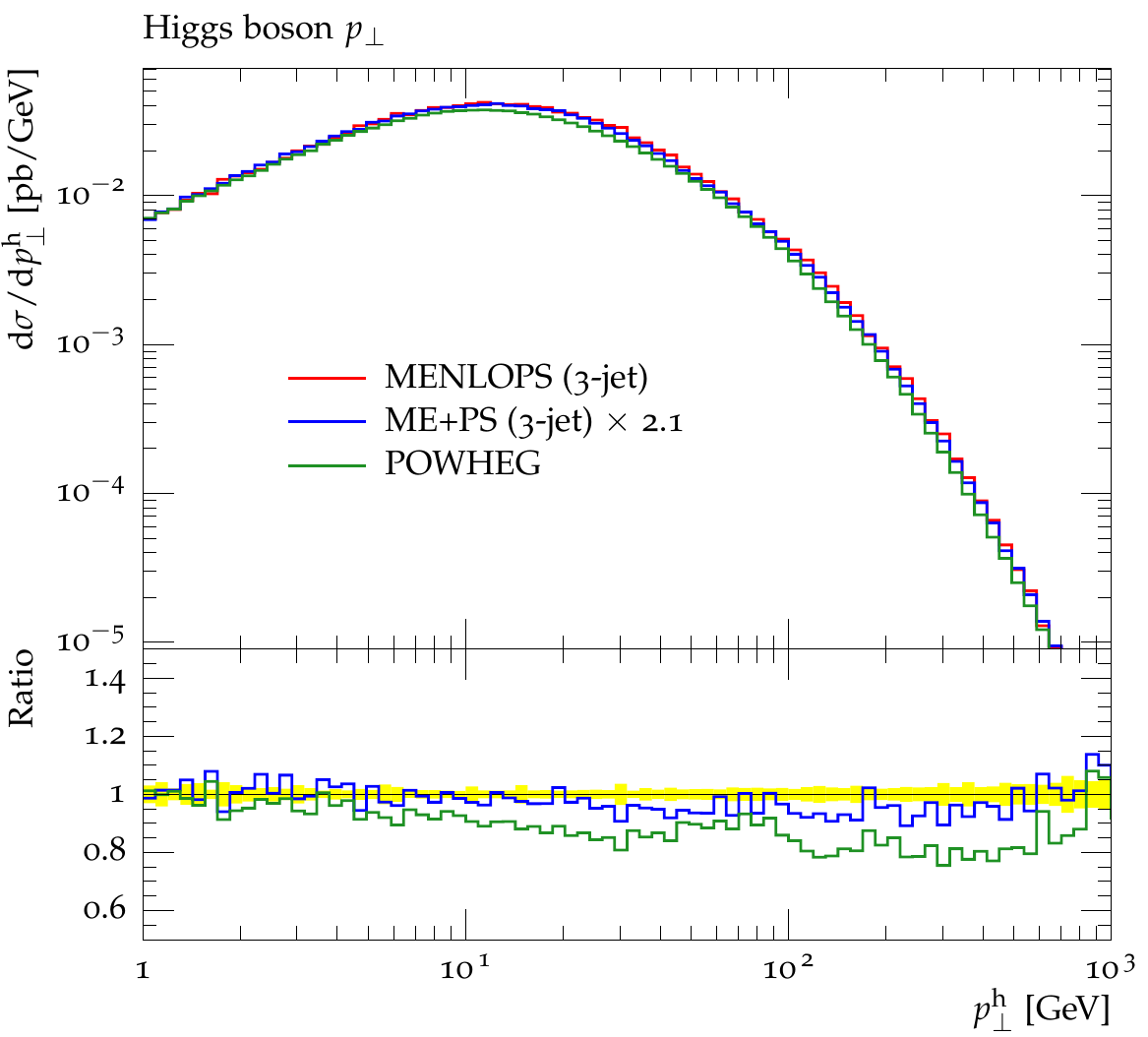}
  \hspace*{0.05\textwidth}
  \includegraphics[width=0.45\textwidth]{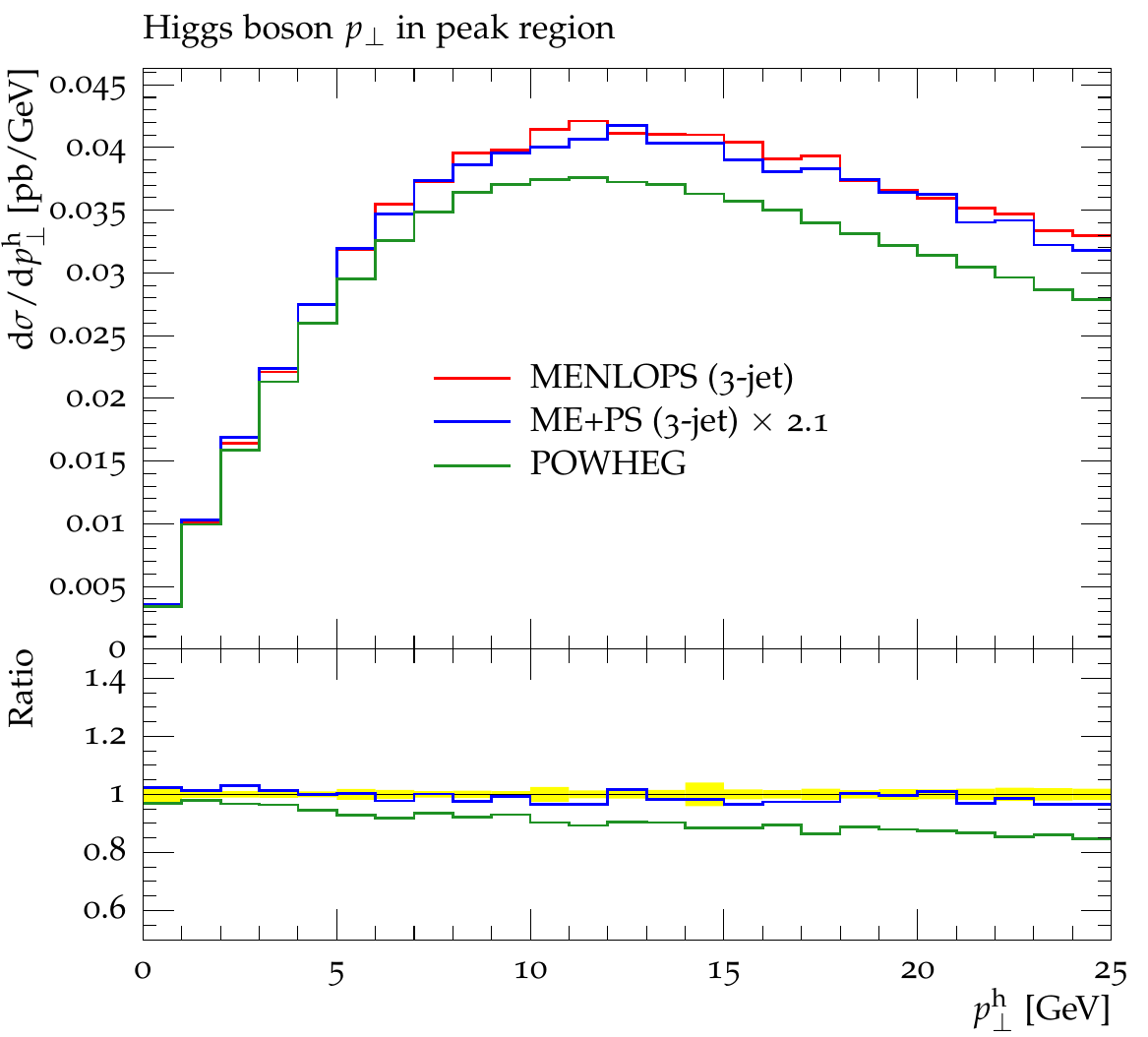}
  \end{center}
  \caption{
  Transverse momentum of the reconstructed Higgs boson in 
  the gluon-fusion process at nominal LHC energies (14 TeV).
  \label{fig:hlhc:hpt_jetsdr12}}
\end{figure}

\begin{figure}[p]
  \begin{center}
  \includegraphics[width=0.45\textwidth]{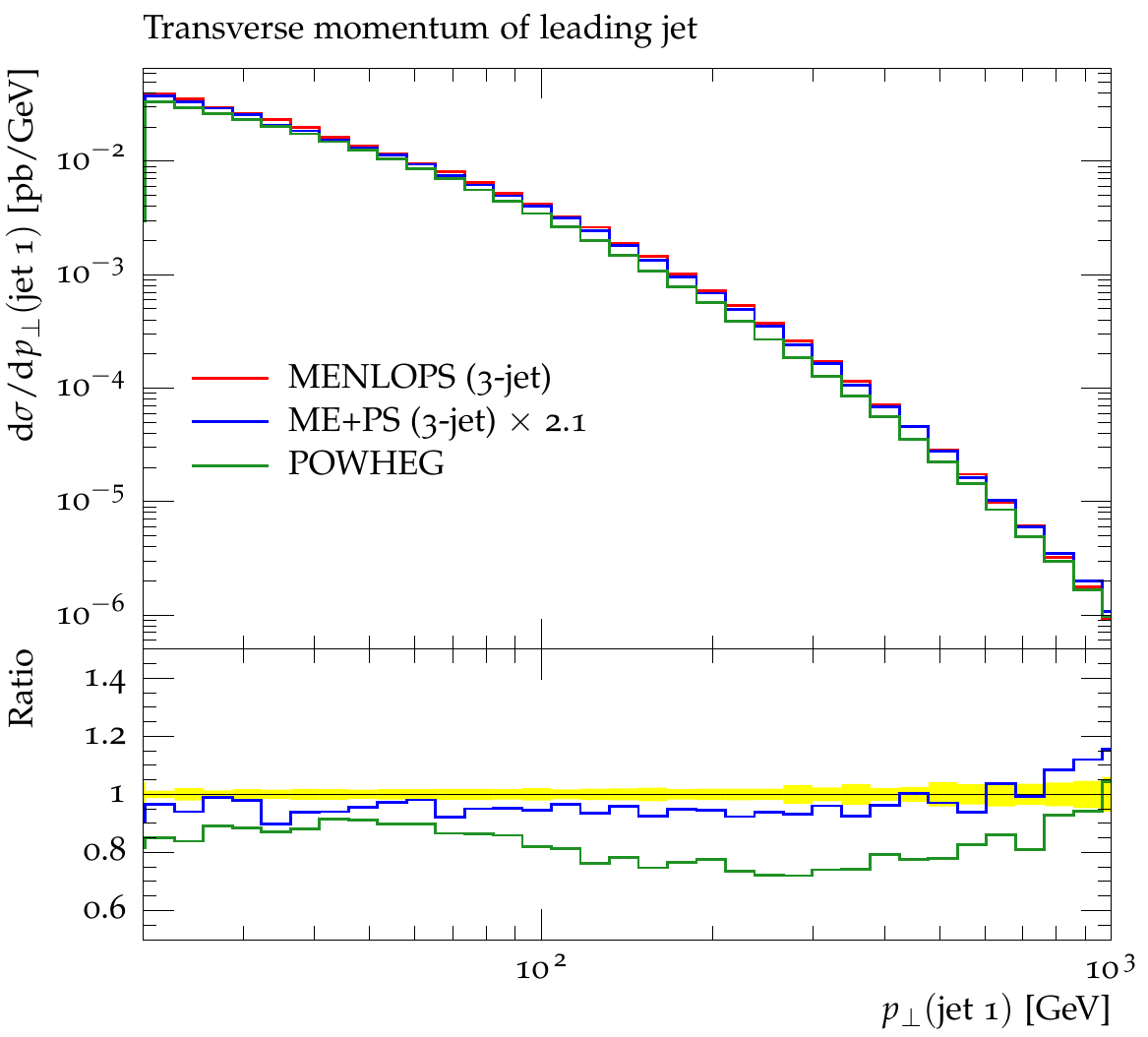}
  \hspace*{0.05\textwidth}
  \includegraphics[width=0.45\textwidth]{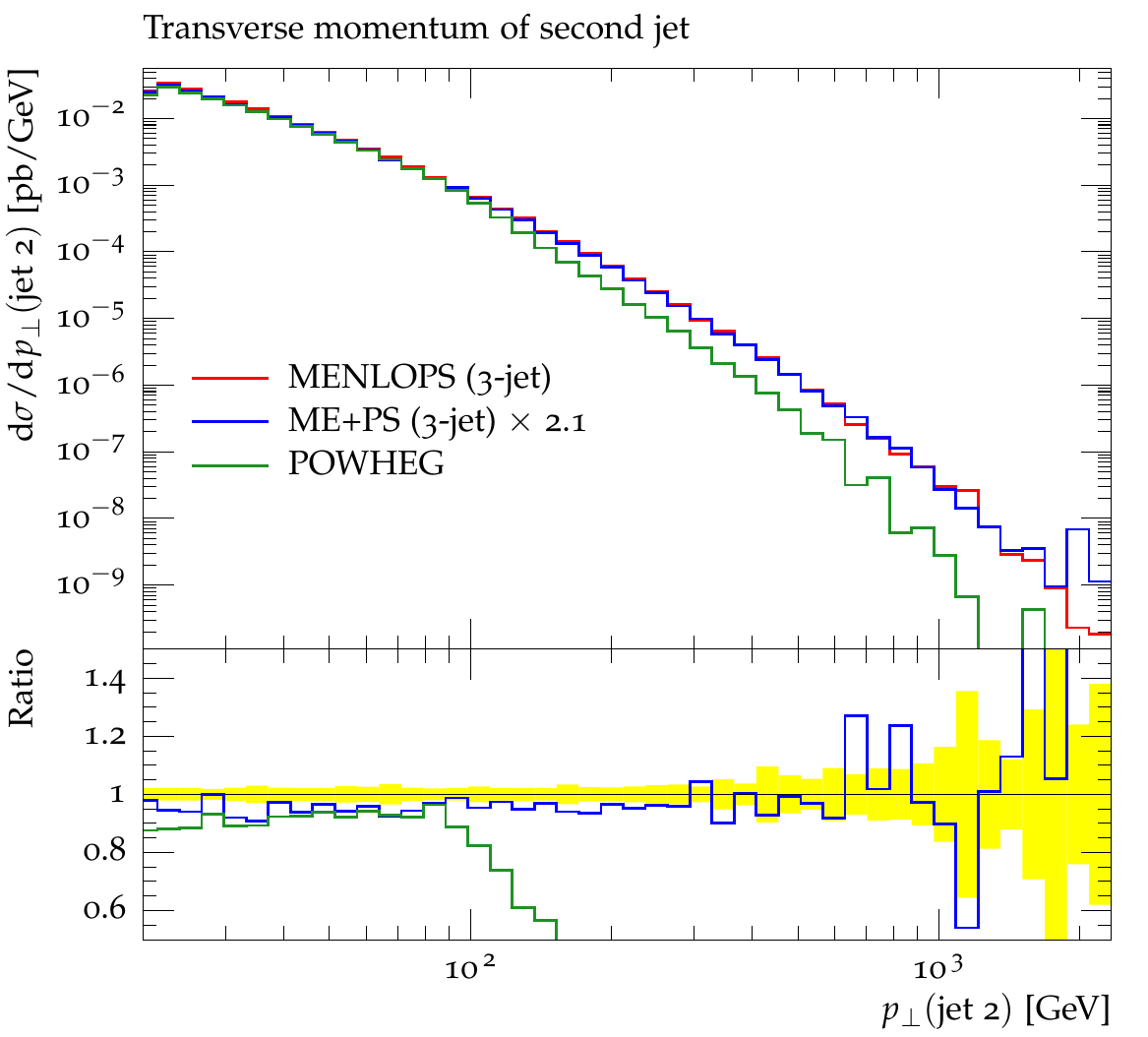}
  \end{center}
  \caption{
  Transverse momentum of the first and second hardest jet in Higgs-boson production via 
  gluon fusion at nominal LHC energies (14 TeV).
  \label{fig:hlhc:jet1pt_jet2pt}}
\end{figure}

\begin{figure}[p]
  \begin{center}
  \includegraphics[width=0.45\textwidth]{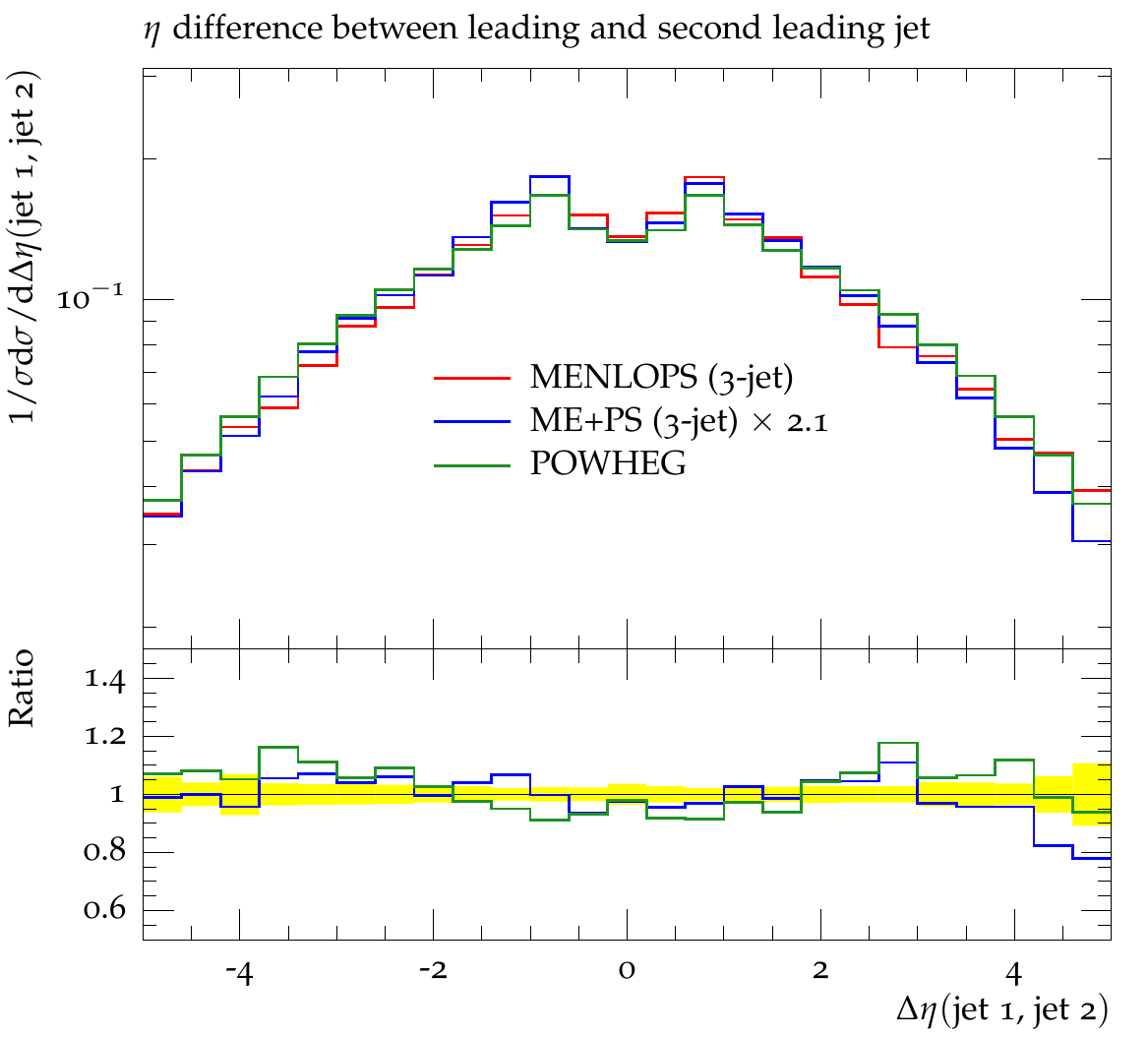}
  \hspace*{0.05\textwidth}
  \includegraphics[width=0.45\textwidth]{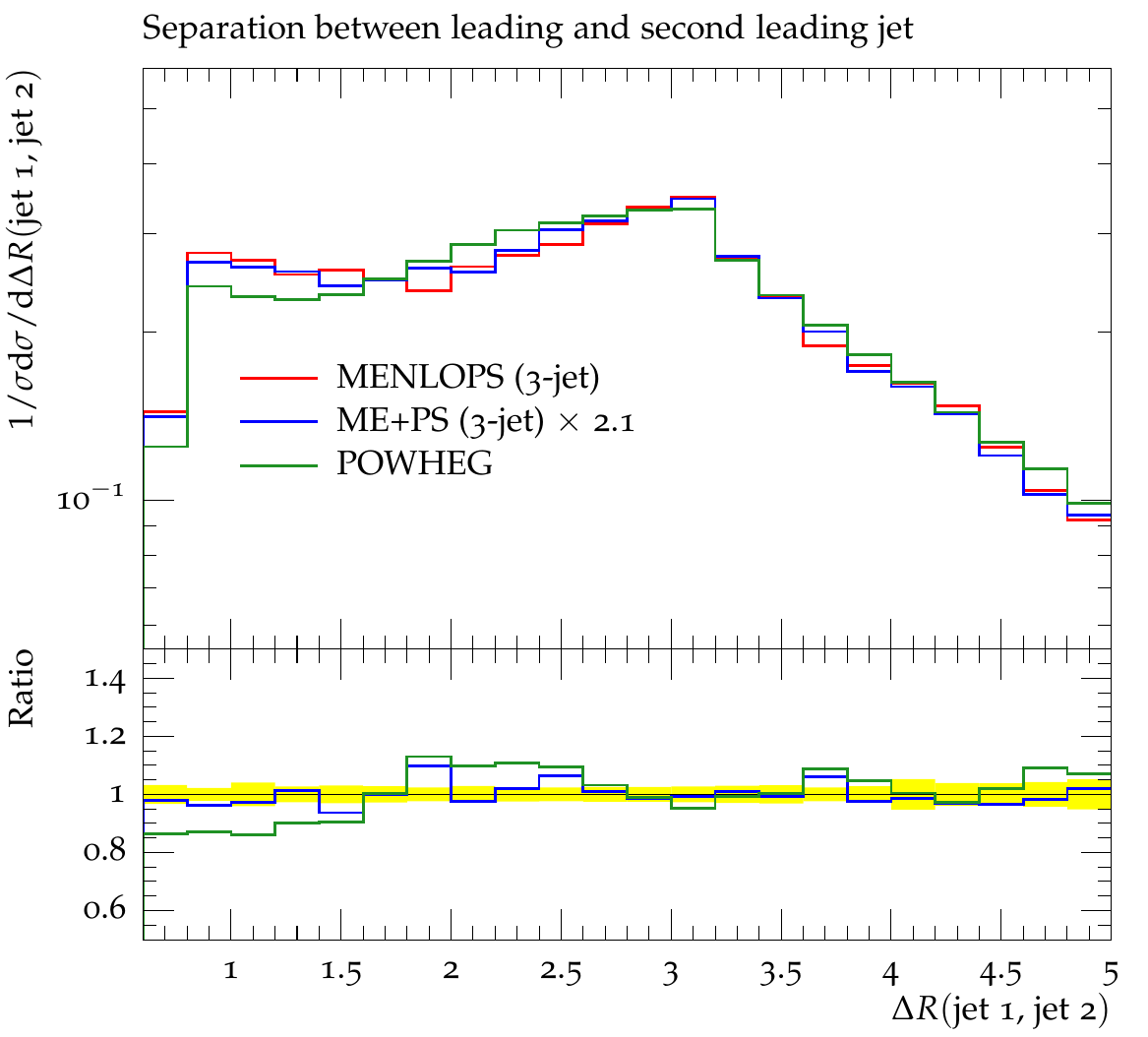}
  \end{center}
  \caption{
  Pseudorapidity difference and angular separation of the first and second hardest 
  jet in Higgs-boson production via gluon fusion at nominal LHC energies (14 TeV).
  \label{fig:hlhc:deta_dR}}
\end{figure}

This section presents predictions for Higgs boson production via gluon fusion at nominal 
LHC energies of $\sqrt{s}=14$ TeV. As NLO corrections to the core process $gg\to h\to \tau^+\tau^-$ 
are rather large, tremendous efforts have been made to perform fully differential calculations at 
NNLO~\cite{Anastasiou:2005qj,*Anastasiou:2007mz,*Anastasiou:2008tj} and several 
predictions have been presented which merged such fixed-order results with resummation 
at next-to-next-to-leading logarithmic accuracy~\cite{Catani:2003zt,*Bozzi:2005wk}. 
In this publication, we have no means for an improvement of the resummed calculation, 
instead we are restricted by the limitations of the parton-shower model. However, 
the systematic inclusion of higher-order tree-level matrix elements through the 
\MENLOPS method can yield a significant improvement of existing NLO predictions,
thus partially closing the gap between full NNLO predictions and Monte-Carlo results.
It was shown, for example, in~\cite{Butterworth:2010ym} that the predictions from 
ME+PS algorithms are often competitive to NNLO results if only the shape, 
not the normalisation, of observable distributions is concerned. 

In our simulations we set $m_H=120$ GeV and we include the decay $h\to\tau^+\tau^-$,
however, the analysis focuses on the properties of QCD radiation associated with 
production of the Higgs boson. The invariant $\tau$-pair mass is restricted to 
$115<m_{\tau\tau}/\mathrm{GeV}<125$ at the matrix-element level. Virtual matrix 
elements are implemented according to~\cite{Dawson:1990zj,*Djouadi:1991tka}.
The decay $h\to \tau^+\tau^-$ is corrected for QED soft-resummation and approximate 
next-to-leading order effects using the YFS approach~\cite{Schonherr:2008av}.

Figure~\ref{fig:hlhc:hpt_jetsdr12} shows the transverse momentum spectrum 
of the reconstructed Higgs boson. We observe that the \POWHEG and \MENLOPS
samples are very consistent in the prediction of this rather inclusive observable.
On the other hand, differences are observed in the results for individual jet 
transverse momentum spectra, cf.\ Fig.~\ref{fig:hlhc:jet1pt_jet2pt}.
They increase with jet multiplicity and with increasing transverse momentum, as can 
be expected, since the higher multiplicity jets are described by the uncorrected
parton shower in the \POWHEG method. Deviations are also found in the prediction of 
the dijet separation in $\eta-\phi$ space, which is shown in Fig.~\ref{fig:hlhc:deta_dR}. 
However, it was previously found that the ME+PS result yields a prediction which 
is very similar to the NNLO result~\cite{Butterworth:2010ym}. This feature is 
naturally retained in the \MENLOPS simulation.


\subsection{\texorpdfstring{$W^+W^-+$}{WW+}jets Production}
\label{Sec:wwjets}

\begin{figure}[p]
  \vspace*{-10mm}
  \begin{center}
  \includegraphics[width=0.45\textwidth]{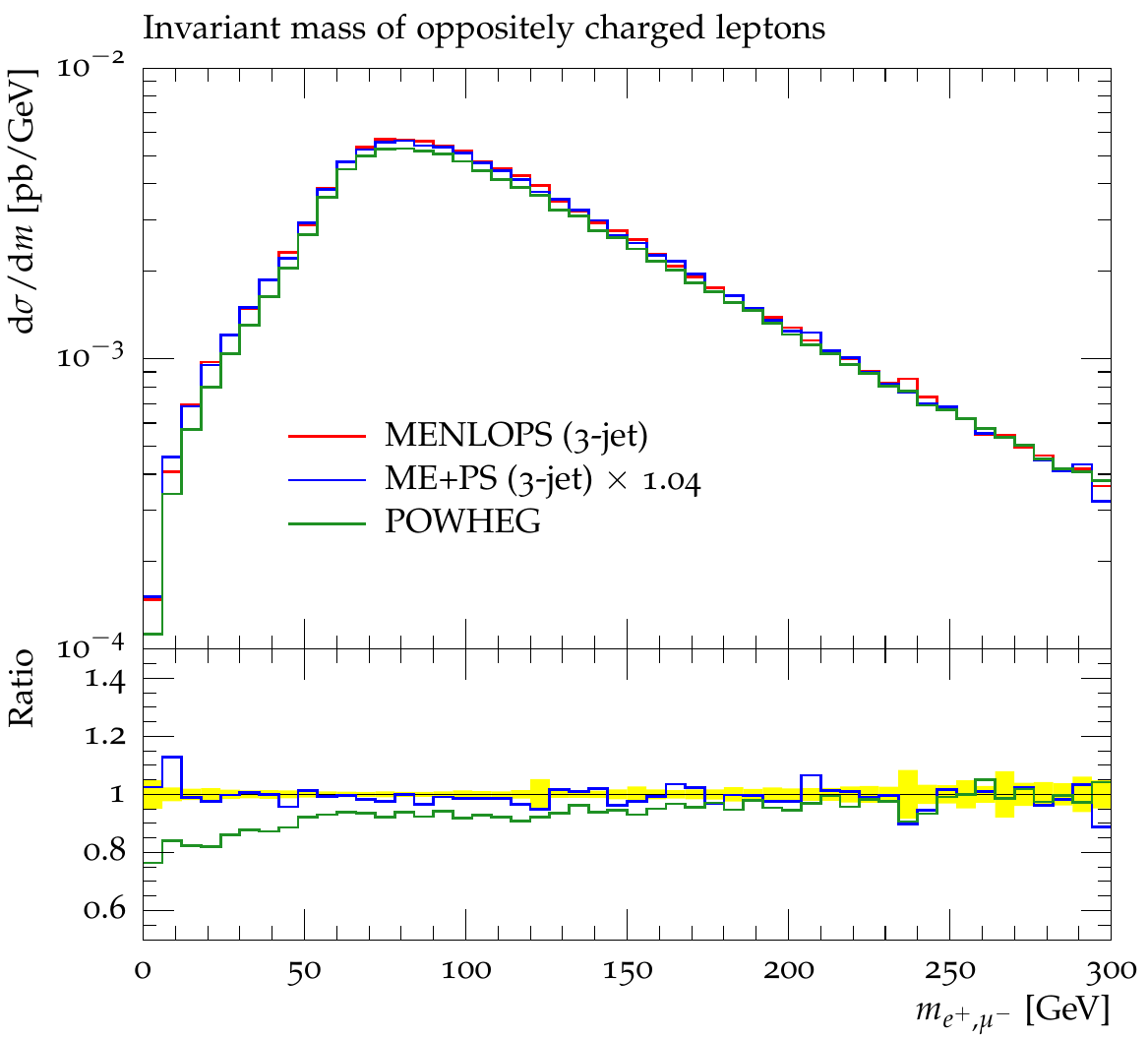}
  \hspace*{0.05\textwidth}
  \includegraphics[width=0.45\textwidth]{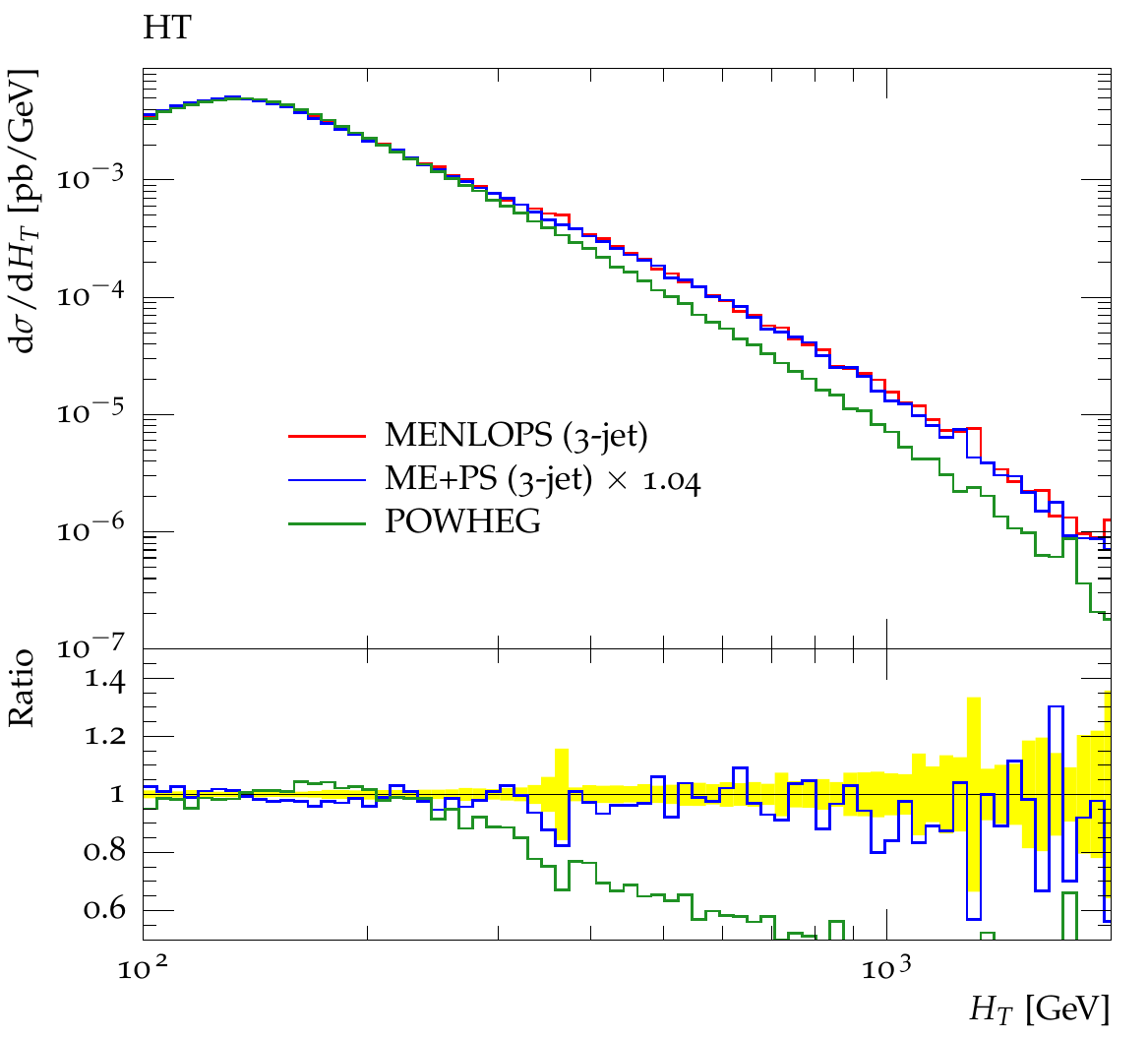}
  \end{center}
  \caption{
  Invariant mass of the electron-muon pair (left) and
  $H_T$ (right) in $W^+W^-$ production at nominal 
  LHC energies (14 TeV).
  \label{fig:wwlhc:ht_emumass}}
\end{figure}

\begin{figure}[p]
  \begin{center}
  \includegraphics[width=0.45\textwidth]{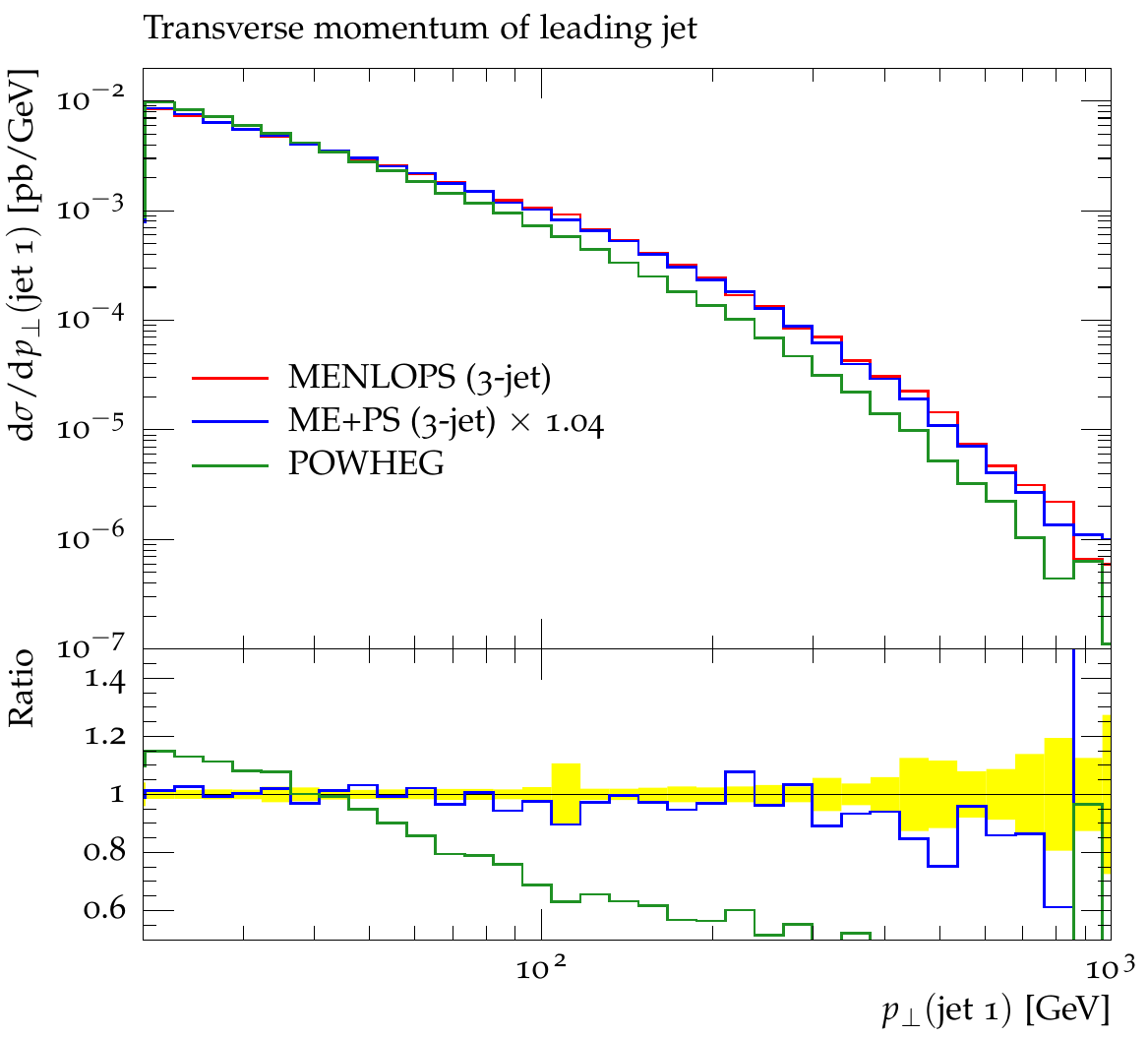}
  \hspace*{0.05\textwidth}
  \includegraphics[width=0.45\textwidth]{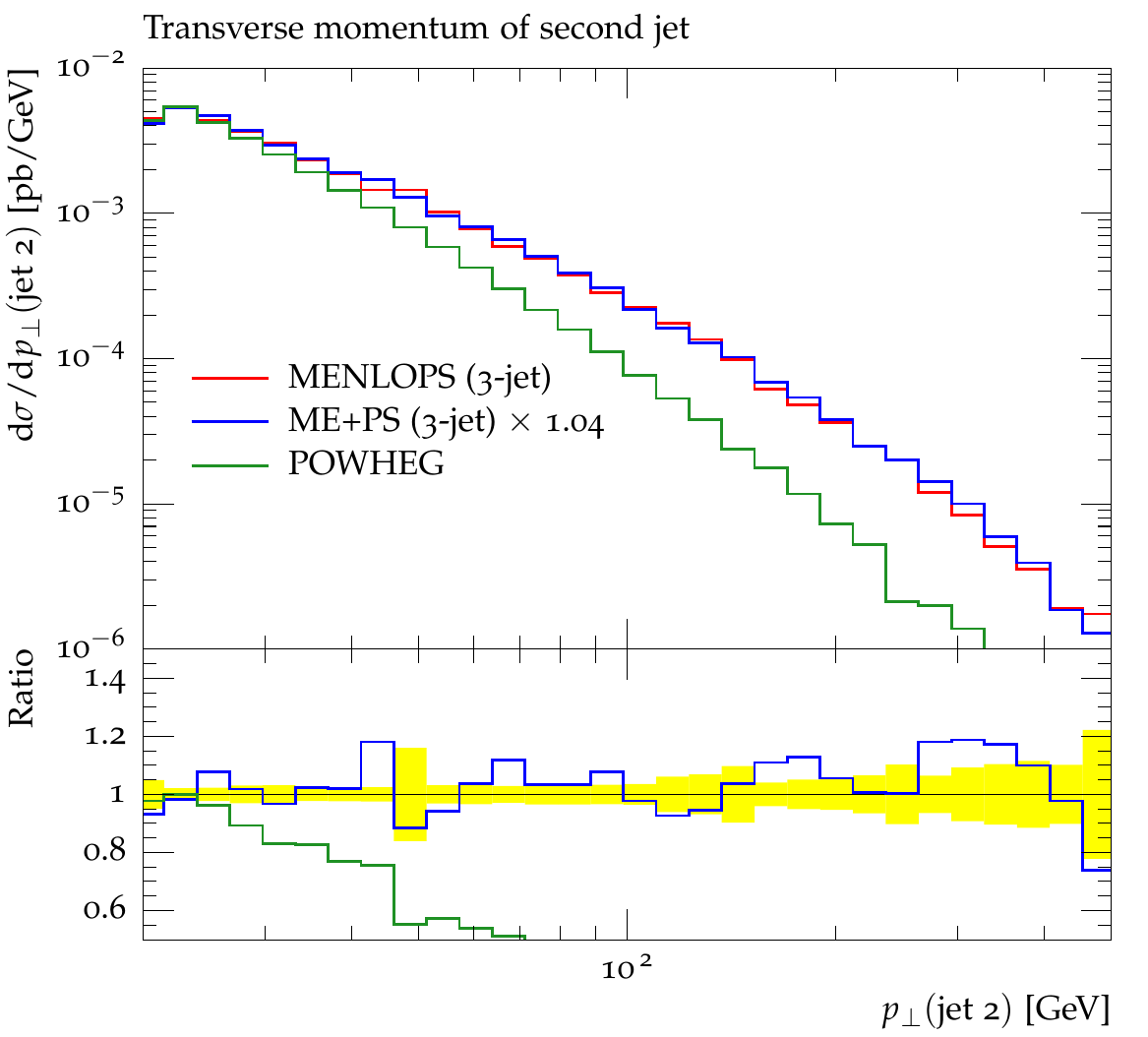}
  \end{center}
  \caption{
  Transverse momentum of the first and second hardest jet in $W^+W^-$ 
  production at nominal LHC energies (14 TeV).
  \label{fig:wwlhc:jet1pt_jet2pt}}
\end{figure}

\begin{figure}[p]
  \begin{center}
  \includegraphics[width=0.45\textwidth]{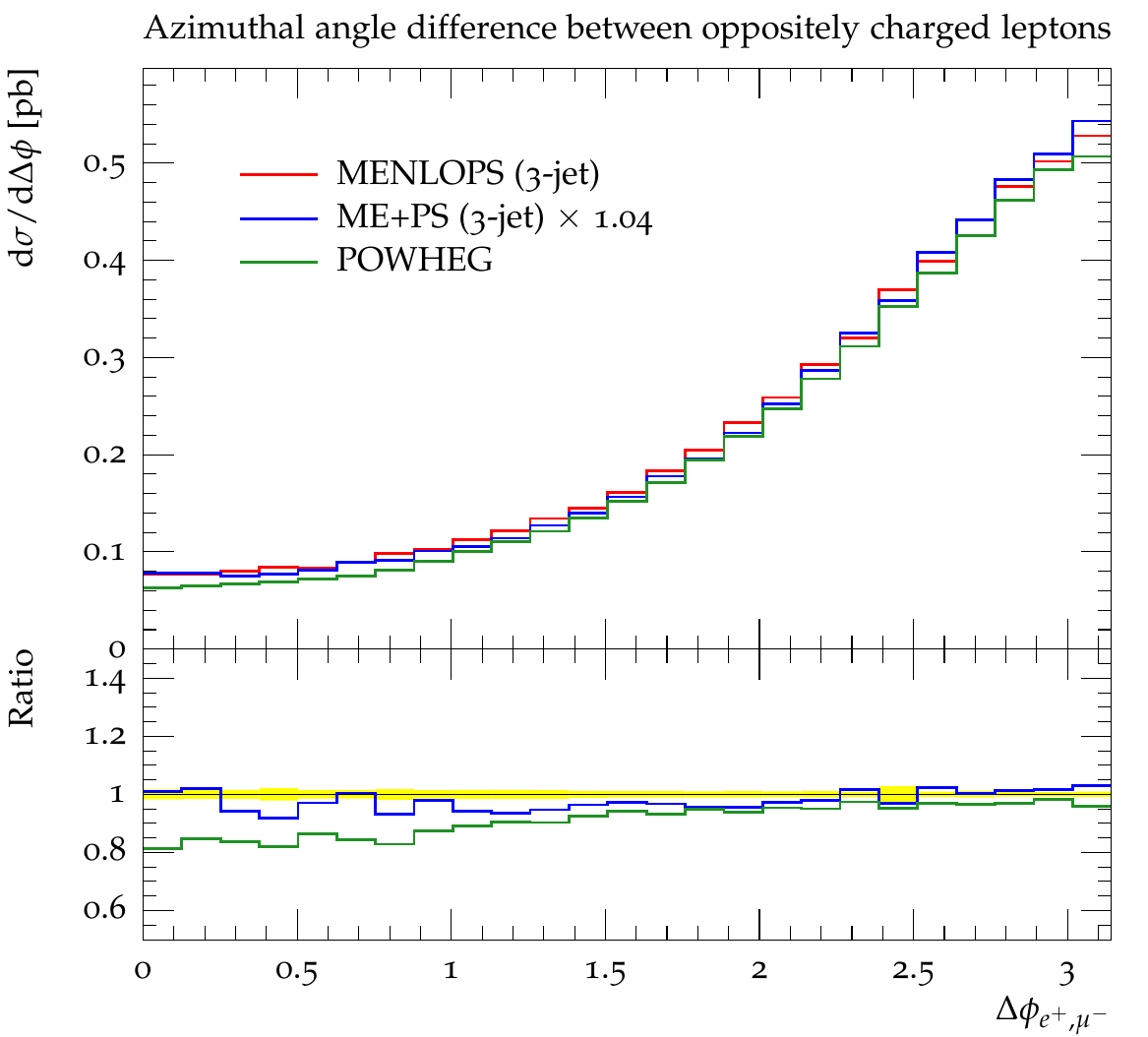}
  \hspace*{0.05\textwidth}
  \includegraphics[width=0.45\textwidth]{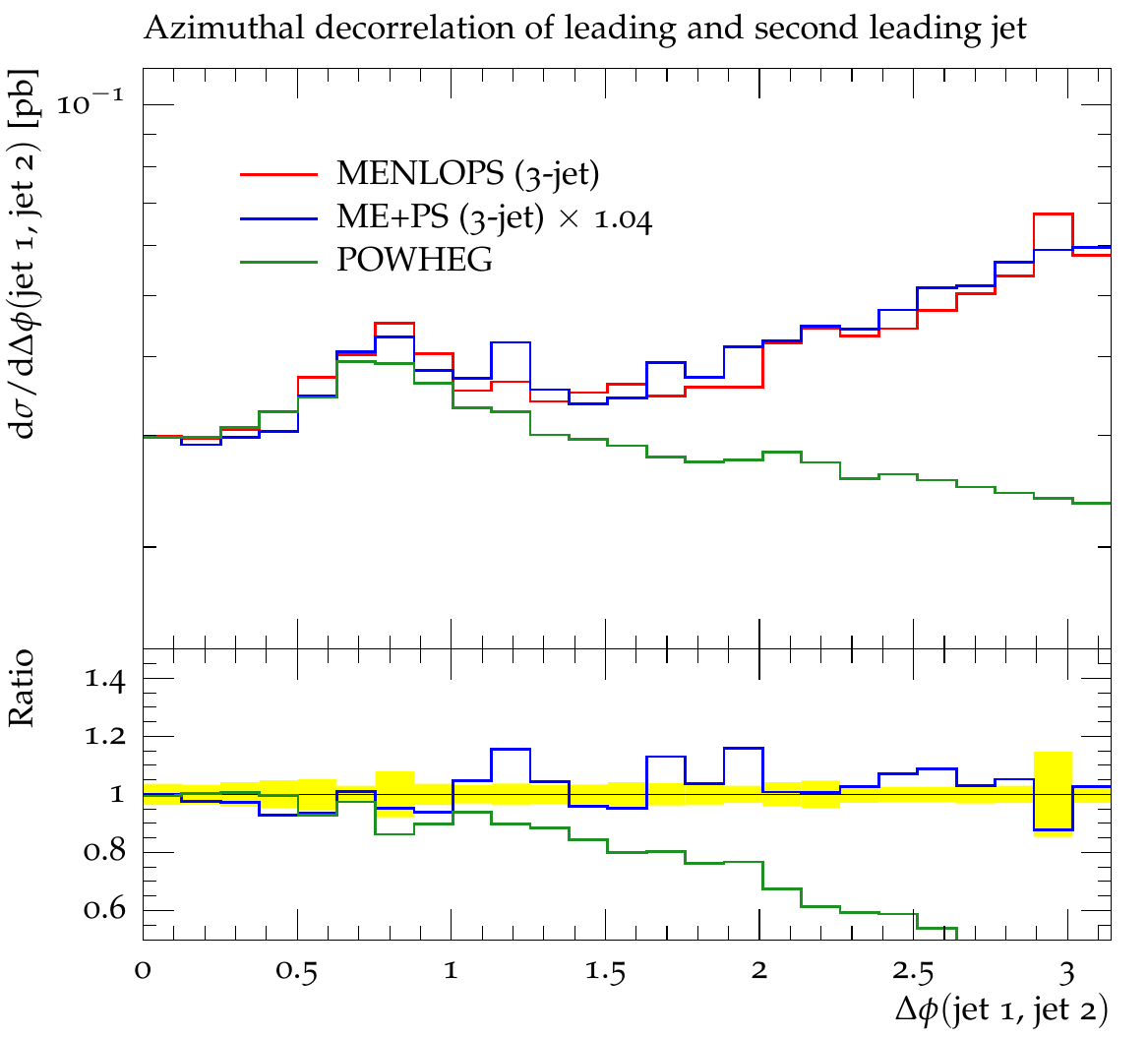}
  \end{center}
  \caption{
  Azimuthal separation of the electron and the muon (left) and of the 
  two hardest jets (right) in $W^+W^-$ production at nominal LHC energies (14 TeV).
  \label{fig:wwlhc:emudphi_j1j2dphi}}
\end{figure}

\begin{figure}[p]
  \vspace*{-10mm}
  \begin{center}
  \includegraphics[width=0.45\textwidth]{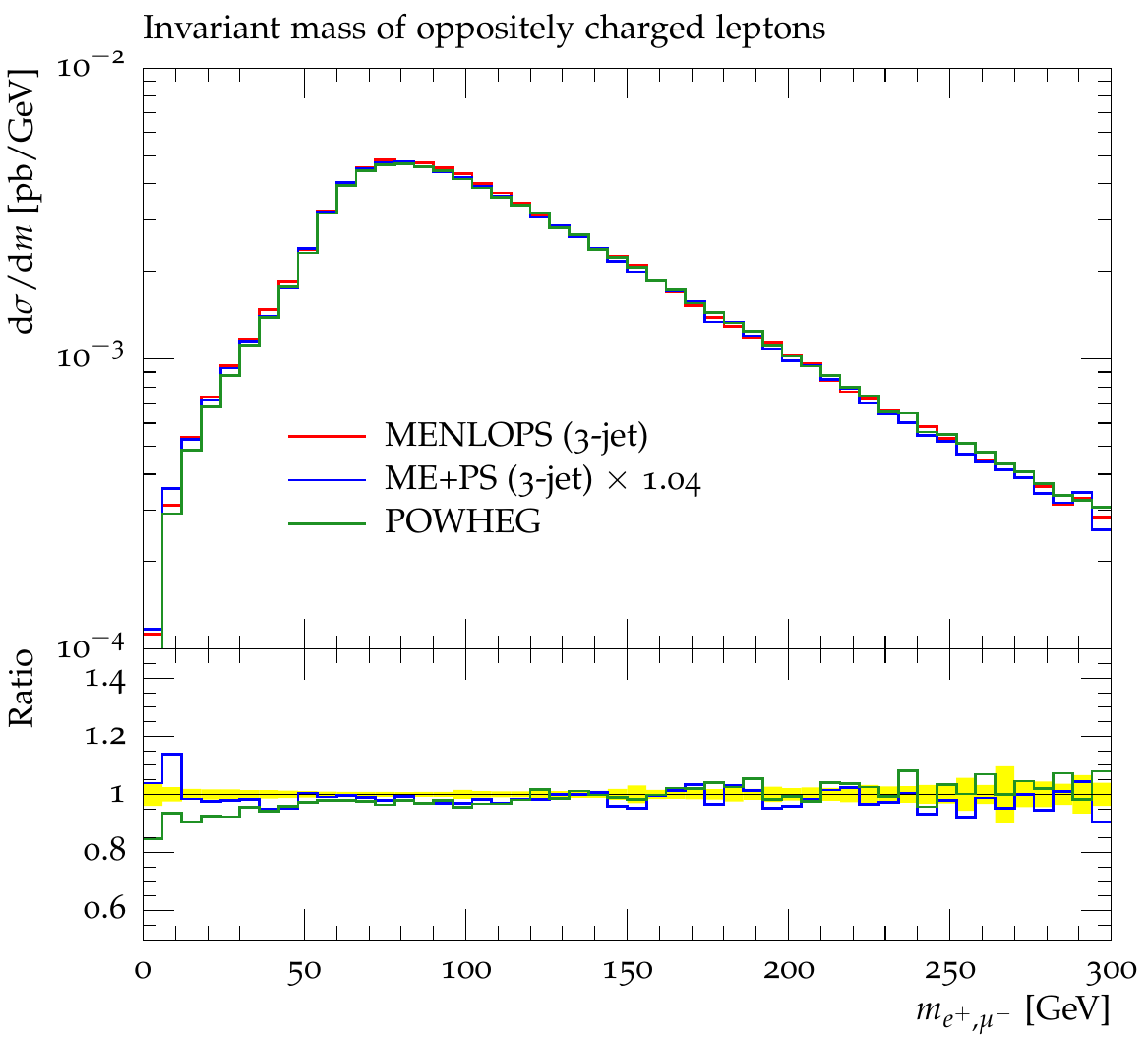}
  \hspace*{0.05\textwidth}
  \includegraphics[width=0.45\textwidth]{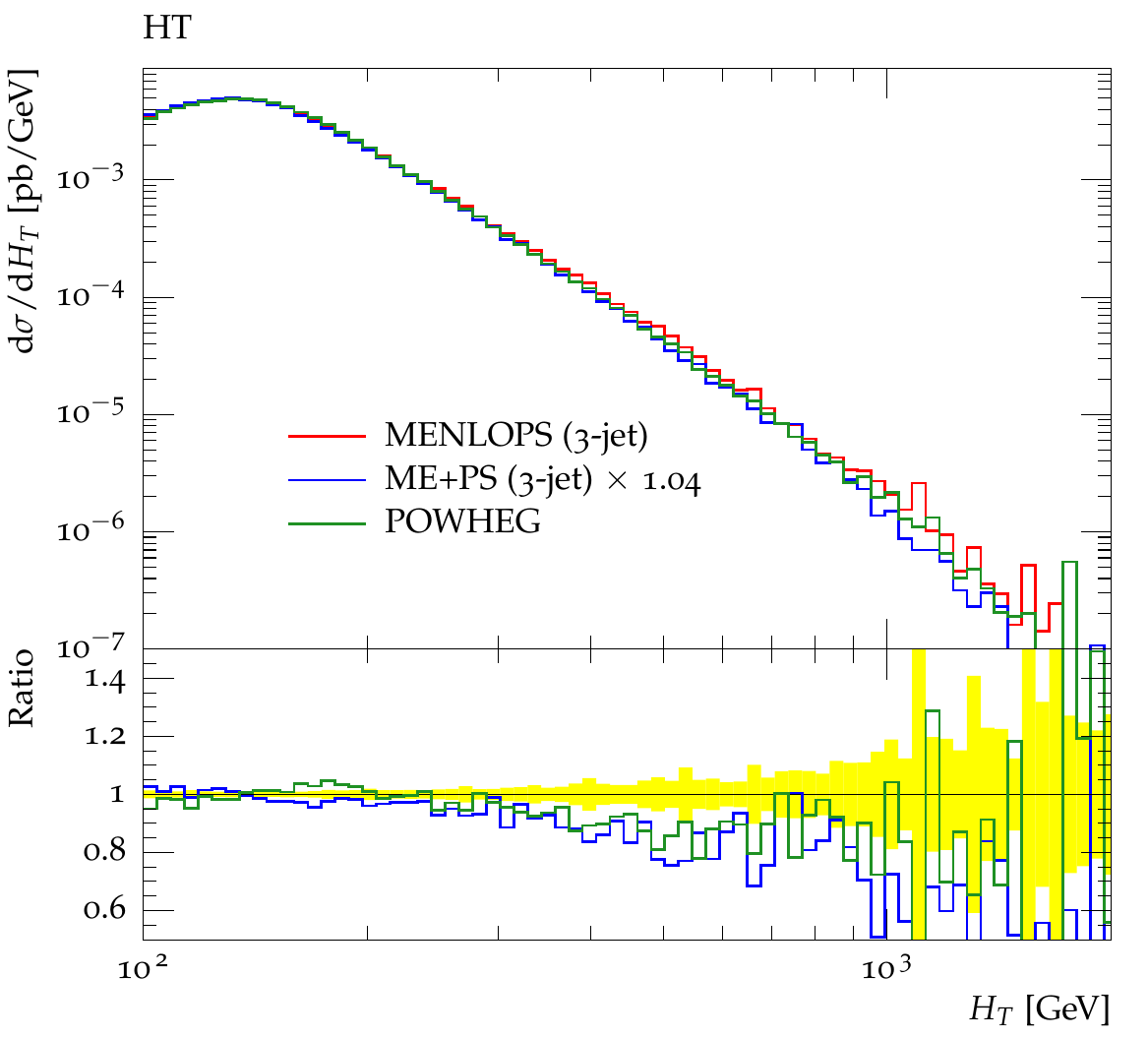}
  \end{center}
  \caption{
  Invariant mass of the electron-muon pair (left) and
  $H_T$ (right) in $W^+W^-$ production at nominal 
  LHC energies (14 TeV)
  after vetoing events with more than one jet with $p_T>20$ GeV.
  \label{fig:wwlhc:ht_emumass_veto}}
\end{figure}

\begin{figure}[p]
  \begin{center}
  \includegraphics[width=0.45\textwidth]{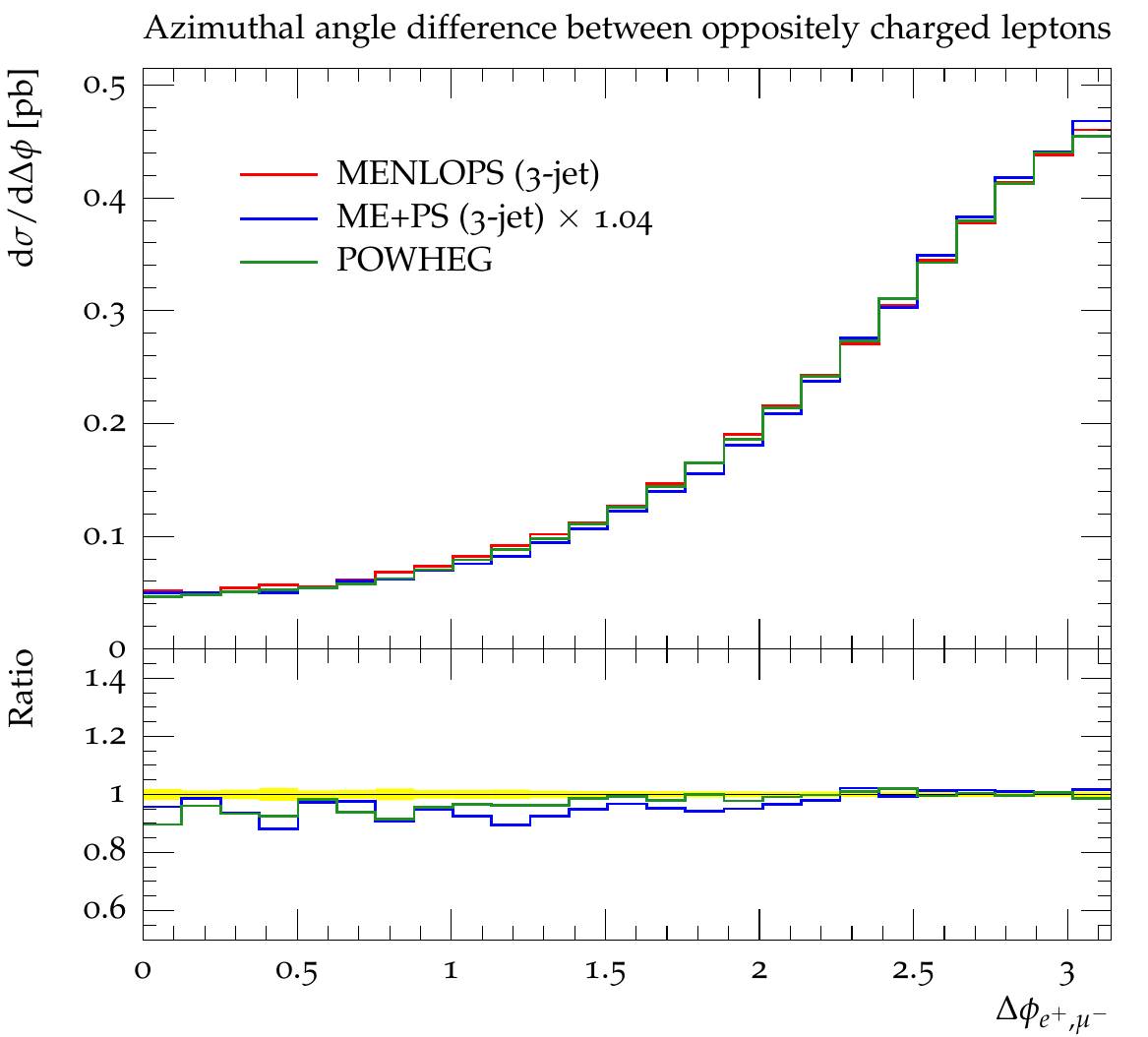}
  \hspace*{0.05\textwidth}
  \includegraphics[width=0.45\textwidth]{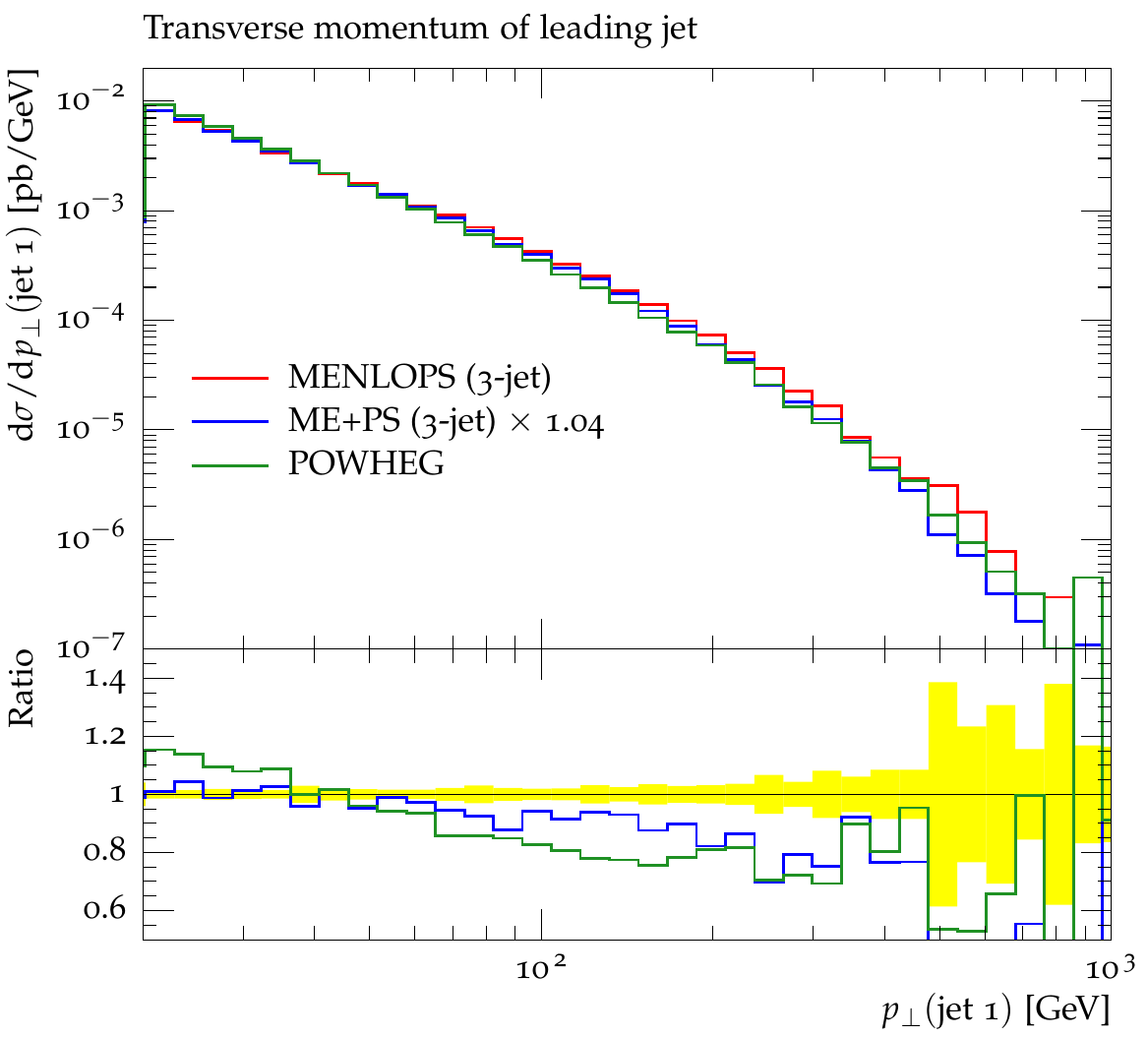}
  \end{center}
  \caption{
  Azimuthal separation of the electron and the muon (left) and transverse 
  momentum of the hardest jet in $W^+W^-$ production at nominal LHC energies 
  (14 TeV) after vetoing events with more than one jet with $p_T>20$ GeV.
  \label{fig:wwlhc:emudphi_jet1pt_veto}}
\end{figure}

In this section we present predictions for the production of the
$W^+[\to\! e^+\nu_e]\;W^-[\to\! \mu^-\bar\nu_\mu]$ final state at nominal LHC 
energies of $\sqrt{s}=14$ TeV. The lepton-neutrino pairs are required to have 
an invariant mass of $m_{\ell\nu}>10$ GeV each. The $W\to \ell\nu$ decays are 
corrected for QED next-to-leading order and soft-resummation effects using 
the YFS approach~\cite{Schonherr:2008av}. Virtual matrix elements are
supplied by \MCFM~\cite{MCFM,Campbell:1999ah,*Dixon:1998py}. Again, this study 
focuses mainly on the properties of QCD radiation accompanying the diboson
production process. Up to three additional jets at $Q_\text{cut}=20$~GeV 
are simulated in both, the \MENLOPS and the ME+PS sample. It is known that 
high-multiplicity matrix elements in the ME+PS approach yield sizable effects
on total event rates and shapes in this reaction~\cite{Gleisberg:2005qq},
a feature which is inherited by the \MENLOPS method. Setting the phase-space separation
criterion to a rather low value compared to the average partonic centre-of-mass
energy will thus always lead to sizable emission-rate differences, which might be
an indication of potentially large higher-order corrections. A similar effect
was observed in a recent analysis of $Z$-boson pair production in association
with a hard jet~\cite{Binoth:2009wk}. While the NLO corrections to this process
are comparably small at Tevatron, they can be rather large at nominal LHC 
energies. Restricting the available final-state phase space by a jet veto,
the corrections were again limited to smaller values, which makes the importance 
of the $ZZ$+2 jets final state explicit. As we include up to three additional
jets in our simulation of $W^+W^-$ production, we observe similar effects.

Figure~\ref{fig:wwlhc:ht_emumass} displays the invariant mass of the lepton pair
and the scalar sum of transverse momenta of the jets, leptons and the missing 
transverse energy, $H_T$. While the former is described very well by the 
next-to-leading order calculation used in the \POWHEG sample and receives 
only mild corrections from higher-order matrix elements, $H_T$ receives sizable 
corrections at rather low values already. The reason for this is easily found 
in the sensitivity of $H_T$ to {\em any} jet activity and thus to higher-order 
matrix element corrections of the parton shower. This can be seen in comparison 
to Fig.~\ref{fig:wwlhc:ht_emumass_veto}, where a veto on additional jet activity
was applied. We exemplify in Fig.~\ref{fig:wwlhc:jet1pt_jet2pt} that the ME+PS 
part of the \MENLOPS simulation predicts significantly harder radiation than 
the \POWHEG subsample. The corresponding corrections naturally amplify the 
deviations between the respective predictions of $H_T$. We show the impact 
of a jet veto on this distribution in the right panel of 
Fig.~\ref{fig:wwlhc:emudphi_jet1pt_veto}.

Figure~\ref{fig:wwlhc:emudphi_j1j2dphi} presents predictions for the azimuthal 
separation of the leptons and the two hardest jets. Again, the former 
receives only comparably small corrections, while higher-order matrix-element
corrections have large impact on the latter. This hints at the importance
to include higher-order matrix elements in Monte-Carlo simulations of 
hadron-collider events if the hadronic centre-of-mass energy is large.
The effect of a jet veto on the azimuthal separation of the leptons is shown 
in the left panel of Fig.~\ref{fig:wwlhc:emudphi_jet1pt_veto}.

\clearpage

\section{Conclusions}
\label{Sec::Conclusions}
In this publication, a parallel development and independent implementation 
of the \MENLOPS algorithm, first discussed in~\cite{Hamilton:2010wh}, has 
been presented.  This new algorithm combines 
the so far most advanced methods to include higher-order corrections to a 
given core process:  The \POWHEG technique, which allows to produce inclusive 
samples for that process with next-to-leading order accuracy, and the ME+PS 
technique, which allows to generate inclusive samples with a leading-order 
cross section, but with the production of additional hard radiation corrected
by higher-order tree-level matrix elements.

Until the work of Hamilton and Nason~\cite{Hamilton:2010wh} and the work 
presented here, these two approaches were considered orthogonal and thus used 
independent from each other, in the regime of their respective strengths and
validity.  With the recent efforts on combining them, the shortcomings of each 
method, i.e.\ the description of higher jet multiplicities in \POWHEG 
and lack of the correct NLO cross section in ME+PS, have been expunged.

We fully confirm the findings of Hamilton and Nason concerning both the 
formalism and the relative improvement in the simulation obtained through it.  
This is even more emphasised here, since the implementation of the \MENLOPS 
method as presented in~\cite{Hamilton:2010wh} seems to suffer from the choice 
of tools.  As already indicated in the introduction, the omission of truncated 
showering in the program used to simulate the ME region may have 
caused a few of the uncertainties.  We are convinced that in total, the 
superior quality of the ME+PS part of the simulation in \Sherpa, including 
the truncated showering, are the only reason behind the 
improved simulation here -- the formalism is identical in both publications.  
The drastically reduced uncertainties stress the great improvement by the 
\MENLOPS method. 

Our results and the ones presented by Hamilton and Nason in fact show a 
significant improvement of many aspects of previous simulations in a variety 
of processes, including here $e^+e^-$ annihilation to hadrons, hadronic final 
states in DIS, jets in association with single vector bosons and with vector 
boson pairs, and the production of Higgs bosons through gluon fusion.  

In the future, the description of many more
processes with this combined NLO matching and multijet merging will become
feasible.  This is possible, because {\em both} the \POWHEG and the ME+PS 
part of the implementation are fully automated in \Sherpa.  

We would like to also point out that the methods developed so far will 
naturally serve as a starting point to promote the ME+PS idea to full NLO, 
in the sense that merging sequences of multijet matrix elements at NLO into 
one inclusive sample becomes feasible.  A first attempt to achieve this from 
a somewhat different angle has been presented in~\cite{Lavesson:2008ah}.

\section*{Acknowledgements}
We would like to thank Jennifer Archibald, Tanju Gleisberg, Steffen
Schumann and Jan Winter for many years of collaboration on the \Sherpa project, 
and for interesting discussions.  We are indebted to Daniel Ma{\^i}tre for
support with linking the \BlackHat code and for useful conversation.  We thank
Keith Ellis for his help in interfacing the MCFM library to \Sherpa.
Special thanks go to Thomas Gehrmann, Thomas Binoth and Gudrun Heinrich for 
numerous fruitful discussions on NLO calculations. We thank Thomas Gehrmann
and Keith Hamilton for valuable comments on the manuscript.

SH acknowledges funding by the Swiss National Science Foundation 
(SNF, contract number 200020-126691) and by the University of Zurich 
(Forschungskredit number 57183003).  MS and FS gratefully acknowledge 
financial support by the MCnet Marie Curie Research Training Network 
(contract number MRTN-CT-2006-035606). MS further acknowledges
financial support by the HEPTOOLS Marie Curie Research Training Network 
(contract number MRTN-CT-2006-035505) and funding by the DFG 
Graduate College 1504. FK would like to thank the theory group at CERN, and
MS would like to thank the Institute for Particle Physics Phenomenology in 
Durham, respectively, for their kind hospitality during various stages of 
this project.

\clearpage
\bibliographystyle{bib/amsunsrt_modp}  
\bibliography{bib/journal}
\end{document}